\documentclass{aa}  

\usepackage{natbib}
\usepackage[english]{babel}
\usepackage{graphicx}
\usepackage{txfonts}
\usepackage[normalem]{ulem}
\usepackage{xcolor}
\usepackage{placeins}
\usepackage{caption}
\usepackage{wrapfig}
\usepackage{float}

\bibpunct{(}{)}{;}{a}{}{,} 
\definecolor{links}{rgb}{0, 0, 255}
\usepackage[colorlinks=true,allcolors=links]{hyperref}

\begin{document} 

\title{The VariableTNG project: how baryonic mechanisms shape galaxy properties}

\author{
Kexin Liu\inst{\ref{inst1},\ref{inst2}}
\and Hong Guo\inst{\ref{inst1}}\fnmsep\thanks{Corresponding author; guohong@shao.ac.cn}
\and Luis C. Ho\inst{\ref{inst3},\ref{inst4}}
\and Tao Wang\inst{\ref{inst5},\ref{inst6}}
\and Yikai Liu\inst{\ref{inst1},\ref{inst2}}
\and Volker Springel\inst{\ref{inst7}}  
}

\institute{
Shanghai Astronomical Observatory, Chinese Academy of Sciences, Shanghai 200030, China.\label{inst1} 
\and University of Chinese Academy of Sciences, Beijing 100049, China.\label{inst2} 
\and Kavli Institute for Astronomy and Astrophysics, Peking University, Beijing 100871, China. \label{inst3}
\and Department of Astronomy, School of Physics, Peking University, Beijing 100871, China \label{inst4}
\and School of Astronomy and Space Science, Nanjing University, Nanjing 210093, China. \label{inst5}
\and Key Laboratory of Modern Astronomy and Astrophysics, Nanjing University, Ministry of Education, Nanjing 210093, China \label{inst6}
\and Max-Planck-Institut f\"{u}r Astrophysik, Karl-Schwarzschild-Stra\ss{}e 1, 85740 Garching bei M\"{u}nchen, Germany \label{inst7}
}

\abstract
    {
    We use 50 Sobol-sampled VariableTNG simulations, varying eight subgrid parameters at fixed cosmology and initial conditions, to determine which baryonic processes regulate galaxies and black holes at $z=6$ and $z=8$. We compare the simulations with recent high-redshift measurements of the stellar mass function, star-forming main sequence, stellar and gas-phase mass–metallicity relations, stellar mass–size relation, and black-hole host and accretion properties. The simulations reproduce several broad trends in these observables, although differences remain in gas-phase metallicity, galaxy size, and the most extreme black-hole populations. Given the substantial uncertainties in both the physical modelling and the observational inference of high-redshift galaxy properties, we regard these differences as diagnostic tensions rather than definitive model failures. Random-forest analyses reveal a clear hierarchy in parameter sensitivity. The abundance of low-mass galaxies is regulated primarily by stellar feedback, particularly the supernova temperature $T_{\rm SN}$ and thermal wind fraction $\tau_{\rm w}$, whereas the sensitivity of the scatter in the star-forming main sequence is weaker and redshift dependent. The high-accretion tail of black-hole growth depends on black-hole seeding and feedback parameters, but also on $T_{\rm SN}$, suggesting that stellar feedback indirectly regulates rapid black-hole growth through its impact on the available gas supply. 
    For an independent VTNG-sized volume, the estimated field-to-field scatter in the integrated SMF abundance exceeds the parameter-driven scatter by factors of approximately 2.3 and 1.6 at $z=6$ and $z=8$, respectively. This volume-dependent scatter limits the detectability of the subgrid response in individual small fields, but does not alter the intrinsic response isolated by the fixed-initial-condition experiment. Our controlled experiment identifies stellar feedback as a common physical link between early low-mass galaxy formation and rapid black-hole growth.
    }

\maketitle

\section{Introduction}

Cosmological hydrodynamical simulations have played a central role in shaping our understanding of galaxy formation and evolution across cosmic time. In the local Universe, several fundamental galaxy relations have already been tightly constrained by observations. Wide-field spectroscopic and multi-wavelength surveys such as SDSS, GAMA, and deep multi-band programs like COSMOS2020 have provided precise measurements of the galaxy stellar mass function (SMF) and the star-formation main sequence (SFMS) \citep{Brinchmann2004,Baldry2012,Moustakas2013,Somerville2015,Weaver2023}. These well-established relations serve as fundamental benchmarks for cosmological hydrodynamical simulations. Early studies such as Illustris and EAGLE reproduced the broad shape of the low-redshift SMF and SFMS but exhibited tensions in the high-mass end and quenched fractions relative to observations \citep{Vogelsberger2014,Schaye2015}. In more recent work, modern simulations such as IllustrisTNG and SIMBA have improved agreement at early times by incorporating updated feedback prescriptions and galaxy-quenching physics \citep{Pillepich2018,Dave2019,Donnari2019,Appleby2020}.

Building on these low-redshift constraints, observational efforts have steadily progressed to higher redshifts, extending observational constraints on the SMF and SFMS to $z\gtrsim4$ with deep imaging and spectroscopy \citep{Song2016,Bhatawdekar2019,Stefanon2021,Venturi2024,Weibel2024,Harvey2025}. Recent high-redshift observations, including those from COSMOS-Web and JWST programs, indicate that while the overall shape of these basic relations remains broadly consistent with their low-redshift counterparts, their normalization and number densities evolve significantly toward earlier cosmic times. Accurately reproducing these evolving trends provides a critical test for cosmological hydrodynamical simulations \citep{Laporte2022,Roberts-Borsani2022,Stanton2026}.

The advent of JWST has dramatically advanced observational constraints on galaxy scaling relations in the first billion years. Deep NIRCam imaging and spectroscopy have provided increasingly informative constraints on SMF, SFMS, and the mass–metallicity relation (MZR) at $z \gtrsim 4$--10, although substantial systematic uncertainties remain. SMF studies now probe down to stellar masses of $\sim10^{7.5}$--$10^{8.5}\,M_\odot$ at $z \sim 4$--9, revealing steep low-mass slopes and rapid stellar mass assembly in the early Universe \citep{Navarro-Carrera2024,Weibel2024}. Analyses of the SFMS indicate an approximately linear relation between stellar mass and star formation rate, but with elevated normalization and increased intrinsic scatter compared to lower redshifts \citep{Clarke2024,Simmonds2025}. Meanwhile, direct metallicity measurements from JWST/NIRSpec have extended the MZR to $z \gtrsim 6$, showing generally lower metallicities at fixed stellar mass together with significant object-to-object variation \citep{Heintz2023,Nakajima2023,Curti2024}. High-resolution imaging has further characterised galaxy sizes and morphologies, revealing compact stellar distributions and rapidly evolving size–mass relations at early times \citep{Allen2025,Yang2025}. In parallel, JWST has uncovered a rapidly growing population of accreting black holes and compact red sources at $z \gtrsim 4$--8, including the so-called ``Little Red Dots'', many of which exhibit spectroscopic signatures of AGN activity \citep[see e.g. ][and references therein]{Inayoshi2025}.

While these measurements establish a broad continuity of galaxy scaling relations across cosmic time, some emerging results pose significant challenges for current cosmological simulations. In particular, SMF measurements at $z \gtrsim 6$ suggest a substantial abundance of massive galaxies with stellar masses approaching $10^{10}$--$10^{11}\,M_\odot$, implying extremely rapid stellar mass assembly within a short cosmic time \citep{Labbe2023,Weibel2024}. Although subsequent analyses have revised some stellar mass estimates and emphasized systematic uncertainties in SED modelling and mass-to-light ratios \citep{Stefanon2021,Wang2025}, matching the observed high-mass end would require star formation efficiencies significantly higher than those typically assumed in cosmological simulations. At the same time, the identification of candidate quiescent or rapidly quenched galaxies at $z \gtrsim 5$ indicates that efficient suppression of star formation must also occur within a few hundred Myr of cosmic history \citep{Carnall2023,Valentino2023,Graff2025}, placing strong constraints on feedback prescriptions and gas depletion timescales. The inferred black hole masses and number densities of compact red AGN candidates further intensify debates about black hole seeding mechanisms and early accretion efficiency \citep{Maiolino2024,Schindler2025}. Reconciling rapid stellar mass build-up, early quenching, accelerated black hole growth, and fast chemical enrichment within a unified feedback framework therefore remains a key challenge for cosmological hydrodynamical models of galaxy formation.

More generally, the sensitivity of galaxy properties to feedback prescriptions reflects a fundamental limitation of cosmological hydrodynamical simulations. Baryonic processes such as star formation, stellar feedback, and black hole accretion occur on scales far below the resolution of cosmological volumes and therefore cannot be resolved from first principles. Instead, they are implemented through subgrid prescriptions \citep{Springel2003, Schaye2008, Vogelsberger2013, Weinberger2017, Nelson2019}. As a result, predictions for galaxy growth, internal structure, and chemical enrichment depend not only on the cosmological framework, but also sensitively on the adopted parameterization and efficiency of these unresolved processes. 
In practice, most flagship simulations adopt a calibrated parameter set tuned to reproduce selected low-redshift observables, such as the stellar mass function, galaxy sizes, or black hole scaling relations \citep{Torrey2014,Hopkins2018,Pillepich2018,Dave2019,Kannan2023,Kannan2025thesan}. Even at low redshift, however, the adopted subgrid models and parameter choices can introduce residual systematic biases in simulated galaxy populations, despite their overall calibration to observables \citep{Crain2015}. These systematic differences are expected to become more pronounced at high redshift. 
Consequently, uncertainties in subgrid parameters are important when interpreting galaxy formation at high redshift.

Recognizing this limitation, several recent projects have explored how variations in physical and numerical assumptions affect galaxy populations. 
The CAMELS project was developed primarily to constrain cosmology while jointly sampling cosmological and astrophysical parameters, including stellar and AGN feedback parameters, across large ensembles of simulations with different initial conditions, enabling the joint impact of cosmology and baryonic physics on galaxy and black hole properties to be quantified \citep{Villaescusa-Navarro2021CAMELS,Ni2022Astrid}.
Similarly, the BRAHMA simulation explores variations in black hole seeding prescriptions to investigate how seed formation physics influences the early growth of black holes and their observable signatures \citep{Bhowmick2024}. These controlled parameter studies demonstrate that variations in feedback efficiency or seeding models can lead to substantial differences in stellar mass assembly, quenching histories, and black hole growth. These results highlight that uncertainties in subgrid parameters can produce substantial variations in predicted galaxy populations, potentially comparable to those arising from differences in numerical resolution or simulation volume, underscoring the necessity of systematically exploring the parameter space when interpreting simulation predictions at early cosmic times.

To explore the impact of baryonic physics, we employ the VariableTNG simulation suite, which is built on the IllustrisTNG galaxy formation model using the moving-mesh code AREPO \citep{Springel2010}. In VariableTNG, eight key parameters regulating stellar and AGN feedback are varied simultaneously within a joint parameter space, while all simulations share identical initial conditions and cosmological parameters. This design efficiently explores the feedback parameter space and minimizes degeneracies associated with environmental or cosmological differences. 

To enable a more direct comparison with observations, we generate mock JWST images based on the simulation outputs. Using radiative transfer calculations with SKIRT \citep{skirt920}, we produce synthetic observations that mimic realistic observing conditions. Galaxy structural properties, such as sizes, are then measured directly from these mock images in a manner consistent with observational analyses. This approach provides a broader set of observables that can be directly compared with data, which offer complementary and more direct constraints beyond global properties of galaxies. In this work, we investigate how variations in stellar and AGN feedback prescriptions influence these basic galaxy properties using the VariableTNG simulation suite.

The paper is organized as follows. In Section~2, we introduce the VariableTNG simulations and describe the generation of mock JWST images and the measurement of galaxy sizes. In Section~3, we present the high-redshift galaxy scaling relations predicted by the 50 VariableTNG simulation runs, including the stellar mass function, the star-forming main sequence, relations involving gas-phase metallicity, correlations between galaxy properties and black hole mass or accretion activity, and the galaxy mass-size relation, at $z=6$ and $z=8$. In Section~4, we investigate the impact of variations in baryonic feedback parameters on galaxy evolution, and assess their role in shaping the observed trends at high redshifts. Finally, we summarize our main conclusions in Section~5. In this paper, we adopt a flat $\Lambda$CDM cosmology with $H_{0}=67.32~\mathrm{km\,s^{-1}\,Mpc^{-1}}$.

\section{Methodology}
\subsection{The VariableTNG simulation}
The VariableTNG (VTNG) simulation suite is built upon the cosmological magneto-hydrodynamical moving-mesh code AREPO \citep{Springel2010,Weinberger2020}, the same numerical framework used in the IllustrisTNG project \citep{Marinacci2018,Naiman2018,Springel2018,Nelson2018,Nelson2019}. Although IllustrisTNG successfully reproduces a broad range of low-redshift galaxy observables, recent high-redshift observations suggest that the adopted feedback prescriptions may require further refinement. The VTNG simulations therefore explore systematic variations in eight baryonic feedback parameters—four governing stellar feedback and four related to AGN feedback—in order to quantify their combined impact on galaxy formation and evolution (see Table~\ref{tab:vtng_params}).
\begin{table*}[t]
    \centering
    \caption{Model parameters varied in the \textsc{VTNG} simulations. Values in \textsc{IllustrisTNG} are shown for comparison.}
    \label{tab:vtng_params}
    \renewcommand{\arraystretch}{1.4} 
    \setlength{\tabcolsep}{6pt} 
    \begin{tabular*}{0.9\textwidth}{@{\extracolsep{\fill}} lccc}
        \hline
        \textbf{Parameters} & \textbf{Label} & \textbf{Value in TNG} & \textbf{Range in VTNG} \\
        \hline
        Max star formation timescale (Gyr) & $t^{\ast}_{0}$ & $2.27$ & $(1.0,\,3.0)$ \\
        Supernova temperature (K) & $T_{\mathrm{SN}}$ & $5.73\times10^{7}$ & $(10^{7},\,10^{8})$ \\
        Thermal wind fraction & $\tau_{\mathrm{w}}$ & $0.1$ & $(0.01,\,0.2)$ \\
        Wind velocity factor & $\kappa_{\mathrm{w}}$ & $7.4$ & $(3,\,10)$ \\
        Seed black hole mass ($M_{\odot}/h$) & $M_{\mathrm{seed}}$ & $8\times10^{5}$ & $(5\times10^{5},\,5\times10^{6})$ \\
        Radio feedback factor & $\epsilon_{\mathrm{m}}$ & $1.0$ & $(0.1,\,10)$ \\
        Black hole feedback factor & $\epsilon_{\mathrm{f,high}}$ & $0.1$ & $(0.01,\,1)$ \\
        Quasar threshold & $\chi_{0}$ & $0.002$ & $(0.001,\,0.004)$ \\
        \hline
    \end{tabular*}
\end{table*}

The stellar feedback parameters regulate star formation and wind-driven outflows \citep{Pillepich2018}. Specifically, $t^{\ast}_{0}$ sets the maximum star formation timescale in the Kennicutt–Schmidt relation; $T_{\mathrm{SN}}$ defines the effective temperature of the supernova-heated phase and therefore the thermal energy injected per unit stellar mass formed; $\tau_{\mathrm{w}}$ specifies the thermal fraction of the wind energy; and $\kappa_{\mathrm{w}}$ determines how the wind velocity scales with the local dark matter velocity dispersion, thereby controlling both the wind mass-loading factor and the overall outflow strength.

The AGN feedback parameters describe the seeding, growth, and energy coupling of supermassive black holes \citep{Weinberger2017}. The parameter $M_{\mathrm{seed}}$ specifies the initial black hole seed mass assigned to halos exceeding $5\times10^{10}\,M_\odot/h$. The parameter $\epsilon_{\mathrm{f,high}}$ sets the feedback efficiency in the high-accretion (quasar) mode, quantifying the fraction of accretion energy coupled to the surrounding gas. The parameter $\epsilon_{\mathrm{m}}$ controls the efficiency of kinetic (radio-mode) feedback at low accretion rates, while $\chi_{0}$ defines the Eddington ratio threshold separating thermal and kinetic feedback modes. Together, these parameters regulate the interplay between star formation, gas outflows, and black hole growth. A more detailed description of the parameter space is provided in \citet{Liuyk2026}.

For the main analysis in this work, we use the original 50 parameter sets selected from the Sobol-sampled VTNG design and evolve each realization from $z\approx127$ using identical initial conditions in a periodic box of $50\,\mathrm{Mpc}/h$ with $1024^{3}$ dark matter particles and gas cells. All simulations adopt the Planck 2018 cosmology with parameters $\Omega_{\mathrm{m}}=0.3158$, $\Omega_{\Lambda}=0.6842$, $\Omega_{\mathrm{b}}=0.04939$, and $h=0.6732$. Further technical details of the hydrodynamical solver and subgrid models are described in the IllustrisTNG publications \citep{Weinberger2017,Pillepich2018}.

\subsection{Radiative Transfer procedure}
To enable a direct comparison with JWST observations, we post-process the VTNG simulations using the Monte Carlo radiative transfer code SKIRT \citep[version 9;][]{skirt15a,skirt15b,skirt920}. Stellar particles are assigned SEDs based on population synthesis models, while young stellar particles are treated with birth-cloud templates to account for nebular emission and local dust attenuation.

The three-dimensional dust distribution is constructed from the simulated gas component by applying a redshift-dependent dust-to-metal mass ratio calibrated to high-redshift galaxy simulations. Dust absorption, scattering, and thermal re-emission are treated self-consistently within the SKIRT framework. For galaxies hosting active black holes, AGN emission is included as a central point source and scaled according to the instantaneous black hole accretion rate.

For the size analysis, we construct mock JWST/NIRCam F150W images at $z=6$ and $z=8$ from the resulting SKIRT data cubes. The images are resampled to the CEERS pixel scale, convolved with the corresponding instrumental PSF, and inserted into multiple real CEERS F150W background cutouts. Galaxy sizes are measured using a non-parametric curve-of-growth method. A detailed description of the radiative-transfer setup, dust modelling, AGN templates, and image-processing procedure is provided in Appendix~\ref{sec:appendixA}.

\section{Results}

Before presenting each scaling relation in detail,
we note that all comparisons between the VTNG median relations and the observational measurements are also evaluated quantitatively using a statistical-distance metric. 
The corresponding $\chi^2_{\rm norm}$ values and mean offsets are
discussed alongside each comparison below, while the full definition and summary are provided in Appendix~\ref{app:chi2_results}.

\subsection{Stellar mass function}
\label{sec:SMF}

\begin{figure*}
    \centering
    \includegraphics[width=0.8\textwidth]{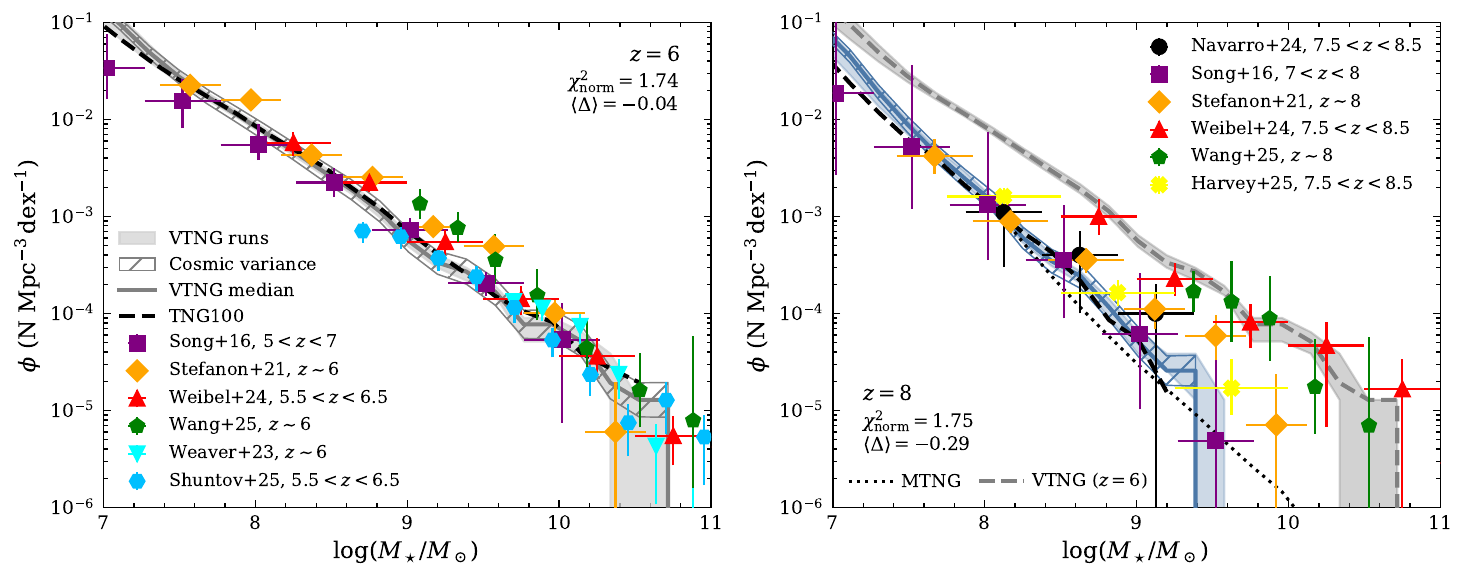}

    \caption{
    Stellar mass functions for the 50 VTNG simulation runs at $z=6$ (left panel) and $z=8$ (right panel). 
    The shaded regions show the 16th--84th percentile range of the SMFs across all runs at fixed stellar mass, while solid lines indicate the median SMF across the 50 VTNG runs.
    The hatched regions show the marginal $1\sigma$ field-to-field cosmic-variance uncertainty in each stellar-mass bin for a VTNG-sized volume, estimated using the lightcone geometry and galaxy-bias model described in Appendix~\ref{app:cosmic_variance}. Cosmic-variance fluctuations are treated as correlated across stellar-mass bins in the statistical-distance calculation. The annotated $\chi^2_{\rm norm}$ and mean offset $\langle\Delta\rangle$ in each panel include this cosmic-variance covariance.
    The black dashed and dotted lines show the SMFs derived from the TNG100 and MillenniumTNG simulations, respectively, and are plotted for reference. In the right panel, the $z=8$ VTNG distribution is shown in blue, with the shaded region indicating the 16th--84th percentile range, while the corresponding $z=6$ distribution is overlaid in grey and its median is shown as a dashed grey line. Coloured symbols represent observational measurements compiled from the literature, including \citet{Song2016}, \citet{Stefanon2021}, \citet{Weaver2023}, \citet{Navarro-Carrera2024}, \citet{Weibel2024}, \citet{Harvey2025}, \citet{Wang2025} and \citet{Shuntov2025}. For consistency, the stellar masses from \citet{Song2016} and \citet{Stefanon2021}, originally reported assuming a Salpeter (\citeyear{Salpete1955}) IMF, are converted to a Chabrier (\citeyear{chabrier03}) IMF by dividing by a factor of 1.7. In each study, only galaxies within the redshift intervals indicated in the legend are included.
    }
    \label{fig:SMF}
\end{figure*}

We begin by comparing the predicted galaxy stellar mass functions (SMFs) in the different VTNG simulation runs with observations at the redshift snapshots of $z=6$ and $z=8$. For each run, stellar masses are taken directly from the simulation outputs and used to compute the SMF using a common set of 21 uniform stellar-mass bins spanning $7 \leq \log(M_\star/M_\odot) < 11.0$, with $\Delta\log(M_\star/M_\odot)=0.19$ dex.
To capture the variation of model predictions, we compute the SMF separately for each of the 50 VTNG runs and summarize the run-to-run distribution at fixed stellar mass. In Fig.~\ref{fig:SMF}, the shaded regions indicate the
16th--84th percentile range across the runs, while the solid
curves show the median SMF across the 50 runs, which we adopt
as the representative VTNG prediction.
This construction quantifies the variation in the SMF arising from different subgrid-parameter choices at fixed initial conditions. Because all VTNG runs share the same realization, this run-to-run scatter does not include field-to-field cosmic variance. We therefore additionally show in Fig.~\ref{fig:SMF} the cosmic-variance uncertainty for a VTNG-sized volume, estimated separately in each stellar-mass bin using the corresponding galaxy bias and an equivalent lightcone geometry, as described in Appendix~\ref{app:cosmic_variance}.

The resulting SMFs at $z=6$ and $z=8$ are compared with recent observational measurements in Fig.~\ref{fig:SMF}. For reference, we also include TNG100 and, at $z=8$, MillenniumTNG \citep{Kannan2023}, to compare the VTNG parameter variations with the original TNG model and to assess the possible impact of simulation volume at the high-mass end. In the $z=8$ panel, the corresponding $z=6$ VTNG distribution is overlaid in grey to illustrate the redshift evolution of the simulated SMF. 
The field-to-field cosmic-variance uncertainty is comparable to or larger than the 16th--84th percentile run-to-run variation over the stellar-mass range. For the integrated abundance over $7.5<\log_{10}(M_\star/M_\odot)<9.5$, the corresponding cosmic-variance scatters are $0.081$ and $0.095$ dex, larger than the parameter-driven scatters by factors of approximately 2.3 and 1.6 at $z=6$ and $z=8$, respectively, as detailed in Appendix~\ref{app:cosmic_variance}.

Overall, the simulated SMF distributions reproduce the broad trends of the current observational estimates at high redshift, although discrepancies remain in detail. For the observational points covered by our per-bin cosmic-variance estimate, including the correlated cosmic-variance contribution reduces $\chi^2_{\rm norm}$ from $2.39$ to $1.74$ at $z=6$ and from $3.67$ to $1.75$ at $z=8$. At the massive end, the marginal cosmic-variance uncertainty reaches approximately $0.15$ dex at $\log(M_\star/M_\odot)\simeq10.0$ at $z=6$ and $0.16$ dex at $\log(M_\star/M_\odot)\simeq9.4$ at $z=8$. At $z=6$, the VTNG predictions are broadly consistent with many of the available observational constraints across the stellar-mass range. At $z=8$, the discrepancy is more localized: the VTNG median remains broadly consistent with the lower-mass measurements, such as those from \citet{Song2016} and \citet{Navarro-Carrera2024}, but lies below several recent JWST-based estimates at the massive end. This is particularly evident in the high-mass measurements of \citet{Weibel2024} and \citet{Wang2025}. Thus, cosmic variance substantially reduces the overall statistical tension, but a residual discrepancy remains toward the massive end.

Limited sampling of intrinsically rare massive galaxies in a $50\,{\rm Mpc}/h$ box may additionally affect the high-mass comparison. Our comparison with the MillenniumTNG (MTNG) simulation, which has a box size of $500 \,{\rm Mpc}/h$ on a side, suggests that even with improved sampling of rare massive systems, the number densities above $\log(M_\star/M_\odot)\sim9$ remain lower than some of the highest estimates from \citet{Weibel2024} and \citet{Wang2025} at $z=8$.
The amplitude of the MTNG is slightly lower than those of the VTNG runs due to the lower resolution of MTNG. We further overlay the $z=6$ VTNG distributions and median relation onto the $z=8$ panel and find that they lie closer to the high-mass end of these measurements at both redshifts. The redshift evolution between $z=6$ and $z=8$ at the massive end inferred from these two measurements is relatively weak. However, there is substantial growth of massive galaxies in the TNG models.

Some of the difference could arise from systematic uncertainties in the observational estimates. At $z \gtrsim 7$, uncertainties in photometric-redshift solutions (see, e.g., Fig.~2 of \citealt{Weibel2024}), stellar-mass estimates, and survey volume may lead to weak evolution in the inferred SMFs. 
Moreover, stellar masses at $z\gtrsim6$ remain sensitive to assumptions in SED modelling, including star-formation histories, dust attenuation, nebular emission, stellar-population templates, photometric-redshift solutions, and the adopted IMF \citep{Steinhardt2023,Pacifici2023}.
These effects are particularly important at the massive end, where small-number statistics \citep{Gehrels1986,BoylanKolchin2010}, cosmic variance \citep{Trenti2008,Moster2011,Jespersen2025_qg,Jespersen2025_cv,Weibel2026_cv,Ji2026PANORAMIC}, and contamination by compact red sources or AGN \citep{Greene2024,Kocevski2025,Kokorev2024} can have a large impact on the inferred SMF.
At $z \gtrsim 8$, MIRI photometry begins to probe blueward of rest-frame $1\,\mathrm{\mu m}$, such that stellar masses for the most massive galaxies may remain overestimated \citep{Papovich2023}.
In this context, even samples that nominally exclude LRDs may remain affected by systematic uncertainties in stellar-mass estimates, as illustrated by the $z=8$ SMF of \citet{Wang2025}, potentially affecting the inferred high-mass end.

\subsection{Star formation main sequence}
\label{sec:sfms}

\begin{figure*}
    \centering
    \includegraphics[width=0.9\textwidth]{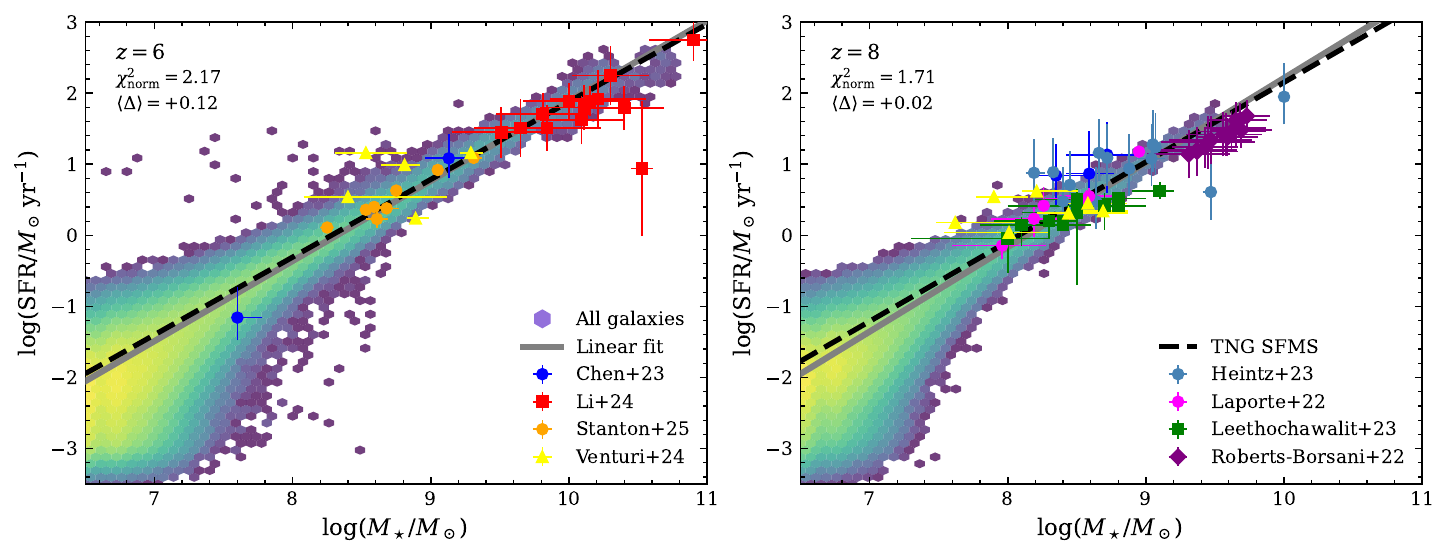}
    \caption{The star-formation main sequence (SFMS) at $z=6$ (left panel) and $z=8$ (right panel). Hexbins show the distribution of all galaxies from the simulations, combined over 50 simulation runs. The solid grey line indicates a linear fit to the simulated SFMS over the mass range $8 < \log(M_\star/M_\odot) < 11$. For reference, the SFMS derived from the TNG100 simulation is overplotted as a black dashed line, using the same fitting mass range. Coloured symbols represent observational measurements selected from the literature within the corresponding redshift intervals for comparison with the simulations. At $z=6$, only galaxies with $4.5 < z < 6.5$ are included, while at $z=8$ we restrict the sample to $7 < z < 9$. The observational data are compiled from multiple studies \citep{Laporte2022,Roberts-Borsani2022,Chen2023,Leethochawalit2023,Li2024,Venturi2024,Stanton2026}.
    The annotated $\chi^2_{\rm norm}$ and mean offset $\langle\Delta\rangle$ quantify the statistical distance between the simulated SFMS and the observational measurements, where $\Delta=\log{\rm SFR}_{\rm sim}-\log{\rm SFR}_{\rm obs}$.
    }
    \label{fig:sfms}
\end{figure*}

Here we examine the star-formation main sequence (SFMS) of galaxies in VTNG runs and compare it with observational constraints at similar redshifts. The SFMS, describing the correlation between stellar mass and star formation rate (SFR) for star-forming galaxies, provides a fundamental benchmark for assessing galaxy growth across cosmic time.

For each simulation run, we follow the methodology of \citet{Liu2025} to define the SFMS as the median SFR in stellar mass bins with a width of 0.2 dex, excluding galaxies with $\log(\mathrm{sSFR}/\,\mathrm{yr}^{-1}) < -11$. The resulting median relation is interpolated to the stellar masses of individual galaxies. A galaxy is classified as star-forming if its SFR lies within 1 dex below the SFMS. All star-forming galaxies from the 50 simulation runs at a given redshift are combined into a single sample. This approach allows us to characterise both the global SFMS and its scatter across the parameter space. 

Figure~\ref{fig:sfms} presents the SFMS predicted by the VTNG simulations at $z=6$ and $z=8$. The hexbins show the distribution of all galaxies combined over the 50 VTNG runs, while the solid
grey line indicates a linear fit to the simulated SFMS over
$8 < \log(M_\star/M_\odot) < 11$. For reference, we overplot the SFMS derived from the TNG100 simulation as a black dashed line, using the same fitting mass range. Observational measurements from the literature are also shown for comparison, with individual studies distinguished by different symbols and colours.

Overall, the simulated SFMS exhibits a well-defined relation with moderate scatter at both redshifts and is broadly consistent with the available observational measurements. At $z=6$, the normalization and slope of the simulated SFMS closely follow most of the observational measurements and the TNG100 relation, especially over the mass range where the observational constraints are densest. At $z=8$, galaxies with $\log(M_\star/M_\odot)\gtrsim 8.5$ populate a dense and continuous sequence that remains consistent with the observed distribution, although the observational datasets show study-to-study differences.

At lower redshifts, the shape of the SFMS at the high-mass end remains debated, with some studies suggesting a flattening trend \citep{Saintonge2016,Li2023}, while others find a more nearly linear relation \citep{Renzini2015}. At higher redshifts, however, the observational constraints remain limited, and the presence of a high-mass flattening is still unclear. In our simulations, we find a mild indication of flattening at the high-mass end ($\log(M_\star/M_\odot)\gtrsim 10$) at $z=6$, but this feature is weak and may be affected by the limited number of massive galaxies. 
Quantitatively, the statistical-distance metric gives $\chi^2_{\rm norm}=2.17$ and 1.71 at $z=6$ and $z=8$, with mean offsets of $+0.12$ and $+0.02$ dex, respectively, indicating that the simulated SFMS broadly follows the observed trend at both redshifts. The full comparison is summarized in Appendix~\ref{app:chi2_results}.
While the SFMS in this work is defined using intrinsic galaxy properties, the existing \textsc{SKIRT} mock-observation pipeline also enables an observationally consistent comparison in colour space. We present this complementary UVJ analysis in Appendix~\ref{app:appendix_uvj}.

\subsection{Mass--metallicity relation}
\label{sec:MZR}
\begin{figure*}
    \centering
    \includegraphics[width=0.9\textwidth]{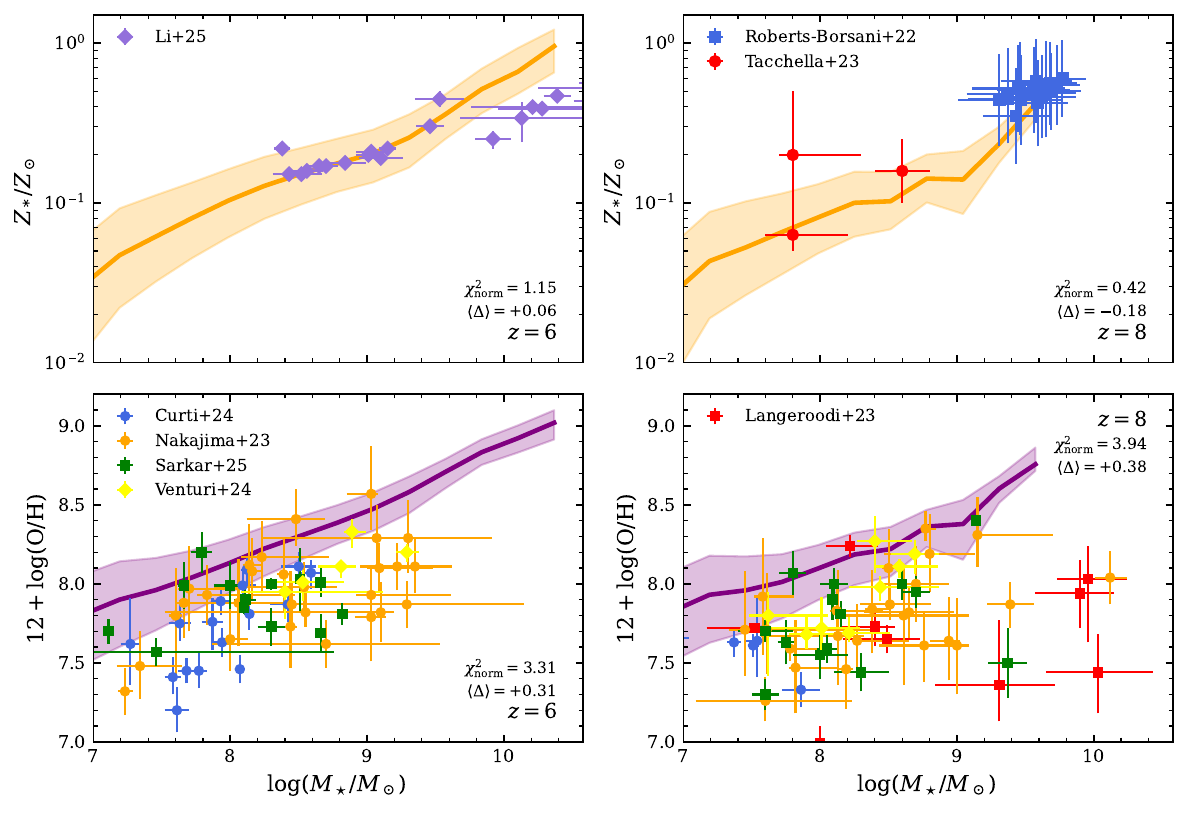}

    \caption{
    Stellar and gas phase metallicity as a function of stellar mass at $z=6$ and $z=8$. The top panels show the stellar metallicity, $Z_*/Z_\odot$, while the bottom panels present $12+\log(\mathrm{O/H})$. 
    Solid curves show the median relations obtained from all simulated galaxies across the VTNG runs, and the shaded regions enclose the 16th--84th percentiles in each stellar-mass bin. 
    Observational constraints are from the literature, including \citet{Roberts-Borsani2022}, \citet{Langeroodi2023}, \citet{Nakajima2023}, \citet{Tacchella2023}, \citet{Curti2024}, \citet{Venturi2024}, \citet{Lirc2026} and \citet{Sarkar2025}.
    The values of $\chi^2_{\rm norm}$ and $\langle\Delta\rangle$ are shown in each panel.
    }
    \label{fig:metal_mass}
\end{figure*}

Figure~\ref{fig:metal_mass} presents the mass--metallicity relations (MZRs) for simulated galaxies at $z=6$ and $z=8$, showing both the mass-weighted stellar metallicity, normalized to $Z_\odot=0.0127$, and the gas-phase oxygen abundance as functions of stellar mass, together with available observational constraints. The stellar metallicity of each galaxy is defined as the mass-weighted metallicity of its stellar particles, where the metallicity of each particle is given by the total metal mass fraction, excluding hydrogen and helium, provided by the simulation.

Observational gas-phase metallicities are typically inferred from nebular emission lines arising from star-forming H\,{\sc ii} regions, and therefore primarily trace the chemical abundance of the star-forming interstellar medium \citep{Torrey2019,Hough2023}. Accordingly, we restrict the calculation to gas cells with non-zero star formation rates and compute the gas-phase oxygen abundance weighted by their SFRs,
\begin{equation}
12+\log_{10}({\rm O/H})
=
12+\log_{10}
\left[
\frac{
\sum_i {\rm SFR}_i
\left(X_{{\rm O},i}/16X_{{\rm H},i}\right)
}{
\sum_i {\rm SFR}_i
}
\right],
\end{equation}
where $X_{{\rm O},i}$ and $X_{{\rm H},i}$ are the oxygen and
hydrogen mass fractions of gas cell $i$, respectively, taken directly from the element abundances tracked by the simulation.

In each panel of Figure~\ref{fig:metal_mass}, the solid curves show the median relations obtained from all simulated galaxies across the VTNG runs. The shaded regions enclose the 16th--84th percentiles of individual galaxies in each stellar-mass bin, and therefore represent the scatter among galaxies in the combined sample rather than the run-to-run variation of the median relations. At both redshifts, the simulations predict a clear positive correlation between metallicity and stellar mass, reflecting the increasing efficiency of metal enrichment in more massive systems. The stellar and gas-phase MZRs show modest evolution between $z=8$ and $z=6$, with a tendency toward lower metallicities at the earlier epoch.

The stellar MZR remains broadly consistent with the available observational constraints, whereas the gas-phase MZR is systematically shifted toward higher metallicities, particularly toward higher stellar masses. Similar behaviour has been reported
in previous comparisons with IllustrisTNG. \citet{Curti2024} found that the IllustrisTNG prediction matches the normalization of their low-mass JWST MZR below $M_\star\sim10^8\,M_\odot$, but predicts a steeper stellar-mass dependence and increasingly diverges from the observed relation toward higher masses. More recently,
\citet{Li2025ASPIRE} also found a higher gas-phase MZR normalization in IllustrisTNG compared with their $z\simeq5$--7 JWST sample.
The absolute normalization of simulated MZRs is sensitive to the
adopted metallicity definition and chemical-enrichment prescriptions, while differences in gas inflows, outflows, and metal retention can further affect the enrichment history
\citep{Torrey2019,Garcia2025,Li2025ASPIRE}.
Observational gas-phase metallicities at these redshifts are also subject to systematic uncertainties associated with nebular metallicity calibrations \citep{Hirschmann2023}. We therefore interpret the offset primarily as a difference in the absolute enrichment, rather than as evidence for a qualitatively different MZR. 
Quantitatively, the gas-phase MZR gives
$\chi^2_{\rm norm}=3.31$ and 3.94, with mean offsets of
$+0.31$ and $+0.38$ dex at $z=6$ and $z=8$, respectively, confirming its systematically high normalization. By contrast, the stellar MZR gives $\chi^2_{\rm norm}=1.15$ and 0.42, with mean
offsets of $+0.06$ and $-0.18$ dex, respectively. The full comparison is summarized in Appendix~\ref{app:chi2_results}.

Beyond the median trends, the simulations predict a substantial scatter in the MZR, particularly toward the low-mass end. 
To clarify the origin of this scatter, we computed the MZR separately for each run and quantified its galaxy-to-galaxy scatter in stellar-mass bins using the 68\% half-width, defined as $w_{68}=(P_{84}-P_{16})/2$. We divided the stellar-mass range into bins of width $0.15$ dex and calculated $w_{68}$ separately for each run in each bin. For each mass bin, we took the median $w_{68}$ across the individual runs and divided it by the $w_{68}$ of the combined galaxy sample in the same bin. We retained mass bins with at least 30 valid runs and averaged the resulting ratios over $7.5 \lesssim \log(M_\star/M_\odot) \lesssim 9.5$. For both the stellar and gas-phase MZRs, the typical within-run scatter amounts to approximately $87\%$ of the combined-sample scatter at $z=6$ and approximately $74\%$ at $z=8$. Thus, most of the total MZR scatter is already present as galaxy-to-galaxy variation within individual simulations.

This behaviour is consistent with earlier analyses of the IllustrisTNG simulations \citep{Torrey2019}, which demonstrated that the scatter in gas-phase metallicity at fixed stellar mass arises from a combination of gas accretion, bursty star formation histories, and feedback-driven outflows. In low-mass systems, where the gravitational potential wells are shallow, variations in recent gas inflow rates and feedback efficiency can lead to rapid dilution or enrichment of the interstellar medium, thereby increasing the dispersion in metallicity.
In this framework, feedback processes do not simply expel metals from galaxies, but regulate the effective metal retention by controlling gas fractions and star formation efficiencies. As a result, the MZR emerges as a statistical, self-regulated relation rather than a strict equilibrium, especially at high redshift where galaxies are rapidly evolving. 

As will be discussed in Section~\ref{sec:discussion}, the MZR diagnostics
explored in our RF analysis do not show robust positive
cross-validated predictive performance. We therefore treat the MZR
primarily as a descriptive comparison with observations rather than
including it in the main RF--SHAP parameter-sensitivity analysis.

\subsection{Mass--size relation}
\label{sec:msr}

\begin{figure*}
\centering
\includegraphics[width=1\textwidth]{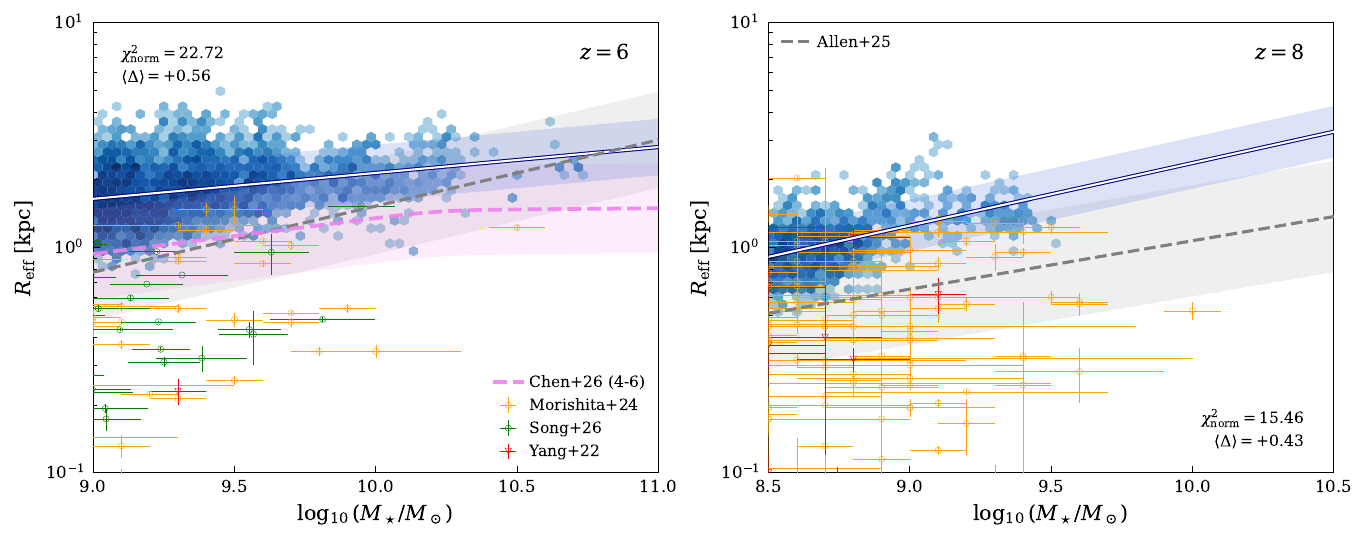}
\caption{
Stellar mass--size relation at $z=6$ (left) and $z=8$ (right). The blue hexagonal density maps show the distribution of galaxies from the 50 VTNG runs, with effective radii measured from mock JWST/NIRCam F150W images using the non-parametric curve-of-growth procedure. Only galaxies with successfully recovered sizes that satisfy all adopted reliability and background-stability criteria are included, accounting for approximately 96\% and 85\% of the input galaxies at $z=6$ and $z=8$, respectively.
The white solid lines show descriptive linear fits to the combined VTNG samples obtained using iterative $3\sigma$ clipping, while the blue shaded regions indicate the robust $1\sigma$ scatter of the residuals about the fits. The fits are included only for visualization.
Open symbols show observational measurements from \citet{Yang2022}, \citet{Morishita2024}, and \citet{Song2026}.
For the $z=6$ panel, observational galaxies are selected over $5 \leq z < 7$, whereas the $z=8$ panel includes galaxies over $7 \leq z < 9$. The grey dashed lines and shaded regions show the median stellar mass--size relations and intrinsic scatter reported by \citet{Allen2025}. In the $z=6$ panel, the purple dashed line and shaded region additionally show the smooth broken-power-law relation and intrinsic scatter for star-forming galaxies at $4 < z < 6$ from \citet{Chen2026}.
The values of $\chi^2_{\rm norm}$ and $\langle\Delta\rangle$ are shown in each panel.
}

\label{fig:MSR}
\end{figure*}

We examine the stellar mass--size relation (MSR) of galaxies at
$z=6$ and $z=8$ using $ \rm R{e}$ measured from mock
JWST/NIRCam F150W images. The mock-image construction and
non-parametric curve-of-growth size-measurement procedure are
described in Appendix~\ref{app:mock_size}. Given the substantial
computational cost of the SKIRT post-processing and mock-image analysis, we select galaxies with $\log(M_\star/M_\odot)>9.0$ at $z=6$ and $\log(M_\star/M_\odot)>8.5$ at $z=8$. Only galaxies with successfully recovered sizes that satisfy adopted reliability and background-stability criteria are included. This selection retains approximately 96\% and 85\% of all galaxies within the selected mass ranges at $z=6$ and $z=8$, respectively. The lower recovery fraction at $z=8$ partly reflects the lower apparent surface brightness of these galaxies, which makes their curve-of-growth measurements more
sensitive to background fluctuations.

Figure~\ref{fig:MSR} presents the resulting MSRs. The hexagonal density maps show the combined galaxy distributions from the 50 VTNG runs. We fit a linear relation to each combined sample using iterative $3\sigma$ clipping to show the overall trend of galaxy size with stellar mass. The fits use galaxies over
$9.0 \leq \log(M_\star/M_\odot) \leq 11.0$ at $z=6$ and $8.5 \leq \log(M_\star/M_\odot) \leq 10.5$ at $z=8$. The blue shaded regions
indicate the robust $1\sigma$ scatter of the residuals about the
fitted relations, estimated from the median absolute deviation.
At both redshifts, the simulated MSRs exhibit a positive stellar-mass dependence, with the median effective radius increasing toward higher stellar masses. Galaxies at $z=8$ are also generally more compact than galaxies of similar stellar mass at $z=6$, consistent with continued structural growth between the two epochs.

For comparison, we include observational measurements from
\citet{Yang2022}, \citet{Morishita2024}, and \citet{Song2026},
together with the median MSRs and intrinsic scatter reported by
\citet{Allen2025}. At $z=6$, we additionally show the smooth
broken power-law relation for star-forming galaxies from
\citet{Chen2026}. Overall, the VTNG relations reproduce the
observed increase in galaxy size with stellar mass. 
At $z=6$, the effective radius continues to increase with stellar mass, although the increase appears relatively weak toward the high-mass end. This behaviour qualitatively resembles the high-mass flattening reported by \citet{Chen2026}. They interpret this bending as an increasing
contribution from compact, massive star-forming galaxies whose size
growth becomes progressively decoupled from halo growth above
$M_\star\sim10^{10}\,M_\odot$, possibly in connection with rapid
bulge and black-hole growth.

The simulated MSR predicts systematically larger effective radii at fixed stellar mass, especially toward the low-mass end. 
The statistical-distance metric presented in Appendix~\ref{app:chi2_results} gives
$\chi^2_{\rm norm}=22.72$ and 15.46 at $z=6$ and $z=8$,
with mean offsets of $+0.56$ and $+0.43$ dex, respectively, confirming that the tension is dominated by a systematic size excess rather than a qualitatively different stellar-mass dependence.

The relationship between rest-frame UV emission and the underlying stellar distribution provides important context for interpreting the size offset. Rest-frame UV light primarily traces recent star
formation, yet observations do not consistently indicate that this
component is more centrally concentrated than the stellar continuum.
At $4.8<z<6.5$, \citet{Matharu2024} find that H$\alpha$ emission is
more extended than the stellar continuum, with an effective-radius
ratio of approximately $1.18$ at fixed stellar mass. This result argues against a general picture in which recent star formation is confined to a more compact region than the stellar continuum, although H$\alpha$ and rest-frame UV emission do not necessarily trace identical spatial distributions.

Meanwhile, \citet{Allen2025} report broadly consistent average rest-frame UV and optical MSRs up to $z\simeq6$, although the UV relation exhibits a larger intrinsic scatter. A significant positive rest-frame optical MSR is also observed for quiescent galaxies at $z\geq3$ \citep{Ito2024}, showing that a positive stellar-mass dependence is present even in systems without substantial ongoing
star formation. Taken together, these observations suggest that the
form of the MSR is not determined solely by the spatial distribution
of recent star formation. Wavelength-dependent light weighting may
modify individual size measurements and their scatter, but is
unlikely by itself to account for the systematic excess of the VTNG
size.

Predictions from other simulations further demonstrate that the
high-redshift MSR is sensitive to the adopted galaxy-formation and
mock-observation models. In FLARES, centrally concentrated dust
attenuation suppresses the intrinsically bright central emission
and can substantially increase the recovered rest-frame UV sizes
\citep{Roper2022}. Mock JWST observations from ASTRID instead
produce a nearly flat MSR at $z\simeq6$ \citep{LaChance2025}. 
In contrast, the high-resolution THESAN-ZOOM simulations predict a positive intrinsic MSR and find H$\alpha$ emission to be systematically more extended than the UV continuum \citep{McClymont2025}. The difference between THESAN-ZOOM and several large-volume simulations has been linked to the treatment of the multiphase interstellar medium and bursty star formation. These contrasting predictions indicate that current high-redshift simulations have not yet reached a consistent picture of either the shape of the MSR or the spatial distribution of recent star formation.
Some simulations predict a relatively flat MSR and centrally
concentrated star formation, whereas others find a clear positive MSR
and more extended star-forming regions. These differences suggest that
the predicted MSR slope and absolute galaxy sizes are sensitive to the
adopted treatments of the ISM, feedback, dust, and radiative transfer.

Finite numerical resolution may additionally contribute to the elevated sizes toward the low-mass end. In the VTNG simulations,
star particles adopt a comoving gravitational softening length of
$\epsilon_{\rm com}=1.25\,{\rm ckpc}/h$. The corresponding physical
softening length is $\epsilon_{\rm phys}=\epsilon_{\rm com}/[h(1+z)]$, giving approximately $0.27$ kpc at $z=6$ and $0.21$ kpc at $z=8$.
In the common stellar-mass interval $9.0\leq\log(M_\star/M_\odot)<9.5$, the median effective radii are $1.72$ and $1.45$ kpc at $z=6$ and $z=8$, respectively. These values correspond to approximately $6.5$ and $7.0$ times the physical softening length. The measured sizes are therefore not directly set by the softening scale. Nevertheless, finite force and mass resolution may still affect the internal structure of low-mass galaxies. In particular, spurious energy transfer from dark matter to stellar particles can artificially increase galaxy sizes below a resolution-dependent mass scale \citep{Ludlow2019,Ludlow2023}. Such numerical heating may contribute to the low-mass size excess, although the present comparison does not constitute a numerical-convergence test. Dedicated higher-resolution simulations would be required to quantify its contribution.

The elevated VTNG sizes may therefore reflect a combination of model-dependent galaxy structure, finite numerical resolution, and differences between the simulated and observational recovery of light-weighted sizes. At these redshifts, the available observational samples remain limited, while the inferred stellar masses, effective radii, and redshifts are subject to substantial uncertainties. These limitations complicate a more detailed interpretation of the size offset.
We therefore present the MSR primarily as a descriptive comparison
with observations and do not apply the RF analysis to investigate
its dependence on the varied subgrid parameters. The relative
contributions of physical modelling, numerical resolution, and
measurement uncertainties cannot be determined from the present
analysis.

\subsection{Black hole -- stellar mass relation}
\label{sec:bh_mass}

Recent JWST observations have revealed a population of compact, red sources at high redshift, often referred to as Little Red Dots (LRDs), along with a growing number of candidate AGN at $z \gtrsim 6$ \citep{Harikane2023, Kokorev2024}. Many of these objects exhibit broad emission lines and high inferred luminosities, suggesting the presence of actively accreting black holes with substantial masses \citep{Juoddzbalis2024, Maiolino2024}. However, their physical nature remains uncertain. In particular, the interpretation of broad-line widths as virial motions has been questioned, as electron scattering may contribute significantly to the observed line profiles, potentially leading to overestimated black-hole masses by up to $\sim 2$ dex \citep{Rusakov2025}. In addition, the lack of X-ray detections and the absence of strong variability in several samples challenge their classification as typical unobscured AGN \citep{Kokubo2025, Tee2025}. These studies highlight that current inferences of black-hole masses and accretion rates at these redshifts are subject to significant uncertainties.

\begin{figure*}
    \centering
    \includegraphics[width=0.9\textwidth]{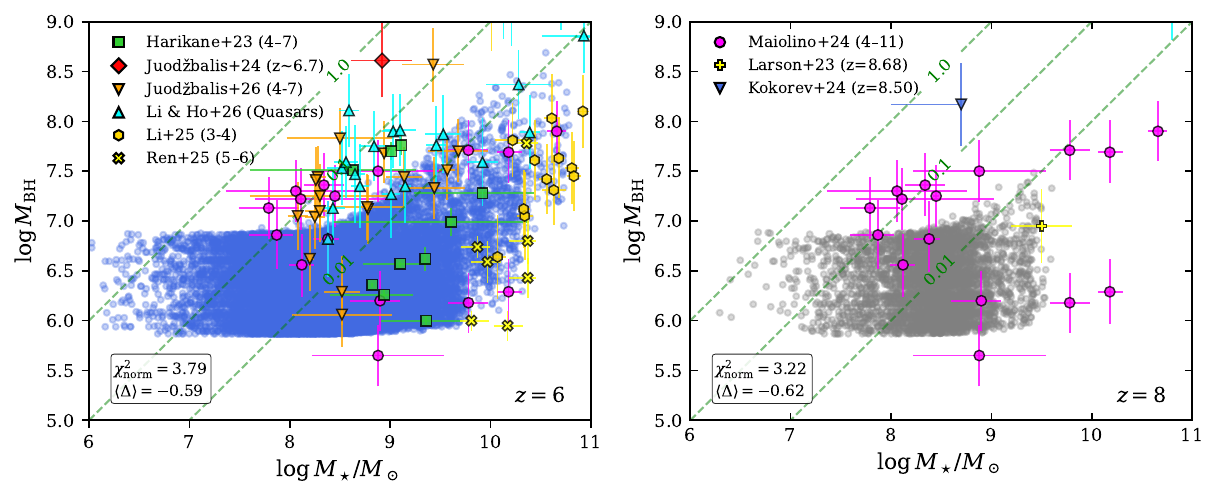}
    \caption{
    Black hole mass as a function of host-galaxy stellar mass at $z=6$ (left) and $z=8$ (right). Individual points show simulated galaxies hosting black holes from all 50 VTNG runs at each redshift. Observational measurements from the literature are overplotted with different symbols and colours, as indicated in the legend. The $z=6$ panel includes broad-line AGN and quasar samples from \citet{Harikane2023}, \citet{Juoddzbalis2024}, \citet{Maiolino2024}, \citet{Lirc2026}, \citet{Lijy2025}, \citet{Ren2025} and \citet{Juodzbalis2026}, while the $z=8$ panel additionally includes AGN at $z>8$ from \citet{Larson2023} and \citet{Kokorev2024}. Dashed green lines denote constant black hole--to--stellar mass ratios,
    $M_{\rm BH}/M_\star = 1$, $0.1$, and $0.01$.
    The values of $\chi^2_{\rm norm}$ and $\langle\Delta\rangle$ are shown in each panel.}
    \label{fig:BH_SMR}
\end{figure*}

Figure~\ref{fig:BH_SMR} presents the relation between black hole mass and stellar mass for simulated galaxies hosting a black hole from 50 VTNG runs at $z=6$ and $z=8$, compared with recent JWST observations. At both redshifts, simulated galaxies populate a broad region in the $\log M_{\rm BH}$--$\log M_\star$ plane, spanning more than one order of magnitude in black hole mass at fixed stellar mass. The majority of the simulated systems lie below the one-to-one relation, with typical mass ratios of $M_{\rm BH}/M_\star$ around $10^{-2}$.

\begin{figure*}
    \centering
    \includegraphics[width=0.9\textwidth]{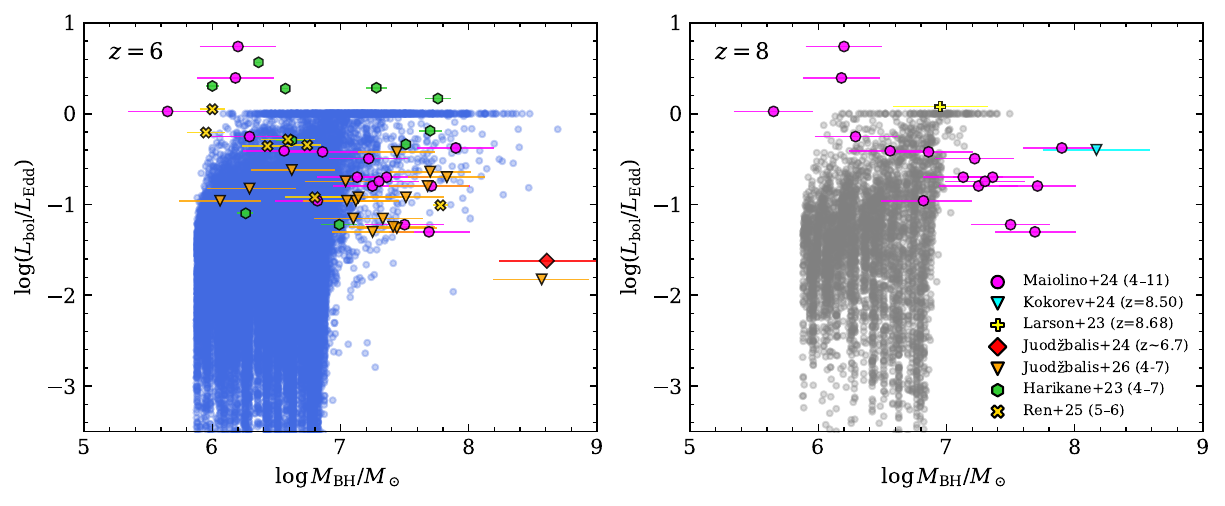}
    \caption{
    The relation between black hole mass and Eddington ratio, $\log(L_{\rm bol}/L_{\rm Edd})$, at $z=6$ (left) and $z=8$ (right). Owing to the heterogeneous and AGN-selected nature of the observational samples, this figure is treated as a descriptive comparison and is not assigned a $\chi^2_{\rm norm}$ value.
    }
    \label{fig:BH_Lbol}
\end{figure*}

At $z=6$, the observational estimates compiled from \citet{Harikane2023}, \citet{Juoddzbalis2024} and \citet{Maiolino2024} overlap substantially with the upper envelope of the simulated distribution. Several observational samples exhibit elevated black hole masses relative to their host stellar masses, with $M_{\rm BH}/M_\star$ exceeding $\sim 10^{-2}$. Such systems become rare as the ratio increases within the simulated population. A similar trend is seen at $z=8$. A small number of JWST-detected AGN at $z \gtrsim 8$ \citep{Larson2023,Kokorev2024} occupy regions of parameter space characterised by high black hole masses relative to their host stellar masses. However, given the systematic uncertainties discussed above, these objects should be interpreted as possible extreme cases rather than definitive evidence for a large population of over-massive black holes.
 
The observed high-redshift AGN samples do not represent an unbiased galaxy population, but are selected from actively accreting systems. In the simulations, these observed systems generally overlap with the upper envelope of the black-hole population. For galaxies with $M_\star \lesssim 10^9\,M_\odot$, black hole masses typically remain close to their seed values and exhibit only limited growth. Above this stellar-mass scale, black holes can undergo more rapid growth, giving rise to the increase in $M_{\rm BH}$ seen in the simulated relation. This behaviour stems from the black-hole growth prescription adopted in the model, where the gas accretion rate follows a Bondi-like scaling $\dot{M}_{\mathrm{BH}} \propto M_{\mathrm{BH}}^{2}$ \citep{Weinberger2017}. As a result, black holes in low-mass galaxies tend to remain close to their seed masses until efficient accretion begins. In the VTNG suite, seed masses of $5\times10^{5}$--$5\times10^{6}\,M_\odot/h$ imply that newly seeded black holes initially accrete relatively inefficiently, delaying rapid black-hole growth at early cosmic times. 
Quantitatively, the $M_{\rm BH}$--$M_\star$ relation gives $\chi^2_{\rm norm}=3.79$ and 3.22 at $z=6$ and $z=8$, with mean offsets of $-0.59$ and $-0.62$ dex, respectively, indicating that the VTNG median lies systematically below the compiled
high-redshift AGN measurements at fixed stellar mass. The full comparison is presented in Appendix~\ref{app:chi2_results}.

We also examine the relation between black-hole mass and bolometric luminosity, expressed as the logarithmic Eddington ratio, as shown in Figure~\ref{fig:BH_Lbol} for $z=6$ and $z=8$. The Eddington ratio is defined as $\log(L_{\rm bol}/L_{\rm Edd})$, where the bolometric luminosity is calculated from the instantaneous black hole accretion rate following Equation~(\ref{eq:Lbol}), and the Eddington luminosity is given by $L_{\rm Edd}=1.26\times10^{38}(M_{\rm BH}/M_\odot)\,{\rm erg\,s^{-1}}$. In simulations, black holes span a wide range of accretion states, with the majority accreting at relatively modest Eddington ratios, typically $\log(L_{\rm bol}/L_{\rm Edd}) \lesssim -1$ at both redshifts. 

Current observationally inferred AGN candidates at $z \gtrsim 7$, including sources from CEERS and UNCOVER, tend to occupy the high-accretion part of this diagram. For black holes of comparable mass, some of these sources have Eddington ratios higher than those commonly realised in the simulations, and in some cases exceeding the Eddington limit. This difference may be attributed to the accretion prescription inherited from the TNG model \citep{Weinberger2017}, where the black hole accretion rate is capped at the Eddington limit ($\lambda_{\rm Edd} \le 1$), and super-Eddington accretion is therefore not permitted.
However, these observational Eddington ratio estimates depend directly on highly uncertain black-hole masses, bolometric correction scalings, and the robustness of AGN source classification. We therefore do not interpret this comparison as a definitive model discrepancy. Instead, it indicates that the most extreme accretion rates are not commonly produced within the present VTNG model assumptions.

In total, the distributions shown in Figure~\ref{fig:BH_SMR} and Figure~\ref{fig:BH_Lbol} indicate that, within the explored VTNG parameter space, both massive black holes and rapidly accreting systems remain rare. Despite spanning a broad range of black hole seed masses, the simulations do not commonly produce the most extreme black-hole masses and accretion rates inferred in some recent high-redshift AGN studies. However, if some black-hole masses are substantially overestimated, or if a fraction of LRDs are not ordinary AGNs, the level of tension would be correspondingly reduced.

In this context, the recent work of \cite{Chaikin2026} has highlighted the potential importance of super-Eddington black hole accretion at high redshifts. Based on controlled simulations with the COLIBRE model, they showed that models allowing super-Eddington growth can more efficiently build up black hole mass at early times and reproduce the observed abundance of massive quiescent galaxies at $z \gtrsim 6$.
Taken together with our results, this suggests that, if the most extreme high-redshift black-hole masses and accretion states are confirmed, variations within the current VTNG parameter space alone may not be sufficient to produce such systems. In that case, rather than invoking larger seed masses alone, processes that enable more rapid early black hole growth, such as super-Eddington accretion, may need to be considered. Incorporating such mechanisms could help to account for the most extreme currently inferred massive black hole and quiescent galaxy populations at high redshift, while remaining subject to future observational confirmation.

\section{Discussion}
\label{sec:discussion}
\begin{figure*}
\centering
\includegraphics[width=1\textwidth]{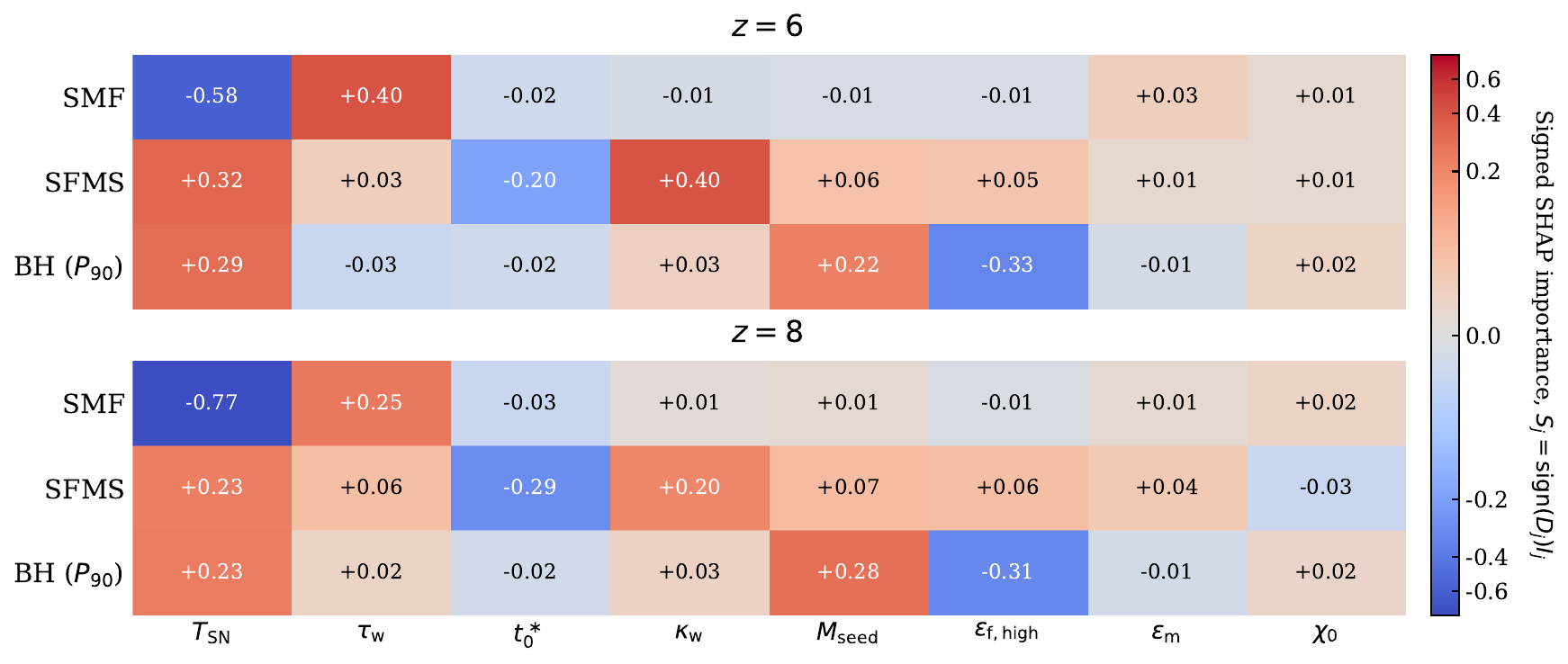}
\caption{
    SHAP-based parameter sensitivity of the eight model parameters for three diagnostics: the integrated number density of the stellar mass function (SMF), the scatter in the star-forming main sequence (SFMS), and the $90^{\rm th}$ percentile of the black-hole Eddington-ratio distribution, $P_{90}$. 
    The SMF diagnostic characterises the normalization of the SMF through its integrated number density, whereas the SFMS diagnostic characterises the scatter about the main sequence rather than its normalization. The upper and lower panels correspond to $z=6$ and $z=8$, respectively. 
    The annotated numbers show the signed SHAP importance, $S_j={\rm sign}(D_j)I_j$. Positive values indicate that larger values of a parameter are associated with higher RF predictions, whereas negative values indicate the opposite trend. The colours visualize the same signed quantity, with red and blue indicating positive and negative effects, respectively. The colour scale is stretched using an inverse-hyperbolic-sine transformation for visualization.
}

\label{fig:RF_shap}
\end{figure*}

The results presented above show that the VTNG simulations reproduce the main observational trends of several galaxy and black-hole scaling relations at high redshifts. The systematic variation of the eight subgrid parameters provides a controlled framework for exploring how changes in baryonic physics translate into measurable signatures in these relations. By sampling a controlled set of the subgrid parameter variations while keeping the numerical setup fixed, the VTNG suite makes it possible to assess how variations in feedback strength, star formation regulation, and black hole growth affect galaxy scaling relations and other statistical properties of galaxy populations.

In principle, the observed galaxy scaling relations can be used to infer the best-fitting physical parameters in the subgrid model. 
To quantify these parameter--observable connections, we use a Random Forest (RF) analysis. For each diagnostic, the RF model takes the eight subgrid parameters as input features and the corresponding summary statistic as the target variable. The role of the RF model is therefore to test whether a given galaxy or black-hole diagnostic is predictable from the explored subgrid parameters. 
The VTNG design aims to have 100 Sobol-sampled parameter sets. The main analysis uses the original 50 simulations for which a complete and internally consistent set of diagnostics, including the SKIRT post-processing, was available. As additional hydrodynamical runs were completed, an enlarged 70-run sample was used in Appendix~\ref{app:rf_robust} to test the robustness of the RF--SHAP results. The Sobol design provides a more space-filling sampling of the explored parameter ranges than purely random sampling and reduces strong clustering in parameter space. Nevertheless, the RF training set remains sparse relative to the dimensionality of the eight-parameter space, and the RF models should therefore be interpreted as tools for identifying broad parameter sensitivities
rather than as high-precision emulators.
We quantify the robustness of each learned relation using five-fold cross-validation and interpret the SHAP rankings only when the RF model achieves positive and reasonably stable cross-validated performance.
Before training each RF model, we rescale the target diagnostic by its run-to-run standard deviation,
\begin{equation}
    \tilde{y}_i = \frac{y_i}{\sigma_y},
\end{equation}
where $\sigma_y$ is the standard deviation of the diagnostic $y$ across the simulation runs. This normalization expresses the RF target relative to its run-to-run variation.

We then use SHAP to interpret the predictions of the trained RF
models. For simulation run $i$, the local SHAP value $\phi_{ij}$ quantifies the contribution of parameter $j$ to the RF prediction relative to the mean model prediction. Because the target has already been normalized by $\sigma_y$, the SHAP values directly measure parameter contributions relative to the run-to-run scatter of the corresponding diagnostic. We characterise the global influence of each parameter using the mean absolute SHAP value,
\begin{equation}
    I_j = \frac{1}{N_{\rm run}}
    \sum_{i=1}^{N_{\rm run}} |\phi_{ij}|.
\end{equation}
Thus, $I_j$ represents the typical amplitude of the RF contribution associated with parameter $j$ relative to the run-to-run standard deviation of that diagnostic.

To describe the direction of each dependence, we compare the mean SHAP contribution in the upper and lower thirds of the marginal distribution of the corresponding parameter,
\begin{equation}
    D_j =
    \langle \phi_{ij} \rangle_{\rm high}
    -
    \langle \phi_{ij} \rangle_{\rm low}.
\end{equation}
A positive value of $D_j$ indicates that larger values of the
parameter are associated with higher RF predictions than smaller values, whereas a negative value indicates the opposite trend.
Because the sampled parameters are varied simultaneously and the VTNG runs share the same cosmological realization, $D_j$ should be interpreted as a descriptive indication of the direction learned by the RF model, rather than as a strict causal response to varying one parameter at a time.

In Fig.~\ref{fig:RF_shap}, the numbers in each cell show the signed SHAP importance, $S_{j}={\rm sign}(D_j) I_j$.
For visualization, the colours are displayed using an inverse-hyperbolic-sine stretch, so that weak but non-zero effects remain visible without allowing the strongest parameters to dominate the full dynamic range. The $|S_j|=I_j$ measures the typical SHAP contribution relative to the run-to-run standard deviation of the diagnostic, while the sign indicates whether larger parameter values tend to increase or decrease the RF prediction.

We first apply this method to the stellar mass function, using the number density of galaxies in the stellar-mass range $7.5 \leq \log(M_\star/M_\odot) < 9.5$, where the SMF is well sampled across all simulation runs. 
Variations in the eight parameters lead to substantial changes in the abundance of low-mass galaxies, and the RF model predicts this abundance with stable cross-validated performance of $R^2 = 0.59 \pm 0.20$ at $z=6$ and $R^2 = 0.80 \pm 0.08$ at $z=8$.
This indicates that the RF model provides a reliable description of the SMF response to changes in subgrid physics, with the higher $R^2$ at $z=8$ suggesting a tighter parameter dependence at earlier times.

The SHAP analysis shows that the SMF is mainly regulated by the supernova feedback temperature, $T_{\rm SN}$, with the thermal wind fraction, $\tau_{\rm w}$, playing a secondary role. 
At both redshifts, $T_{\rm SN}$ has the largest SHAP importance among all parameters for the SMF. The SHAP directions, as quantified by the signs of $D_j$, indicate that increasing $T_{\rm SN}$ tends to lower the predicted SMF number density, while increasing $\tau_{\rm w}$ tends to raise it. This is consistent with the physical expectation that stronger supernova heating suppresses the formation and survival of low-mass galaxies by making stellar feedback more effective at removing or heating gas. The positive dependence on $\tau_{\rm w}$ reflects the way in which variations in the wind-energy partition and recycling efficiency modulate the amount of gas available for continued star formation. The remaining parameters have much smaller SHAP importance, indicating that black-hole-related processes do not directly control the abundance of galaxies in this stellar-mass range within the explored parameter space. Overall, these results demonstrate that stellar feedback dominates the regulation of the high-redshift SMF, while black hole physics plays a negligible role for the low-mass galaxy abundance considered here.

Because all VTNG runs share the same initial conditions, the RF--SHAP analysis isolates the response of the galaxy population to changes in the subgrid parameters at fixed large-scale structure. This controlled design holds field-to-field fluctuations fixed by construction.
We therefore compare the parameter-driven variation with the expected cosmic variance for the same integrated SMF abundance used as the RF target, $7.5<\log(M_\star/M_\odot)<9.5$. The cosmic-variance scatter is $0.081$ dex at $z=6$ and $0.095$ dex at $z=8$, larger than the parameter-driven scatter by factors of approximately 2.3 and 1.6, respectively, as discussed in Appendix~\ref{app:cosmic_variance}.

To test the impact of cosmic variance on the RF predictability, 
we generate 200 Monte Carlo realizations of the integrated SMF abundance by drawing stochastic field-to-field fluctuations from a negative-binomial model with the expected cosmic-variance amplitude, and repeat the RF analysis for each realization using the same five-fold cross-validation procedure as in the fiducial analysis. The median mean cross-validated $R^2$ decreases to $-0.185$, with a 16th--84th percentile range of $[-0.325,-0.039]$, at $z=6$, and to $-0.024$, with a range of $[-0.202,0.099]$, at $z=8$. Thus, once field-to-field fluctuations appropriate for an independent VTNG-sized volume are included, the parameter-driven SMF signal is no longer robustly recoverable from the integrated abundance alone. 
This does not invalidate the controlled comparison among the VTNG runs. Rather, it indicates that the SHAP-inferred SMF response should be interpreted as the intrinsic response to subgrid-parameter variations at fixed initial conditions, while cosmic variance may make this signal difficult to recover when applying these small-volume simulations as emulators to fit the observed SMFs. This limitation is volume dependent and is expected to become less important for larger survey volumes, where the contribution from cosmic variance is reduced.

While the SMF probes the abundance of galaxies, the scatter in the SFMS provides complementary information on the stability of star formation at fixed stellar mass. We quantify the scatter in the SFMS as $\sigma_{\mathrm{MS}} = \mathrm{std}\!\left(\log \mathrm{SFR} - \log \mathrm{SFR}_{\mathrm{MS}}\right)$, where $\log {\rm SFR}_{\rm MS}$ is defined by a linear fit to the star-forming population in stellar-mass bins. The scatter is measured over the stellar-mass ranges $8.0 < \log(M_\star/M_\odot) < 9.5$ at $z=6$ and $8.0 < \log(M_\star/M_\odot) < 9.0$ at $z=8$, ensuring both sufficient statistics and a robust estimate of $\sigma_{\rm MS}$ in each simulation run. Across the 50 VTNG runs, the median scatter in the SFMS is 
$\sigma_{\rm MS}=0.131^{+0.012}_{-0.010}$ dex at $z=6$ and
$\sigma_{\rm MS}=0.108^{+0.008}_{-0.008}$ dex at $z=8$, where the uncertainties denote the run-to-run 16--84 percentile range.

Applying the same RF and SHAP analysis to $\sigma_{\rm MS}$, we find a strong redshift dependence in how well this quantity can be predicted from the explored subgrid parameters.
At $z=6$, the RF model yields $R^2 = 0.58 \pm 0.12$, indicating that variations in the subgrid parameters significantly regulate the burstiness of star formation. 
At $z=8$, however, the RF model shows weaker cross-validated results, with $R^2 = -0.03 \pm 0.50$, consistent with no robust predictive performance. This implies that the SFMS scatter at $z=8$ is not reliably predictable from the explored subgrid parameters. Consistent with the reliable $z=6$ prediction, the SHAP importance at $z=6$ is dominated by the wind velocity factor $\kappa_{\rm w}$ and $T_{\rm SN}$, with the maximum star-formation timescale $t_0^\ast$ also contributing. The SHAP directions indicate that larger $T_{\rm SN}$ and larger $\kappa_{\rm w}$ tend to increase $\sigma_{\rm MS}$, whereas larger $t_0^\ast$ tends to reduce it.

This behaviour supports a physical picture in which stronger stellar feedback and more efficient winds drive burstier star-formation histories. A larger $\kappa_{\rm w}$ increases the efficiency of stellar-driven outflows, temporarily removing or heating star-forming gas and suppressing star formation, followed by later gas re-accretion that can trigger renewed star-formation episodes. This feedback-driven cycle naturally increases the scatter of the star-formation main sequence. Similarly, a higher $T_{\rm SN}$ leads to more energetic supernova feedback and a higher gas-density threshold for star formation \citep{Springel2003}, causing more violent regulation of dense star-forming gas and increasing short-timescale fluctuations in the star-formation rate. In contrast, a larger $t_0^\ast$ effectively lengthens the gas consumption timescale and smooths the conversion of gas into stars, thereby reducing short-timescale variability and lowering $\sigma_{\rm MS}$. 
The weaker RF performance at $z=8$ indicates that the run-to-run variation in $\sigma_{\rm MS}$ is not robustly predicted by the explored subgrid parameters. At $z=8$, the SFMS sample is dominated by smaller and less well-resolved galaxies, for which instantaneous SFRs and offsets from the main sequence are more susceptible to particle discreteness, stochastic star formation, and short-timescale feedback cycles. This is consistent with previous studies of galaxy-to-galaxy chaos and predictability in IllustrisTNG \citep{2019Genel_scaling,Chuang2024}. We therefore present the $z=8$ SFMS-scatter ranking for completeness only and focus our physical interpretation on the better-constrained $z=6$ result.

To quantify the response of black hole accretion to variations in subgrid physics, we adopt the $90^{\rm th}$ percentile of $\log(L/L_{\rm Edd})$, denoted as $P_{90}$, as a tracer of the high-accretion tail of the black hole population. By construction, $P_{90}$ characterises the typical accretion level reached by the most rapidly accreting $\sim 10\%$ of black holes in each simulation run. The RF model predicts this diagnostic reliably, with $R^2 = 0.59 \pm 0.12$ at $z=6$ and $R^2 = 0.56 \pm 0.14$ at $z=8$.

The SHAP analysis shows that $P_{90}$ is primarily regulated by three parameters at both redshifts: the high-accretion black hole feedback factor $\epsilon_{\rm f,high}$, the seed black hole mass $M_{\rm seed}$, and $T_{\rm SN}$. At $z=6$, $\epsilon_{\rm f,high}$ and $T_{\rm SN}$ have comparable SHAP importance, followed by $M_{\rm seed}$. At $z=8$, $\epsilon_{\rm f,high}$ and $M_{\rm seed}$ become the two most important parameters, with $T_{\rm SN}$ remaining the dominant stellar-feedback contribution. 
The SHAP directions indicate that $P_{90}$ tends to increase with increasing $T_{\rm SN}$ and $M_{\rm seed}$, but tends to decrease with increasing $\epsilon_{\rm f,high}$. The positive dependence on $M_{\rm seed}$ is expected, since larger initial black hole masses allow black holes to reach higher accretion luminosities more easily at early times. The negative dependence on $\epsilon_{\rm f,high}$ indicates that stronger high-accretion feedback suppresses the upper tail of the Eddington-ratio distribution by more efficiently regulating gas supply in the vicinity of rapidly growing black holes. The positive dependence on $T_{\rm SN}$ suggests that stellar feedback also affects the high-accretion tail indirectly, either by changing the gas supply available to black holes or by altering the stellar-mass normalization against which black hole growth is compared. In contrast, parameters associated with galactic winds, radio-mode feedback, and feedback-mode transitions exhibit much smaller SHAP importance. These results suggest that the most luminous black hole accretion episodes are jointly regulated by the interplay between early gas regulation by stellar feedback, the initial black hole seed mass, and the strength of black hole feedback during high-accretion phases.
Given that the growth of massive black holes is closely linked to AGN feedback and galaxy quenching, related challenges have also been reported in other simulations. Using FLARES, \citet{Turner2026} performed targeted searches for massive quiescent galaxy analogues at $z\gtrsim7$ and analysed their formation histories. They found that such systems are extremely rare or absent, suggesting that their formation remains difficult to achieve within current simulation models. A similar behaviour is present in our results, as we do not find a significant population of massive quiescent galaxies in the SFR distributions at $z=6$ and $z=8$ (Fig.~\ref{fig:sfms}).

Unlike the scalar diagnostics considered above, the MZR and MSR are not included in our main RF--SHAP interpretation. For the MZR, we explored several diagnostics, including values of $12+\log({\rm O/H})$ at fixed stellar mass, the slope of the relation, and global offsets relative to the median MZR. 
In all cases, the RF models show no robust positive cross-validated predictive performance for these diagnostics.
This lack of predictability may reflect the physical picture that metallicity at fixed stellar mass is regulated by the coupled effects of metal production, gas inflows, and metal-rich outflows, whose relative importance depends sensitively on recent accretion and star-formation histories.
For the MSR, we do not perform a dedicated RF--SHAP analysis, as discussed in Section~\ref{sec:msr}. Instead, the combined sample from the 50 VTNG runs is presented as a descriptive comparison with observations. The simulated relation reproduces the positive stellar-mass dependence of the observed MSR but predicts
systematically larger effective radii at fixed stellar mass.
We therefore present the MZR and MSR primarily as descriptive comparisons, without drawing parameter-sensitivity conclusions from either relation.

It is useful to place our RF--SHAP results in the context of the recent \textsc{sapphire} framework of \citet{Pandya2026}, which characterises the sensitivity of galaxy scaling relations to astrophysical parameters using a differentiable semi-analytic model. Although \textsc{sapphire} focuses on a lower-redshift semi-analytic model, it independently identifies supernova energy loading as a key parameter regulating galaxy evolution. This is qualitatively consistent with our finding that the supernova feedback parameter $T_{\rm SN}$ is the dominant driver of the high-redshift SMF response. Physically, both results point to the same conclusion: stellar feedback controls the efficiency with which gas is converted into stars, and therefore strongly regulates the normalization and scatter of galaxy scaling relations. In our simulations, this regulation is seen most clearly in the abundance of low-mass galaxies at $z=6$--$8$, while the impact on the SFMS scatter is robust only at $z=6$. The comparison therefore suggests that the dominant role of supernova feedback is not specific to a particular modelling framework, but reflects a more general feature of galaxy formation models. In contrast, our RF--SHAP analysis is applied directly to a suite of cosmological hydrodynamical simulations at $z=6$--$8$, allowing us to capture non-linear baryonic processes in a fully numerical setting and to connect the parameter response to high-redshift observables. The drawback is that our approach is limited by the finite number of simulations, the non-orthogonality of the sampled parameter set, and the use of a single cosmological realization. Therefore, our SHAP directions should be interpreted as descriptive trends of the RF-learned model surface rather than as exact causal derivatives. Nevertheless, the agreement between these independent approaches strengthens the conclusion that supernova feedback is a key regulator of galaxy scaling relations.

Taken together, the RF--SHAP analysis reveals a hierarchy in how subgrid physics regulates different galaxy and black hole diagnostics. The abundance of low-mass galaxies, as quantified by the integrated SMF number density, is primarily controlled by supernova feedback, indicating that stellar feedback sets the overall efficiency of galaxy formation. The scatter of the SFMS, which traces the burstiness of star formation at fixed stellar mass, is sensitive to both the strength of stellar feedback and the efficiency and variability of stellar-driven winds, although this dependence is robust only at $z=6$ in our current analysis. As expected, the impact of black hole feedback in our analysis of $P_{90}$ is primarily seen in the high-accretion regime. 

\section{Conclusion}

In this work, we investigate how variations in subgrid physics influence galaxy and black hole properties using the VTNG simulation suite. We first compare the simulation predictions with recent high-redshift observations for several key galaxy scaling relations, including the stellar mass function (SMF), the star-forming main sequence (SFMS), the mass--metallicity relation (MZR), the black hole--stellar mass relation, and the stellar mass--size relation. Overall, the simulations recover several broad qualitative trends in the current high-redshift observations, although the normalization and scatter vary across different parameter combinations. Our analysis is based on 50 VTNG runs that sample different combinations of the eight subgrid parameters. By systematically exploring the parameter space, we assess how variations in the underlying subgrid models affect the predicted galaxy and black hole properties. Our main conclusions are summarized as follows.

\begin{enumerate}

\item
The VTNG simulations recover the broad trends seen in current observational constraints at $z=6$ and $z=8$. In particular, the SMFs and SFMSs broadly reproduce the observed stellar-mass dependence and overall normalization, although moderate tension remains over some stellar-mass ranges.
The stellar MZR remains broadly compatible with current
measurements, whereas the gas-phase MZR is systematically shifted toward higher metallicities. The simulated MSR follows the observed stellar-mass dependence but is systematically offset toward larger effective radii.
The distributions of black hole masses and accretion properties broadly overlap with current observations, although the most massive and rapidly accreting black holes reported by recent JWST observations remain difficult to reproduce in the simulations within the adopted black-hole seeding and Eddington-limited accretion model. 
Taken together, these results capture several broad trends in galaxy and black hole growth while highlighting remaining tensions with current high-redshift observations.

\item
Based on the RF--SHAP analysis, we identify the subgrid parameters that most strongly influence several commonly used galaxy and black-hole diagnostics. In the VTNG suite with fixed initial conditions, the abundance of low-mass galaxies, quantified by the integrated SMF number density over $7.5 \leq \log(M_\star/M_\odot) < 9.5$, is one of the more predictable diagnostics ($R^2\sim0.6$--$0.8$) and is primarily regulated by the supernova feedback temperature $T_{\rm SN}$ and the thermal wind fraction $\tau_{\rm w}$. 
However, the estimated field-to-field scatter is larger than the parameter-driven scatter by factors of approximately 2.3 and 1.6 at $z=6$ and $z=8$, respectively. Injecting this variance into the RF target reduces the median cross-validated $R^2$ to $-0.185$ and $-0.024$, respectively. 
This does not invalidate the controlled comparison among the VTNG runs, but highlights the impact of finite-volume field-to-field fluctuations when connecting the intrinsic parameter response to observed SMFs. Since cosmic variance decreases with increasing survey volume, this limitation becomes less important for larger observational volumes.

\item
For the remaining diagnostics, the RF--SHAP analysis reveals a more varied dependence on the subgrid parameters. The scatter of the SFMS at $z=6$ is moderately predictable and is mainly influenced by $T_{\rm SN}$, the wind velocity factor $\kappa_{\rm w}$, and the maximum star-formation timescale $t_0^\ast$, whereas the SFMS scatter at $z=8$ is not robustly predictable from the explored parameter variations. The high-accretion tail of black hole growth, characterised by $P_{90}$, is regulated by a combination of stellar feedback and black-hole physics, with the dominant parameters being $T_{\rm SN}$, the seed black hole mass $M_{\rm seed}$, and the high-accretion black hole feedback factor $\epsilon_{\rm f,high}$. In contrast, the MZR diagnostics explored here do not show robust positive cross-validated predictive performance, which may reflect the interplay of multiple physical processes rather than regulation by a single dominant subgrid parameter, while the MSR is not included in the RF--SHAP analysis. Nevertheless, the simulations predict substantial scatter in these relations, which can serve as a reference range for future observational measurements.

\end{enumerate}

We note that the VTNG simulations are subject to limitations associated with both the simulation volume and numerical resolution, which may affect the sampling of rare massive systems and the modelling of small-scale gas physics. Within the explored VTNG parameter ranges and numerical setup, massive black holes, rapidly accreting systems, and massive quiescent galaxies remain rare at high redshift.
A recent comparison further shows that galaxy-formation models differ substantially in their predictions for massive and quiescent galaxies at high redshift, with several widely used simulations and empirical models underpredicting the abundance of massive quiescent galaxies by $\gtrsim 1$ dex \citep{Ji2026PANORAMIC}. 
We therefore do not interpret the rarity of these systems in VTNG as a definitive model failure, given the remaining uncertainties in the observationally inferred stellar masses, black-hole masses, AGN classifications, and simulation limitations. However, if the most extreme JWST-inferred systems are confirmed, additional physical ingredients may be required. For example, more rapid early black-hole growth channels, such as super-Eddington accretion, may be needed to reproduce the most massive and luminous AGN candidates, while improved treatments of early gas supply, feedback, and quenching may be required for massive quiescent galaxies. 
Beyond the scaling relations considered here, synthetic UV luminosity functions, galaxy clustering, and field-to-field variance provide promising complementary observables for constraining the VTNG feedback parameters and for separating parameter-driven variations from environmental fluctuations.
Future simulations combining larger cosmological volumes, higher resolution, and improved physical models will therefore be essential for sampling rare systems and for enabling more robust comparisons with the rapidly expanding JWST observations.

\begin{acknowledgements}
We thank the anonymous reviewer for the helpful suggestions that significantly improve the presentation of this paper. This work is supported by the National SKA Program of China (2025SKA0150100), the National Natural Science Foundation of China (12595313), and the CAS Project for Young Scientists in Basic Research (YSBR-092). LCH was supported by the National Science Foundation of China (12233001) and the China Manned Space Program (CMS-CSST-2025-A09). We thank Longyue Chen for helpful discussions.
\end{acknowledgements}

\bibliographystyle{aa} 
\bibliography{ref} 

@ARTICLE{Springel2003,
       author = {{Springel}, Volker and {Hernquist}, Lars},
        title = "{Cosmological smoothed particle hydrodynamics simulations: a hybrid multiphase model for star formation}",
      journal = {\mnras},
         year = 2003,
        month = feb,
       volume = {339},
       number = {2},
        pages = {289-311},
          doi = {10.1046/j.1365-8711.2003.06206.x},
archivePrefix = {arXiv},
       eprint = {astro-ph/0206393},
 primaryClass = {astro-ph},
       adsurl = {https://ui.adsabs.harvard.edu/abs/2003MNRAS.339..289S}
}

@ARTICLE{Nelson2018,
       author = {{Nelson}, Dylan and {Pillepich}, Annalisa and {Springel}, Volker and {Weinberger}, Rainer and {Hernquist}, Lars and {Pakmor}, R{\"u}diger and {Genel}, Shy and {Torrey}, Paul and {Vogelsberger}, Mark and {Kauffmann}, Guinevere and {Marinacci}, Federico and {Naiman}, Jill},
        title = "{First results from the IllustrisTNG simulations: the galaxy colour bimodality}",
      journal = {\mnras},
         year = 2018,
        month = mar,
       volume = {475},
       number = {1},
        pages = {624-647},
          doi = {10.1093/mnras/stx3040},
archivePrefix = {arXiv},
       eprint = {1707.03395},
 primaryClass = {astro-ph.GA},
       adsurl = {https://ui.adsabs.harvard.edu/abs/2018MNRAS.475..624N}
}

@ARTICLE{Nelson2019,
       author = {{Nelson}, Dylan and {Springel}, Volker and {Pillepich}, Annalisa and {Rodriguez-Gomez}, Vicente and {Torrey}, Paul and {Genel}, Shy and {Vogelsberger}, Mark and {Pakmor}, Ruediger and {Marinacci}, Federico and {Weinberger}, Rainer and {Kelley}, Luke and {Lovell}, Mark and {Diemer}, Benedikt and {Hernquist}, Lars},
        title = "{The IllustrisTNG simulations: public data release}",
      journal = {Computational Astrophysics and Cosmology},
         year = 2019,
        month = may,
       volume = {6},
       number = {1},
          eid = {2},
        pages = {2},
          doi = {10.1186/s40668-019-0028-x},
archivePrefix = {arXiv},
       eprint = {1812.05609},
 primaryClass = {astro-ph.GA},
       adsurl = {https://ui.adsabs.harvard.edu/abs/2019ComAC...6....2N}
}

@ARTICLE{Springel2018,
       author = {{Springel}, Volker and {Pakmor}, R{\"u}diger and {Pillepich}, Annalisa and {Weinberger}, Rainer and {Nelson}, Dylan and {Hernquist}, Lars and {Vogelsberger}, Mark and {Genel}, Shy and {Torrey}, Paul and {Marinacci}, Federico and {Naiman}, Jill},
        title = "{First results from the IllustrisTNG simulations: matter and galaxy clustering}",
      journal = {\mnras},
         year = 2018,
        month = mar,
       volume = {475},
       number = {1},
        pages = {676-698},
          doi = {10.1093/mnras/stx3304},
archivePrefix = {arXiv},
       eprint = {1707.03397},
 primaryClass = {astro-ph.GA},
       adsurl = {https://ui.adsabs.harvard.edu/abs/2018MNRAS.475..676S}
}

@ARTICLE{Naiman2018,
       author = {{Naiman}, Jill P. and {Pillepich}, Annalisa and {Springel}, Volker and {Ramirez-Ruiz}, Enrico and {Torrey}, Paul and {Vogelsberger}, Mark and {Pakmor}, R{\"u}diger and {Nelson}, Dylan and {Marinacci}, Federico and {Hernquist}, Lars and {Weinberger}, Rainer and {Genel}, Shy},
        title = "{First results from the IllustrisTNG simulations: a tale of two elements - chemical evolution of magnesium and europium}",
      journal = {\mnras},
         year = 2018,
        month = jun,
       volume = {477},
       number = {1},
        pages = {1206-1224},
          doi = {10.1093/mnras/sty618},
archivePrefix = {arXiv},
       eprint = {1707.03401},
 primaryClass = {astro-ph.GA},
       adsurl = {https://ui.adsabs.harvard.edu/abs/2018MNRAS.477.1206N}
}

@ARTICLE{Marinacci2018,
       author = {{Marinacci}, Federico and {Vogelsberger}, Mark and {Pakmor}, R{\"u}diger and {Torrey}, Paul and {Springel}, Volker and {Hernquist}, Lars and {Nelson}, Dylan and {Weinberger}, Rainer and {Pillepich}, Annalisa and {Naiman}, Jill and {Genel}, Shy},
        title = "{First results from the IllustrisTNG simulations: radio haloes and magnetic fields}",
      journal = {\mnras},
         year = 2018,
        month = nov,
       volume = {480},
       number = {4},
        pages = {5113-5139},
          doi = {10.1093/mnras/sty2206},
archivePrefix = {arXiv},
       eprint = {1707.03396},
 primaryClass = {astro-ph.CO},
       adsurl = {https://ui.adsabs.harvard.edu/abs/2018MNRAS.480.5113M}
}

@ARTICLE{Vogelsberger2013,
       author = {{Vogelsberger}, Mark and {Genel}, Shy and {Sijacki}, Debora and {Torrey}, Paul and {Springel}, Volker and {Hernquist}, Lars},
        title = "{A model for cosmological simulations of galaxy formation physics}",
      journal = {\mnras},
         year = 2013,
        month = dec,
       volume = {436},
       number = {4},
        pages = {3031-3067},
          doi = {10.1093/mnras/stt1789},
archivePrefix = {arXiv},
       eprint = {1305.2913},
 primaryClass = {astro-ph.CO},
       adsurl = {https://ui.adsabs.harvard.edu/abs/2013MNRAS.436.3031V}
}

@ARTICLE{Weinberger2017,
       author = {{Weinberger}, Rainer and {Springel}, Volker and {Hernquist}, Lars and {Pillepich}, Annalisa and {Marinacci}, Federico and {Pakmor}, R{\"u}diger and {Nelson}, Dylan and {Genel}, Shy and {Vogelsberger}, Mark and {Naiman}, Jill and {Torrey}, Paul},
        title = "{Simulating galaxy formation with black hole driven thermal and kinetic feedback}",
      journal = {\mnras},
         year = 2017,
        month = mar,
       volume = {465},
       number = {3},
        pages = {3291-3308},
          doi = {10.1093/mnras/stw2944},
archivePrefix = {arXiv},
       eprint = {1607.03486},
 primaryClass = {astro-ph.GA},
       adsurl = {https://ui.adsabs.harvard.edu/abs/2017MNRAS.465.3291W}
}

@ARTICLE{Pillepich2018,
       author = {{Pillepich}, Annalisa and {Springel}, Volker and {Nelson}, Dylan and {Genel}, Shy and {Naiman}, Jill and {Pakmor}, R{\"u}diger and {Hernquist}, Lars and {Torrey}, Paul and {Vogelsberger}, Mark and {Weinberger}, Rainer and {Marinacci}, Federico},
        title = "{Simulating galaxy formation with the IllustrisTNG model}",
      journal = {\mnras},
         year = 2018,
        month = jan,
       volume = {473},
       number = {3},
        pages = {4077-4106},
          doi = {10.1093/mnras/stx2656},
archivePrefix = {arXiv},
       eprint = {1703.02970},
 primaryClass = {astro-ph.GA},
       adsurl = {https://ui.adsabs.harvard.edu/abs/2018MNRAS.473.4077P}
}

@ARTICLE{Trcka22,
       author = {{Tr{\v{c}}ka}, Ana and {Baes}, Maarten and {Camps}, Peter and {Kapoor}, Anand Utsav and {Nelson}, Dylan and {Pillepich}, Annalisa and {Barrientos}, Daniela and {Hernquist}, Lars and {Marinacci}, Federico and {Vogelsberger}, Mark},
        title = "{UV to submillimetre luminosity functions of TNG50 galaxies}",
      journal = {\mnras},
         year = 2022,
        month = nov,
       volume = {516},
       number = {3},
        pages = {3728-3749},
          doi = {10.1093/mnras/stac2277},
archivePrefix = {arXiv},
       eprint = {2208.06424},
 primaryClass = {astro-ph.GA},
       adsurl = {https://ui.adsabs.harvard.edu/abs/2022MNRAS.516.3728T}
}

@ARTICLE{skirt920,
       author = {{Camps}, P. and {Baes}, M.},
        title = "{SKIRT 9: Redesigning an advanced dust radiative transfer code to allow kinematics, line transfer and polarization by aligned dust grains}",
      journal = {Astronomy and Computing},
         year = 2020,
        month = apr,
       volume = {31},
          eid = {100381},
        pages = {100381},
          doi = {10.1016/j.ascom.2020.100381},
archivePrefix = {arXiv},
       eprint = {2003.00721},
 primaryClass = {astro-ph.GA},
       adsurl = {https://ui.adsabs.harvard.edu/abs/2020A&C....3100381C}
}

@ARTICLE{skirt15a,
       author = {{Baes}, M. and {Camps}, P.},
        title = "{SKIRT: The design of a suite of input models for Monte Carlo radiative transfer simulations}",
      journal = {Astronomy and Computing},
         year = 2015,
        month = sep,
       volume = {12},
        pages = {33-44},
          doi = {10.1016/j.ascom.2015.05.006},
archivePrefix = {arXiv},
       eprint = {1505.07708},
 primaryClass = {astro-ph.IM},
       adsurl = {https://ui.adsabs.harvard.edu/abs/2015A&C....12...33B}
}

@ARTICLE{skirt15b,
       author = {{Camps}, P. and {Baes}, M.},
        title = "{SKIRT: An advanced dust radiative transfer code with a user-friendly architecture}",
      journal = {Astronomy and Computing},
         year = 2015,
        month = mar,
       volume = {9},
        pages = {20-33},
          doi = {10.1016/j.ascom.2014.10.004},
archivePrefix = {arXiv},
       eprint = {1410.1629},
 primaryClass = {astro-ph.IM},
       adsurl = {https://ui.adsabs.harvard.edu/abs/2015A&C.....9...20C}
}

@ARTICLE{bc03,
       author = {{Bruzual}, G. and {Charlot}, S.},
        title = "{Stellar population synthesis at the resolution of 2003}",
      journal = {\mnras},
         year = 2003,
        month = oct,
       volume = {344},
       number = {4},
        pages = {1000-1028},
          doi = {10.1046/j.1365-8711.2003.06897.x},
archivePrefix = {arXiv},
       eprint = {astro-ph/0309134},
 primaryClass = {astro-ph},
       adsurl = {https://ui.adsabs.harvard.edu/abs/2003MNRAS.344.1000B}
}

@ARTICLE{chabrier03,
       author = {{Chabrier}, Gilles},
        title = "{Galactic Stellar and Substellar Initial Mass Function}",
      journal = {\pasp},
         year = 2003,
        month = jul,
       volume = {115},
       number = {809},
        pages = {763-795},
          doi = {10.1086/376392},
archivePrefix = {arXiv},
       eprint = {astro-ph/0304382},
 primaryClass = {astro-ph},
       adsurl = {https://ui.adsabs.harvard.edu/abs/2003PASP..115..763C}
}

@ARTICLE{tng50atlas24,
       author = {{Baes}, Maarten and {Gebek}, Andrea and {Tr{\v{c}}ka}, Ana and {Camps}, Peter and {van der Wel}, Arjen and {Abdurro'uf} and {Andreadis}, Nick and {Tulu}, Sena Bokona and {Emana}, Abdissa Tassama and {Fritz}, Jacopo and {Kelly}, Raymond and {Kova{\v{c}}i{\'c}}, Inja and {La Marca}, Antonio and {Martorano}, Marco and {Mosenkov}, Aleksandr and {Nersesian}, Angelos and {Rodriguez-Gomez}, Vicente and {Tortora}, Crescenzo and {Vander Meulen}, Bert and {Wang}, Lingyu},
        title = "{The TNG50-SKIRT Atlas: Post-processing methodology and first data release}",
      journal = {\aap},
         year = 2024,
        month = mar,
       volume = {683},
          eid = {A181},
        pages = {A181},
          doi = {10.1051/0004-6361/202348418},
archivePrefix = {arXiv},
       eprint = {2401.04224},
 primaryClass = {astro-ph.GA},
       adsurl = {https://ui.adsabs.harvard.edu/abs/2024A&A...683A.181B}
}

@ARTICLE{groves08,
       author = {{Groves}, Brent and {Dopita}, Michael A. and {Sutherland}, Ralph S. and {Kewley}, Lisa J. and {Fischera}, J{\"o}rg and {Leitherer}, Claus and {Brandl}, Bernhard and {van Breugel}, Wil},
        title = "{Modeling the Pan-Spectral Energy Distribution of Starburst Galaxies. IV. The Controlling Parameters of the Starburst SED}",
      journal = {\apjs},
         year = 2008,
        month = jun,
       volume = {176},
       number = {2},
        pages = {438-456},
          doi = {10.1086/528711},
archivePrefix = {arXiv},
       eprint = {0712.1824},
 primaryClass = {astro-ph},
       adsurl = {https://ui.adsabs.harvard.edu/abs/2008ApJS..176..438G}
}

@ARTICLE{Kapoor21,
       author = {{Kapoor}, Anand Utsav and {Camps}, Peter and {Baes}, Maarten and {Tr{\v{c}}ka}, Ana and {Grand}, Robert J.~J. and {van der Wel}, Arjen and {Cortese}, Luca and {De Looze}, Ilse and {Barrientos}, Daniela},
        title = "{High-resolution synthetic UV-submm images for simulated Milky Way-type galaxies from the Auriga project}",
      journal = {\mnras},
         year = 2021,
        month = oct,
       volume = {506},
       number = {4},
        pages = {5703-5720},
          doi = {10.1093/mnras/stab2043},
archivePrefix = {arXiv},
       eprint = {2107.06595},
 primaryClass = {astro-ph.GA},
       adsurl = {https://ui.adsabs.harvard.edu/abs/2021MNRAS.506.5703K}
}

@ARTICLE{vogelsberger20_modelc,
       author = {{Vogelsberger}, Mark and {Nelson}, Dylan and {Pillepich}, Annalisa and {Shen}, Xuejian and {Marinacci}, Federico and {Springel}, Volker and {Pakmor}, R{\"u}diger and {Tacchella}, Sandro and {Weinberger}, Rainer and {Torrey}, Paul and {Hernquist}, Lars},
        title = "{High-redshift JWST predictions from IllustrisTNG: dust modelling and galaxy luminosity functions}",
      journal = {\mnras},
         year = 2020,
        month = mar,
       volume = {492},
       number = {4},
        pages = {5167-5201},
          doi = {10.1093/mnras/staa137},
archivePrefix = {arXiv},
       eprint = {1904.07238},
 primaryClass = {astro-ph.GA},
       adsurl = {https://ui.adsabs.harvard.edu/abs/2020MNRAS.492.5167V}
}

@ARTICLE{themis17,
       author = {{Jones}, A.~P. and {K{\"o}hler}, M. and {Ysard}, N. and {Bocchio}, M. and {Verstraete}, L.},
        title = "{The global dust modelling framework THEMIS}",
      journal = {\aap},
         year = 2017,
        month = jun,
       volume = {602},
          eid = {A46},
        pages = {A46},
          doi = {10.1051/0004-6361/201630225},
archivePrefix = {arXiv},
       eprint = {1703.00775},
 primaryClass = {astro-ph.GA},
       adsurl = {https://ui.adsabs.harvard.edu/abs/2017A&A...602A..46J}
}

@ARTICLE{saftly13,
       author = {{Camps}, P. and {Baes}, M. and {Saftly}, W.},
        title = "{Using 3D Voronoi grids in radiative transfer simulations}",
      journal = {\aap},
         year = 2013,
        month = dec,
       volume = {560},
          eid = {A35},
        pages = {A35},
          doi = {10.1051/0004-6361/201322281},
archivePrefix = {arXiv},
       eprint = {1310.1854},
 primaryClass = {astro-ph.IM},
       adsurl = {https://ui.adsabs.harvard.edu/abs/2013A&A...560A..35C}
}

@ARTICLE{Antonucci1993,
       author = {{Antonucci}, Robert},
        title = "{Unified models for active galactic nuclei and quasars.}",
      journal = {\araa},
         year = 1993,
        month = jan,
       volume = {31},
        pages = {473-521},
          doi = {10.1146/annurev.aa.31.090193.002353},
       adsurl = {https://ui.adsabs.harvard.edu/abs/1993ARA&A..31..473A}
}

@ARTICLE{Stalevski2012,
       author = {{Stalevski}, Marko and {Fritz}, Jacopo and {Baes}, Maarten and {Nakos}, Theodoros and {Popovi{\'c}}, Luka {\v{C}}.},
        title = "{3D radiative transfer modelling of the dusty tori around active galactic nuclei as a clumpy two-phase medium}",
      journal = {\mnras},
         year = 2012,
        month = mar,
       volume = {420},
       number = {4},
        pages = {2756-2772},
          doi = {10.1111/j.1365-2966.2011.19775.x},
archivePrefix = {arXiv},
       eprint = {1109.1286},
 primaryClass = {astro-ph.CO},
       adsurl = {https://ui.adsabs.harvard.edu/abs/2012MNRAS.420.2756S}
}

@ARTICLE{Stalevski2016,
       author = {{Stalevski}, Marko and {Ricci}, Claudio and {Ueda}, Yoshihiro and {Lira}, Paulina and {Fritz}, Jacopo and {Baes}, Maarten},
        title = "{The dust covering factor in active galactic nuclei}",
      journal = {\mnras},
         year = 2016,
        month = may,
       volume = {458},
       number = {3},
        pages = {2288-2302},
          doi = {10.1093/mnras/stw444},
archivePrefix = {arXiv},
       eprint = {1602.06954},
 primaryClass = {astro-ph.GA},
       adsurl = {https://ui.adsabs.harvard.edu/abs/2016MNRAS.458.2288S}
}

@ARTICLE{Boquien2019,
       author = {{Boquien}, M. and {Burgarella}, D. and {Roehlly}, Y. and {Buat}, V. and {Ciesla}, L. and {Corre}, D. and {Inoue}, A.~K. and {Salas}, H.},
        title = "{CIGALE: a python Code Investigating GALaxy Emission}",
      journal = {\aap},
         year = 2019,
        month = feb,
       volume = {622},
          eid = {A103},
        pages = {A103},
          doi = {10.1051/0004-6361/201834156},
archivePrefix = {arXiv},
       eprint = {1811.03094},
 primaryClass = {astro-ph.GA},
       adsurl = {https://ui.adsabs.harvard.edu/abs/2019A&A...622A.103B}
}

@ARTICLE{Yang2020,
       author = {{Yang}, G. and {Boquien}, M. and {Buat}, V. and {Burgarella}, D. and {Ciesla}, L. and {Duras}, F. and {Stalevski}, M. and {Brandt}, W.~N. and {Papovich}, C.},
        title = "{X-CIGALE: Fitting AGN/galaxy SEDs from X-ray to infrared}",
      journal = {\mnras},
         year = 2020,
        month = jan,
       volume = {491},
       number = {1},
        pages = {740-757},
          doi = {10.1093/mnras/stz3001},
archivePrefix = {arXiv},
       eprint = {2001.08263},
 primaryClass = {astro-ph.GA},
       adsurl = {https://ui.adsabs.harvard.edu/abs/2020MNRAS.491..740Y}
}

@ARTICLE{Vijayan2022,
       author = {{Vijayan}, Aswin P. and {Wilkins}, Stephen M. and {Lovell}, Christopher C. and {Thomas}, Peter A. and {Camps}, Peter and {Baes}, Maarten and {Trayford}, James and {Kuusisto}, Jussi and {Roper}, William J.},
        title = "{First Light And Reionisation Epoch Simulations (FLARES) - III. The properties of massive dusty galaxies at cosmic dawn}",
      journal = {\mnras},
         year = 2022,
        month = apr,
       volume = {511},
       number = {4},
        pages = {4999-5017},
          doi = {10.1093/mnras/stac338},
archivePrefix = {arXiv},
       eprint = {2108.00830},
 primaryClass = {astro-ph.GA},
       adsurl = {https://ui.adsabs.harvard.edu/abs/2022MNRAS.511.4999V}
}

@ARTICLE{Roper2022,
       author = {{Roper}, William J. and {Lovell}, Christopher C. and {Vijayan}, Aswin P. and {Marshall}, Madeline A. and {Irodotou}, Dimitrios and {Kuusisto}, Jussi K. and {Thomas}, Peter A. and {Wilkins}, Stephen M.},
        title = "{First Light And Reionisation Epoch Simulations (FLARES) - IV. The size evolution of galaxies at z {\ensuremath{\geq}} 5}",
      journal = {\mnras},
         year = 2022,
        month = aug,
       volume = {514},
       number = {2},
        pages = {1921-1939},
          doi = {10.1093/mnras/stac1368},
archivePrefix = {arXiv},
       eprint = {2203.12627},
 primaryClass = {astro-ph.GA},
       adsurl = {https://ui.adsabs.harvard.edu/abs/2022MNRAS.514.1921R}
}

@ARTICLE{Punyasheel2025,
       author = {{Punyasheel}, Paurush and {Vijayan}, Aswin P. and {Greve}, Thomas R. and {Roper}, William J. and {Algera}, Hiddo and {Gillman}, Steven and {Gullberg}, Bitten and {Irodotou}, Dimitrios and {Lovell}, Christopher C. and {Seeyave}, Louise T.~C. and {Thomas}, Peter A. and {Wilkins}, Stephen M.},
        title = "{First Light And Reionisation Epoch Simulations (FLARES): XVI. Size evolution of massive dusty galaxies at cosmic dawn from the ultraviolet to infrared}",
      journal = {\aap},
         year = 2025,
        month = apr,
       volume = {696},
          eid = {A234},
        pages = {A234},
          doi = {10.1051/0004-6361/202452040},
archivePrefix = {arXiv},
       eprint = {2408.11037},
 primaryClass = {astro-ph.GA},
       adsurl = {https://ui.adsabs.harvard.edu/abs/2025A&A...696A.234P}
}

@INPROCEEDINGS{Perrin2012,
       author = {{Perrin}, Marshall D. and {Soummer}, R{\'e}mi and {Elliott}, Erin M. and {Lallo}, Matthew D. and {Sivaramakrishnan}, Anand},
        title = "{Simulating point spread functions for the James Webb Space Telescope with WebbPSF}",
    booktitle = {Space Telescopes and Instrumentation 2012: Optical, Infrared, and Millimeter Wave},
         year = 2012,
       editor = {{Clampin}, Mark C. and {Fazio}, Giovanni G. and {MacEwen}, Howard A. and {Oschmann}, Jr., Jacobus M.},
       series = {Society of Photo-Optical Instrumentation Engineers (SPIE) Conference Series},
       volume = {8442},
        month = sep,
          eid = {84423D},
        pages = {84423D},
          doi = {10.1117/12.925230},
       adsurl = {https://ui.adsabs.harvard.edu/abs/2012SPIE.8442E..3DP}
}

@software{Bradley2016,
       author = {{Bradley}, Larry and {Sipocz}, Brigitta and {Robitaille}, Thomas and {Tollerud}, Erik and {Deil}, Christoph and {Vin{\'\i}cius}, Z{\`e} and {Barbary}, Kyle and {G{\"u}nther}, Hans Moritz and {Bostroem}, Azalee and {Droettboom}, Michael and {Bray}, Erik and {Bratholm}, Lars Andersen and {Pickering}, T.~E. and {Craig}, Matt and {Pascual}, Sergio and {Greco}, Johnny and {Donath}, Axel and {Kerzendorf}, Wolfgang and {Littlefair}, Stuart and {Barentsen}, Geert and {D'Eugenio}, Francesco and {Weaver}, Benjamin Alan},
        title = "{Photutils: Photometry tools}",
         howpublished = {Astrophysics Source Code Library, record ascl:1609.011},
         year = 2016,
        month = sep,
          eid = {ascl:1609.011},
        archivePrefix = {ascl},
       eprint = {1609.011},
       adsurl = {https://ui.adsabs.harvard.edu/abs/2016ascl.soft09011B}
}

@ARTICLE{Navarro-Carrera2024,
       author = {{Navarro-Carrera}, Rafael and {Rinaldi}, Pierluigi and {Caputi}, Karina I. and {Iani}, Edoardo and {Kokorev}, Vasily and {van Mierlo}, Sophie E.},
        title = "{Constraints on the Faint End of the Galaxy Stellar Mass Function at z ≃ 4{\textendash}8 from Deep JWST Data}",
      journal = {\apj},
         year = 2024,
        month = feb,
       volume = {961},
       number = {2},
          eid = {207},
        pages = {207},
          doi = {10.3847/1538-4357/ad0df6},
archivePrefix = {arXiv},
       eprint = {2305.16141},
 primaryClass = {astro-ph.GA},
       adsurl = {https://ui.adsabs.harvard.edu/abs/2024ApJ...961..207N}
}

@ARTICLE{Weaver2023,
       author = {{Weaver}, J.~R. and {Davidzon}, I. and {Toft}, S. and {Ilbert}, O. and {McCracken}, H.~J. and {Gould}, K.~M.~L. and {Jespersen}, C.~K. and {Steinhardt}, C. and {Lagos}, C.~D.~P. and {Capak}, P.~L. and {Casey}, C.~M. and {Chartab}, N. and {Faisst}, A.~L. and {Hayward}, C.~C. and {Kartaltepe}, J.~S. and {Kauffmann}, O.~B. and {Koekemoer}, A.~M. and {Kokorev}, V. and {Laigle}, C. and {Liu}, D. and {Long}, A. and {Magdis}, G.~E. and {McPartland}, C.~J.~R. and {Milvang-Jensen}, B. and {Mobasher}, B. and {Moneti}, A. and {Peng}, Y. and {Sanders}, D.~B. and {Shuntov}, M. and {Sneppen}, A. and {Valentino}, F. and {Zalesky}, L. and {Zamorani}, G.},
        title = "{COSMOS2020: The galaxy stellar mass function. The assembly and star formation cessation of galaxies at 0.2< z {\ensuremath{\leq}} 7.5}",
      journal = {\aap},
         year = 2023,
        month = sep,
       volume = {677},
          eid = {A184},
        pages = {A184},
          doi = {10.1051/0004-6361/202245581},
archivePrefix = {arXiv},
       eprint = {2212.02512},
 primaryClass = {astro-ph.GA},
       adsurl = {https://ui.adsabs.harvard.edu/abs/2023A&A...677A.184W}
}

@ARTICLE{Harvey2025,
       author = {{Harvey}, Thomas and {Conselice}, Christopher J. and {Adams}, Nathan J. and {Austin}, Duncan and {Juod{\v{z}}balis}, Ignas and {Trussler}, James and {Li}, Qiong and {Ormerod}, Katherine and {Ferreira}, Leonardo and {Lovell}, Christopher C. and {Duan}, Qiao and {Westcott}, Lewi and {Harris}, Honor and {Bhatawdekar}, Rachana and {Coe}, Dan and {Cohen}, Seth H. and {Caruana}, Joseph and {Cheng}, Cheng and {Driver}, Simon P. and {Frye}, Brenda and {Furtak}, Lukas J. and {Grogin}, Norman A. and {Hathi}, Nimish P. and {Holwerda}, Benne W. and {Jansen}, Rolf A. and {Koekemoer}, Anton M. and {Marshall}, Madeline A. and {Nonino}, Mario and {Vijayan}, Aswin P. and {Wilkins}, Stephen M. and {Windhorst}, Rogier and {Willmer}, Christopher N.~A. and {Yan}, Haojing and {Zitrin}, Adi},
        title = "{EPOCHS. IV. SED Modeling Assumptions and Their Impact on the Stellar Mass Function at 6.5 {\ensuremath{\leq}} z {\ensuremath{\leq}} 13.5 Using PEARLS and Public JWST Observations}",
      journal = {\apj},
         year = 2025,
        month = jan,
       volume = {978},
       number = {1},
          eid = {89},
        pages = {89},
          doi = {10.3847/1538-4357/ad8c29},
archivePrefix = {arXiv},
       eprint = {2403.03908},
 primaryClass = {astro-ph.GA},
       adsurl = {https://ui.adsabs.harvard.edu/abs/2025ApJ...978...89H}
}

@ARTICLE{Weibel2024,
       author = {{Weibel}, Andrea and {Oesch}, Pascal A. and {Barrufet}, Laia and {Gottumukkala}, Rashmi and {Ellis}, Richard S. and {Santini}, Paola and {Weaver}, John R. and {Allen}, Natalie and {Bouwens}, Rychard and {Bowler}, Rebecca A.~A. and {Brammer}, Gabe and {Carnall}, Adam C. and {Cullen}, Fergus and {Dayal}, Pratika and {Dickinson}, Mark and {Donnan}, Callum T. and {Dunlop}, James S. and {Giavalisco}, Mauro and {Grogin}, Norman A. and {Illingworth}, Garth D. and {Koekemoer}, Anton M. and {Labbe}, Ivo and {Marchesini}, Danilo and {McLeod}, Derek J. and {McLure}, Ross J. and {Naidu}, Rohan P. and {P{\'e}rez-Gonz{\'a}lez}, Pablo G. and {Shuntov}, Marko and {Stefanon}, Mauro and {Toft}, Sune and {Xiao}, Mengyuan},
        title = "{Galaxy build-up in the first 1.5 Gyr of cosmic history: insights from the stellar mass function at z   4-9 from JWST NIRCam observations}",
      journal = {\mnras},
         year = 2024,
        month = sep,
       volume = {533},
       number = {2},
        pages = {1808-1838},
          doi = {10.1093/mnras/stae1891},
archivePrefix = {arXiv},
       eprint = {2403.08872},
 primaryClass = {astro-ph.GA},
       adsurl = {https://ui.adsabs.harvard.edu/abs/2024MNRAS.533.1808W}
}

@ARTICLE{Stefanon2021,
       author = {{Stefanon}, Mauro and {Bouwens}, Rychard J. and {Labb{\'e}}, Ivo and {Illingworth}, Garth D. and {Gonzalez}, Valentino and {Oesch}, Pascal A.},
        title = "{Galaxy Stellar Mass Functions from z   10 to z   6 using the Deepest Spitzer/Infrared Array Camera Data: No Significant Evolution in the Stellar-to-halo Mass Ratio of Galaxies in the First Gigayear of Cosmic Time}",
      journal = {\apj},
         year = 2021,
        month = nov,
       volume = {922},
       number = {1},
          eid = {29},
        pages = {29},
          doi = {10.3847/1538-4357/ac1bb6},
archivePrefix = {arXiv},
       eprint = {2103.16571},
 primaryClass = {astro-ph.GA},
       adsurl = {https://ui.adsabs.harvard.edu/abs/2021ApJ...922...29S}
}

@ARTICLE{Song2016,
       author = {{Song}, Mimi and {Finkelstein}, Steven L. and {Ashby}, Matthew L.~N. and {Grazian}, A. and {Lu}, Yu and {Papovich}, Casey and {Salmon}, Brett and {Somerville}, Rachel S. and {Dickinson}, Mark and {Duncan}, K. and {Faber}, Sandy M. and {Fazio}, Giovanni G. and {Ferguson}, Henry C. and {Fontana}, Adriano and {Guo}, Yicheng and {Hathi}, Nimish and {Lee}, Seong-Kook and {Merlin}, Emiliano and {Willner}, S.~P.},
        title = "{The Evolution of the Galaxy Stellar Mass Function at z = 4-8: A Steepening Low-mass-end Slope with Increasing Redshift}",
      journal = {\apj},
         year = 2016,
        month = jul,
       volume = {825},
       number = {1},
          eid = {5},
        pages = {5},
          doi = {10.3847/0004-637X/825/1/5},
archivePrefix = {arXiv},
       eprint = {1507.05636},
 primaryClass = {astro-ph.GA},
       adsurl = {https://ui.adsabs.harvard.edu/abs/2016ApJ...825....5S}
}

@ARTICLE{Liu2025,
       author = {{Liu}, Kexin and {Guo}, Hong and {Wang}, Sen and {Xu}, Dandan and {Lu}, Shengdong and {Cui}, Weiguang and {Dav{\'e}}, Romeel},
        title = "{Disparate effects of circumgalactic medium angular momentum in IllustrisTNG and SIMBA}",
      journal = {\aap},
         year = 2025,
        month = jan,
       volume = {693},
          eid = {A48},
        pages = {A48},
          doi = {10.1051/0004-6361/202452248},
archivePrefix = {arXiv},
       eprint = {2409.09379},
 primaryClass = {astro-ph.GA},
       adsurl = {https://ui.adsabs.harvard.edu/abs/2025A&A...693A..48L}
}

@ARTICLE{Roberts-Borsani2022,
       author = {{Roberts-Borsani}, Guido and {Morishita}, Takahiro and {Treu}, Tommaso and {Leethochawalit}, Nicha and {Trenti}, Michele},
        title = "{The Physical Properties of Luminous z {\ensuremath{\gtrsim}} 8 Galaxies and Implications for the Cosmic Star Formation Rate Density from 0.35 deg$^{2}$ of (Pure-)Parallel HST Observations}",
      journal = {\apj},
         year = 2022,
        month = mar,
       volume = {927},
       number = {2},
          eid = {236},
        pages = {236},
          doi = {10.3847/1538-4357/ac4803},
archivePrefix = {arXiv},
       eprint = {2106.06544},
 primaryClass = {astro-ph.GA},
       adsurl = {https://ui.adsabs.harvard.edu/abs/2022ApJ...927..236R}
}

@ARTICLE{Chen2023,
       author = {{Chen}, Zuyi and {Stark}, Daniel P. and {Endsley}, Ryan and {Topping}, Michael and {Whitler}, Lily and {Charlot}, St{\'e}phane},
        title = "{JWST/NIRCam observations of stars and H II regions in z ≃ 6-8 galaxies: properties of star-forming complexes on 150 pc scales}",
      journal = {\mnras},
         year = 2023,
        month = feb,
       volume = {518},
       number = {4},
        pages = {5607-5619},
          doi = {10.1093/mnras/stac3476},
archivePrefix = {arXiv},
       eprint = {2207.12657},
 primaryClass = {astro-ph.GA},
       adsurl = {https://ui.adsabs.harvard.edu/abs/2023MNRAS.518.5607C}
}

@ARTICLE{Leethochawalit2023,
       author = {{Leethochawalit}, N. and {Trenti}, M. and {Santini}, P. and {Yang}, L. and {Merlin}, E. and {Castellano}, M. and {Fontana}, A. and {Treu}, T. and {Mason}, C. and {Glazebrook}, K. and {Jones}, T. and {Vulcani}, B. and {Nanayakkara}, T. and {Marchesini}, D. and {Mascia}, S. and {Morishita}, T. and {Roberts-Borsani}, G. and {Bonchi}, A. and {Paris}, D. and {Boyett}, K. and {Strait}, V. and {Calabr{\`o}}, A. and {Pentericci}, L. and {Bradac}, M. and {Wang}, X. and {Scarlata}, C.},
        title = "{Early Results from GLASS-JWST. X. Rest-frame UV-optical Properties of Galaxies at 7 < z < 9}",
      journal = {\apjl},
         year = 2023,
        month = jan,
       volume = {942},
       number = {2},
          eid = {L26},
        pages = {L26},
          doi = {10.3847/2041-8213/ac959b},
archivePrefix = {arXiv},
       eprint = {2207.11135},
 primaryClass = {astro-ph.GA},
       adsurl = {https://ui.adsabs.harvard.edu/abs/2023ApJ...942L..26L}
}

@ARTICLE{Laporte2022,
       author = {{Laporte}, N. and {Zitrin}, A. and {Dole}, H. and {Roberts-Borsani}, G. and {Furtak}, L.~J. and {Witten}, C.},
        title = "{A lensed protocluster candidate at z = 7.66 identified in JWST observations of the galaxy cluster SMACS0723{\ensuremath{-}}7327}",
      journal = {\aap},
         year = 2022,
        month = nov,
       volume = {667},
          eid = {L3},
        pages = {L3},
          doi = {10.1051/0004-6361/202244719},
archivePrefix = {arXiv},
       eprint = {2208.04930},
 primaryClass = {astro-ph.GA},
       adsurl = {https://ui.adsabs.harvard.edu/abs/2022A&A...667L...3L}
}

@ARTICLE{Stanton2026,
       author = {{Stanton}, T.~M. and {Cullen}, F. and {Carnall}, A.~C. and {Scholte}, D. and {Arellano-C{\'o}rdova}, K.~Z. and {Shapley}, A.~E. and {McLeod}, D.~J. and {Donnan}, C.~T. and {Begley}, R. and {Dav{\'e}}, R. and {Dunlop}, J.~S. and {McLure}, R.~J. and {Rowlands}, K. and {Bondestam}, C. and {Hamadouche}, M.~L. and {Leung}, H.-H. and {Stevenson}, S.~D. and {Taylor}, E.},
        title = "{The JWST EXCELS Survey: gas-phase metallicity evolution at 2 < z < 8}",
      journal = {\mnras},
         year = 2026,
        month = apr,
       volume = {547},
       number = {4},
          eid = {stag449},
        pages = {stag449},
          doi = {10.1093/mnras/stag449},
archivePrefix = {arXiv},
       eprint = {2511.00705},
 primaryClass = {astro-ph.GA},
       adsurl = {https://ui.adsabs.harvard.edu/abs/2026MNRAS.547ag449S}
}

@ARTICLE{Li2024,
       author = {{Li}, Juno and {Da Cunha}, Elisabete and {Gonz{\'a}lez-L{\'o}pez}, Jorge and {Aravena}, Manuel and {De Looze}, Ilse and {F{\"o}rster Schreiber}, N.~M. and {Herrera-Camus}, Rodrigo and {Spilker}, Justin and {Tadaki}, Ken-ichi and {Barcos-Munoz}, Loreto and {Battisti}, Andrew J. and {Birkin}, Jack E. and {Bowler}, Rebecca A.~A. and {Davies}, Rebecca and {D{\'\i}az-Santos}, Tanio and {Ferrara}, Andrea and {Fisher}, Deanne B. and {Hodge}, Jacqueline and {Ikeda}, Ryota and {Killi}, Meghana and {Lee}, Lilian and {Liu}, Daizhong and {Lutz}, Dieter and {Mitsuhashi}, Ikki and {Naab}, Thorsten and {Posses}, Ana and {Rela{\~n}o}, Monica and {Solimano}, Manuel and {{\"U}bler}, Hannah and {van der Giessen}, Stefan Anthony and {Villanueva}, Vicente},
        title = "{The ALMA-CRISTAL Survey: Spatially Resolved Star Formation Activity and Dust Content in 4 < z < 6 Star-forming Galaxies}",
      journal = {\apj},
         year = 2024,
        month = nov,
       volume = {976},
       number = {1},
          eid = {70},
        pages = {70},
          doi = {10.3847/1538-4357/ad7fee},
archivePrefix = {arXiv},
       eprint = {2409.10961},
 primaryClass = {astro-ph.GA},
       adsurl = {https://ui.adsabs.harvard.edu/abs/2024ApJ...976...70L}
}

@ARTICLE{Venturi2024,
       author = {{Venturi}, G. and {Carniani}, S. and {Parlanti}, E. and {Kohandel}, M. and {Curti}, M. and {Pallottini}, A. and {Vallini}, L. and {Arribas}, S. and {Bunker}, A.~J. and {Cameron}, A.~J. and {Castellano}, M. and {Ferrara}, A. and {Fontana}, A. and {Gallerani}, S. and {Gelli}, V. and {Maiolino}, R. and {Ntormousi}, E. and {Pacifici}, C. and {Pentericci}, L. and {Salvadori}, S. and {Vanzella}, E.},
        title = "{Gas-phase metallicity gradients in galaxies at z {\ensuremath{\sim}} 6{\textendash}8}",
      journal = {\aap},
         year = 2024,
        month = nov,
       volume = {691},
          eid = {A19},
        pages = {A19},
          doi = {10.1051/0004-6361/202449855},
archivePrefix = {arXiv},
       eprint = {2403.03977},
 primaryClass = {astro-ph.GA},
       adsurl = {https://ui.adsabs.harvard.edu/abs/2024A&A...691A..19V}
}

@ARTICLE{Torrey2019,
       author = {{Torrey}, Paul and {Vogelsberger}, Mark and {Marinacci}, Federico and {Pakmor}, R{\"u}diger and {Springel}, Volker and {Nelson}, Dylan and {Naiman}, Jill and {Pillepich}, Annalisa and {Genel}, Shy and {Weinberger}, Rainer and {Hernquist}, Lars},
        title = "{The evolution of the mass-metallicity relation and its scatter in IllustrisTNG}",
      journal = {\mnras},
         year = 2019,
        month = apr,
       volume = {484},
       number = {4},
        pages = {5587-5607},
          doi = {10.1093/mnras/stz243},
archivePrefix = {arXiv},
       eprint = {1711.05261},
 primaryClass = {astro-ph.GA},
       adsurl = {https://ui.adsabs.harvard.edu/abs/2019MNRAS.484.5587T}
}

@ARTICLE{Tacchella2023,
       author = {{Tacchella}, Sandro and {Johnson}, Benjamin D. and {Robertson}, Brant E. and {Carniani}, Stefano and {D'Eugenio}, Francesco and {Kumari}, Nimisha and {Maiolino}, Roberto and {Nelson}, Erica J. and {Suess}, Katherine A. and {{\"U}bler}, Hannah and {Williams}, Christina C. and {Adebusola}, Alabi and {Alberts}, Stacey and {Arribas}, Santiago and {Bhatawdekar}, Rachana and {Bonaventura}, Nina and {Bowler}, Rebecca A.~A. and {Bunker}, Andrew J. and {Cameron}, Alex J. and {Curti}, Mirko and {Egami}, Eiichi and {Eisenstein}, Daniel J. and {Frye}, Brenda and {Hainline}, Kevin and {Helton}, Jakob M. and {Ji}, Zhiyuan and {Looser}, Tobias J. and {Lyu}, Jianwei and {Perna}, Michele and {Rawle}, Timothy and {Rieke}, George and {Rieke}, Marcia and {Saxena}, Aayush and {Sandles}, Lester and {Shivaei}, Irene and {Simmonds}, Charlotte and {Sun}, Fengwu and {Willmer}, Christopher N.~A. and {Willott}, Chris J. and {Witstok}, Joris},
        title = "{JWST NIRCam + NIRSpec: interstellar medium and stellar populations of young galaxies with rising star formation and evolving gas reservoirs}",
      journal = {\mnras},
         year = 2023,
        month = jul,
       volume = {522},
       number = {4},
        pages = {6236-6249},
          doi = {10.1093/mnras/stad1408},
archivePrefix = {arXiv},
       eprint = {2208.03281},
 primaryClass = {astro-ph.GA},
       adsurl = {https://ui.adsabs.harvard.edu/abs/2023MNRAS.522.6236T}
}

@ARTICLE{Curti2024,
       author = {{Curti}, Mirko and {Maiolino}, Roberto and {Curtis-Lake}, Emma and {Chevallard}, Jacopo and {Carniani}, Stefano and {D'Eugenio}, Francesco and {Looser}, Tobias J. and {Scholtz}, Jan and {Charlot}, Stephane and {Cameron}, Alex and {{\"U}bler}, Hannah and {Witstok}, Joris and {Boyett}, Kristian and {Laseter}, Isaac and {Sandles}, Lester and {Arribas}, Santiago and {Bunker}, Andrew and {Giardino}, Giovanna and {Maseda}, Michael V. and {Rawle}, Tim and {Rodr{\'\i}guez Del Pino}, Bruno and {Smit}, Renske and {Willott}, Chris J. and {Eisenstein}, Daniel J. and {Hausen}, Ryan and {Johnson}, Benjamin and {Rieke}, Marcia and {Robertson}, Brant and {Tacchella}, Sandro and {Williams}, Christina C. and {Willmer}, Christopher and {Baker}, William M. and {Bhatawdekar}, Rachana and {Egami}, Eiichi and {Helton}, Jakob M. and {Ji}, Zhiyuan and {Kumari}, Nimisha and {Perna}, Michele and {Shivaei}, Irene and {Sun}, Fengwu},
        title = "{JADES: Insights into the low-mass end of the mass-metallicity-SFR relation at 3 < z < 10 from deep JWST/NIRSpec spectroscopy}",
      journal = {\aap},
         year = 2024,
        month = apr,
       volume = {684},
          eid = {A75},
        pages = {A75},
          doi = {10.1051/0004-6361/202346698},
archivePrefix = {arXiv},
       eprint = {2304.08516},
 primaryClass = {astro-ph.GA},
       adsurl = {https://ui.adsabs.harvard.edu/abs/2024A&A...684A..75C}
}

@ARTICLE{Sarkar2025,
       author = {{Sarkar}, Arnab and {Chakraborty}, Priyanka and {Vogelsberger}, Mark and {McDonald}, Michael and {Torrey}, Paul and {Garcia}, Alex M. and {Khullar}, Gourav and {Ferland}, Gary J. and {Forman}, William and {Wolk}, Scott and {Schneider}, Benjamin and {Bautz}, Mark and {Miller}, Eric and {Grant}, Catherine and {ZuHone}, John},
        title = "{Unveiling the Cosmic Chemistry: Revisiting the Mass{\textendash}Metallicity Relation with JWST/NIRSpec at 4 < z < 10}",
      journal = {\apj},
         year = 2025,
        month = jan,
       volume = {978},
       number = {2},
          eid = {136},
        pages = {136},
          doi = {10.3847/1538-4357/ad8f32},
archivePrefix = {arXiv},
       eprint = {2408.07974},
 primaryClass = {astro-ph.GA},
       adsurl = {https://ui.adsabs.harvard.edu/abs/2025ApJ...978..136S}
}

@ARTICLE{Langeroodi2023,
       author = {{Langeroodi}, Danial and {Hjorth}, Jens and {Chen}, Wenlei and {Kelly}, Patrick L. and {Williams}, Hayley and {Lin}, Yu-Heng and {Scarlata}, Claudia and {Zitrin}, Adi and {Broadhurst}, Tom and {Diego}, Jose M. and {Huang}, Xiaosheng and {Filippenko}, Alexei V. and {Foley}, Ryan J. and {Jha}, Saurabh and {Koekemoer}, Anton M. and {Oguri}, Masamune and {Perez-Fournon}, Ismael and {Pierel}, Justin and {Poidevin}, Frederick and {Strolger}, Lou},
        title = "{Evolution of the Mass-Metallicity Relation from Redshift z {\ensuremath{\approx}} 8 to the Local Universe}",
      journal = {\apj},
         year = 2023,
        month = nov,
       volume = {957},
       number = {1},
          eid = {39},
        pages = {39},
          doi = {10.3847/1538-4357/acdbc1},
archivePrefix = {arXiv},
       eprint = {2212.02491},
 primaryClass = {astro-ph.GA},
       adsurl = {https://ui.adsabs.harvard.edu/abs/2023ApJ...957...39L}
}

@ARTICLE{Nakajima2023,
       author = {{Nakajima}, Kimihiko and {Ouchi}, Masami and {Isobe}, Yuki and {Harikane}, Yuichi and {Zhang}, Yechi and {Ono}, Yoshiaki and {Umeda}, Hiroya and {Oguri}, Masamune},
        title = "{JWST Census for the Mass-Metallicity Star Formation Relations at z = 4-10 with Self-consistent Flux Calibration and Proper Metallicity Calibrators}",
      journal = {\apjs},
         year = 2023,
        month = dec,
       volume = {269},
       number = {2},
          eid = {33},
        pages = {33},
          doi = {10.3847/1538-4365/acd556},
archivePrefix = {arXiv},
       eprint = {2301.12825},
 primaryClass = {astro-ph.GA},
       adsurl = {https://ui.adsabs.harvard.edu/abs/2023ApJS..269...33N}
}

@ARTICLE{Maiolino2024,
       author = {{Maiolino}, Roberto and {Scholtz}, Jan and {Curtis-Lake}, Emma and {Carniani}, Stefano and {Baker}, William and {de Graaff}, Anna and {Tacchella}, Sandro and {{\"U}bler}, Hannah and {D'Eugenio}, Francesco and {Witstok}, Joris and {Curti}, Mirko and {Arribas}, Santiago and {Bunker}, Andrew J. and {Charlot}, St{\'e}phane and {Chevallard}, Jacopo and {Eisenstein}, Daniel J. and {Egami}, Eiichi and {Ji}, Zhiyuan and {Jones}, Gareth C. and {Lyu}, Jianwei and {Rawle}, Tim and {Robertson}, Brant and {Rujopakarn}, Wiphu and {Perna}, Michele and {Sun}, Fengwu and {Venturi}, Giacomo and {Williams}, Christina C. and {Willott}, Chris},
        title = "{JADES: The diverse population of infant black holes at 4 < z < 11: Merging, tiny, poor, but mighty}",
      journal = {\aap},
         year = 2024,
        month = nov,
       volume = {691},
          eid = {A145},
        pages = {A145},
          doi = {10.1051/0004-6361/202347640},
archivePrefix = {arXiv},
       eprint = {2308.01230},
 primaryClass = {astro-ph.GA},
       adsurl = {https://ui.adsabs.harvard.edu/abs/2024A&A...691A.145M}
}

@ARTICLE{Harikane2023,
       author = {{Harikane}, Yuichi and {Zhang}, Yechi and {Nakajima}, Kimihiko and {Ouchi}, Masami and {Isobe}, Yuki and {Ono}, Yoshiaki and {Hatano}, Shun and {Xu}, Yi and {Umeda}, Hiroya},
        title = "{A JWST/NIRSpec First Census of Broad-line AGNs at z = 4-7: Detection of 10 Faint AGNs with M $_{BH}$ {}10$^{6}$-{}10$^{8}$ M $_{{\ensuremath{\odot}}}$ and Their Host Galaxy Properties}",
      journal = {\apj},
         year = 2023,
        month = dec,
       volume = {959},
       number = {1},
          eid = {39},
        pages = {39},
          doi = {10.3847/1538-4357/ad029e},
archivePrefix = {arXiv},
       eprint = {2303.11946},
 primaryClass = {astro-ph.GA},
       adsurl = {https://ui.adsabs.harvard.edu/abs/2023ApJ...959...39H}
}

@ARTICLE{Juoddzbalis2024,
       author = {{Juod{\v{z}}balis}, Ignas and {Maiolino}, Roberto and {Baker}, William M. and {Tacchella}, Sandro and {Scholtz}, Jan and {D'Eugenio}, Francesco and {Witstok}, Joris and {Schneider}, Raffaella and {Trinca}, Alessandro and {Valiante}, Rosa and {DeCoursey}, Christa and {Curti}, Mirko and {Carniani}, Stefano and {Chevallard}, Jacopo and {de Graaff}, Anna and {Arribas}, Santiago and {Bennett}, Jake S. and {Bourne}, Martin A. and {Bunker}, Andrew J. and {Charlot}, St{\'e}phane and {Jiang}, Brian and {Koudmani}, Sophie and {Perna}, Michele and {Robertson}, Brant and {Sijacki}, Debora and {{\"U}bler}, Hannah and {Williams}, Christina C. and {Willott}, Chris},
        title = "{A dormant overmassive black hole in the early Universe}",
      journal = {\nat},
         year = 2024,
        month = dec,
       volume = {636},
       number = {8043},
        pages = {594-597},
          doi = {10.1038/s41586-024-08210-5},
archivePrefix = {arXiv},
       eprint = {2403.03872},
 primaryClass = {astro-ph.GA},
       adsurl = {https://ui.adsabs.harvard.edu/abs/2024Natur.636..594J}
}

@ARTICLE{Lirc2026,
       author = {{Li}, Ruancun and {Ho}, Luis C. and {Chen}, Chang-Hao},
        title = "{The Dichotomy in the Nuclear and Host-galaxy Properties of High-redshift Quasars}",
      journal = {\apj},
         year = 2026,
        month = mar,
       volume = {999},
       number = {1},
          eid = {70},
        pages = {70},
          doi = {10.3847/1538-4357/ae38b3},
archivePrefix = {arXiv},
       eprint = {2505.12867},
 primaryClass = {astro-ph.GA},
       adsurl = {https://ui.adsabs.harvard.edu/abs/2026ApJ...999...70L}
}

@ARTICLE{Larson2023,
       author = {{Larson}, Rebecca L. and {Finkelstein}, Steven L. and {Kocevski}, Dale D. and {Hutchison}, Taylor A. and {Trump}, Jonathan R. and {Arrabal Haro}, Pablo and {Bromm}, Volker and {Cleri}, Nikko J. and {Dickinson}, Mark and {Fujimoto}, Seiji and {Kartaltepe}, Jeyhan S. and {Koekemoer}, Anton M. and {Papovich}, Casey and {Pirzkal}, Nor and {Tacchella}, Sandro and {Zavala}, Jorge A. and {Bagley}, Micaela and {Behroozi}, Peter and {Champagne}, Jaclyn B. and {Cole}, Justin W. and {Jung}, Intae and {Morales}, Alexa M. and {Yang}, Guang and {Zhang}, Haowen and {Zitrin}, Adi and {Amor{\'\i}n}, Ricardo O. and {Burgarella}, Denis and {Casey}, Caitlin M. and {Ch{\'a}vez Ortiz}, {\'O}scar A. and {Cox}, Isabella G. and {Chworowsky}, Katherine and {Fontana}, Adriano and {Gawiser}, Eric and {Grazian}, Andrea and {Grogin}, Norman A. and {Harish}, Santosh and {Hathi}, Nimish P. and {Hirschmann}, Michaela and {Holwerda}, Benne W. and {Juneau}, St{\'e}phanie and {Leung}, Gene C.~K. and {Lucas}, Ray A. and {McGrath}, Elizabeth J. and {P{\'e}rez-Gonz{\'a}lez}, Pablo G. and {Rigby}, Jane R. and {Seill{\'e}}, Lise-Marie and {Simons}, Raymond C. and {de La Vega}, Alexander and {Weiner}, Benjamin J. and {Wilkins}, Stephen M. and {Yung}, L.~Y. Aaron and {Ceers Team}},
        title = "{A CEERS Discovery of an Accreting Supermassive Black Hole 570 Myr after the Big Bang: Identifying a Progenitor of Massive z > 6 Quasars}",
      journal = {\apjl},
         year = 2023,
        month = aug,
       volume = {953},
       number = {2},
          eid = {L29},
        pages = {L29},
          doi = {10.3847/2041-8213/ace619},
archivePrefix = {arXiv},
       eprint = {2303.08918},
 primaryClass = {astro-ph.GA},
       adsurl = {https://ui.adsabs.harvard.edu/abs/2023ApJ...953L..29L}
}

@ARTICLE{Kokorev2024,
       author = {{Kokorev}, Vasily and {Caputi}, Karina I. and {Greene}, Jenny E. and {Dayal}, Pratika and {Trebitsch}, Maxime and {Cutler}, Sam E. and {Fujimoto}, Seiji and {Labb{\'e}}, Ivo and {Miller}, Tim B. and {Iani}, Edoardo and {Navarro-Carrera}, Rafael and {Rinaldi}, Pierluigi},
        title = "{A Census of Photometrically Selected Little Red Dots at 4 < z < 9 in JWST Blank Fields}",
      journal = {\apj},
         year = 2024,
        month = jun,
       volume = {968},
       number = {1},
          eid = {38},
        pages = {38},
          doi = {10.3847/1538-4357/ad4265},
archivePrefix = {arXiv},
       eprint = {2401.09981},
 primaryClass = {astro-ph.GA},
       adsurl = {https://ui.adsabs.harvard.edu/abs/2024ApJ...968...38K}
}

@ARTICLE{Morishita2024,
       author = {{Morishita}, Takahiro and {Stiavelli}, Massimo and {Chary}, Ranga-Ram and {Trenti}, Michele and {Bergamini}, Pietro and {Chiaberge}, Marco and {Leethochawalit}, Nicha and {Roberts-Borsani}, Guido and {Shen}, Xuejian and {Treu}, Tommaso},
        title = "{Enhanced Subkiloparsec-scale Star Formation: Results from a JWST Size Analysis of 341 Galaxies at 5 < z < 14}",
      journal = {\apj},
         year = 2024,
        month = mar,
       volume = {963},
       number = {1},
          eid = {9},
        pages = {9},
          doi = {10.3847/1538-4357/ad1404},
archivePrefix = {arXiv},
       eprint = {2308.05018},
 primaryClass = {astro-ph.GA},
       adsurl = {https://ui.adsabs.harvard.edu/abs/2024ApJ...963....9M}
}

@ARTICLE{Allen2025,
       author = {{Allen}, Natalie and {Oesch}, Pascal A. and {Toft}, Sune and {Matharu}, Jasleen and {McPartland}, Conor J.~R. and {Weibel}, Andrea and {Brammer}, Gabe and {Bowler}, Rebecca A.~A. and {Ito}, Kei and {Gottumukkala}, Rashmi and {Rizzo}, Francesca and {Valentino}, Francesco and {Varadaraj}, Rohan G. and {Weaver}, John R. and {Whitaker}, Katherine E.},
        title = "{Galaxy size and mass build-up in the first 2 Gyr of cosmic history from multi-wavelength JWST NIRCam imaging}",
      journal = {\aap},
         year = 2025,
        month = jun,
       volume = {698},
          eid = {A30},
        pages = {A30},
          doi = {10.1051/0004-6361/202452690},
archivePrefix = {arXiv},
       eprint = {2410.16354},
 primaryClass = {astro-ph.GA},
       adsurl = {https://ui.adsabs.harvard.edu/abs/2025A&A...698A..30A}
}

@ARTICLE{Song2026,
       author = {{Song}, Qi and {Liu}, F.~S. and {Ren}, Jian and {Zhao}, Pinsong and {Cui}, Qifan and {Li}, Yubin and {Mo}, Hao and {Luo}, Yuchong and {Wang}, Guanghuan and {Li}, Nan and {Yesuf}, Hassen M. and {Wang}, Weichen and {Zhang}, Xin and {Meng}, Xianmin and {Fu}, Mingxiang and {Zhang}, Bingqing and {Ling}, Chenxiaoji},
        title = "{The Size Evolution and the Size─Mass Relation of Ly{\ensuremath{\alpha}} Emitters across 3 {\ensuremath{\lesssim}} z< 7 as Observed by JWST}",
      journal = {\apj},
         year = 2026,
        month = jan,
       volume = {997},
       number = {1},
          eid = {126},
        pages = {126},
          doi = {10.3847/1538-4357/ae24e4},
archivePrefix = {arXiv},
       eprint = {2508.05052},
 primaryClass = {astro-ph.GA},
       adsurl = {https://ui.adsabs.harvard.edu/abs/2026ApJ...997..126S}
}

@ARTICLE{Yang2022,
       author = {{Yang}, L. and {Morishita}, T. and {Leethochawalit}, N. and {Castellano}, M. and {Calabr{\`o}}, A. and {Treu}, T. and {Bonchi}, A. and {Fontana}, A. and {Mason}, C. and {Merlin}, E. and {Paris}, D. and {Trenti}, M. and {Roberts-Borsani}, G. and {Bradac}, M. and {Vanzella}, E. and {Vulcani}, B. and {Marchesini}, D. and {Ding}, X. and {Nanayakkara}, T. and {Birrer}, S. and {Glazebrook}, K. and {Jones}, T. and {Boyett}, K. and {Santini}, P. and {Strait}, V. and {Wang}, X.},
        title = "{Early Results from GLASS-JWST. V: The First Rest-frame Optical Size-Luminosity Relation of Galaxies at z > 7}",
      journal = {\apjl},
         year = 2022,
        month = oct,
       volume = {938},
       number = {2},
          eid = {L17},
        pages = {L17},
          doi = {10.3847/2041-8213/ac8803},
archivePrefix = {arXiv},
       eprint = {2207.13101},
 primaryClass = {astro-ph.GA},
       adsurl = {https://ui.adsabs.harvard.edu/abs/2022ApJ...938L..17Y}
}

@ARTICLE{Springel2010,
       author = {{Springel}, Volker},
        title = "{E pur si muove: Galilean-invariant cosmological hydrodynamical simulations on a moving mesh}",
      journal = {\mnras},
         year = 2010,
        month = jan,
       volume = {401},
       number = {2},
        pages = {791-851},
          doi = {10.1111/j.1365-2966.2009.15715.x},
archivePrefix = {arXiv},
       eprint = {0901.4107},
 primaryClass = {astro-ph.CO},
       adsurl = {https://ui.adsabs.harvard.edu/abs/2010MNRAS.401..791S}
}

@ARTICLE{Weinberger2020,
       author = {{Weinberger}, Rainer and {Springel}, Volker and {Pakmor}, R{\"u}diger},
        title = "{The AREPO Public Code Release}",
      journal = {\apjs},
         year = 2020,
        month = jun,
       volume = {248},
       number = {2},
          eid = {32},
        pages = {32},
          doi = {10.3847/1538-4365/ab908c},
archivePrefix = {arXiv},
       eprint = {1909.04667},
 primaryClass = {astro-ph.IM},
       adsurl = {https://ui.adsabs.harvard.edu/abs/2020ApJS..248...32W}
}

@ARTICLE{Inayoshi2025,
       author = {{Inayoshi}, Kohei and {Ho}, Luis C.},
        title = "{A critical evaluation of the physical nature of the little red dots}",
      journal = {Nature Astronomy},
         year = 2026,
        month = sep,
       volume = {10},
        pages = {1261--1272},
          doi = {10.1038/s41550-026-02934-2},
       adsurl = {https://ui.adsabs.harvard.edu/abs/2026NatAs..10.1261I}
}

@ARTICLE{McClymont2025,
       author = {{McClymont}, William and {Tacchella}, Sandro and {Smith}, Aaron and {Kannan}, Rahul and {Puchwein}, Ewald and {Borrow}, Josh and {Garaldi}, Enrico and {Keating}, Laura and {Vogelsberger}, Mark and {Zier}, Oliver and {Shen}, Xuejian and {Popovic}, Filip},
        title = "{The THESAN-ZOOM project: central starbursts and inside-out quenching govern galaxy sizes in the early Universe}",
      journal = {\mnras},
         year = 2025,
        month = dec,
       volume = {544},
       number = {2},
        pages = {1732-1747},
          doi = {10.1093/mnras/staf1861},
archivePrefix = {arXiv},
       eprint = {2503.04894},
 primaryClass = {astro-ph.GA},
       adsurl = {https://ui.adsabs.harvard.edu/abs/2025MNRAS.544.1732M}
}

@ARTICLE{Kannan2025thesan,
       author = {{Kannan}, Rahul and {Puchwein}, Ewald and {Smith}, Aaron and {Borrow}, Josh and {Garaldi}, Enrico and {Keating}, Laura and {Vogelsberger}, Mark and {Zier}, Oliver and {McClymont}, William and {Shen}, Xuejian and {Popovic}, Filip and {Tacchella}, Sandro and {Hernquist}, Lars and {Springel}, Volker},
        title = "{Introducing the THESAN-ZOOM project: radiation-hydrodynamic simulations of high-redshift galaxies with a multi-phase interstellar medium}",
      journal = {The Open Journal of Astrophysics},
         year = 2025,
        month = oct,
       volume = {8},
          eid = {153},
        pages = {153},
          doi = {10.33232/001c.145804},
archivePrefix = {arXiv},
       eprint = {2502.20437},
 primaryClass = {astro-ph.GA},
       adsurl = {https://ui.adsabs.harvard.edu/abs/2025OJAp....8E.153K}
}

@ARTICLE{Hopkins2018,
       author = {{Hopkins}, Philip F. and {Wetzel}, Andrew and {Kere{\v{s}}}, Du{\v{s}}an and {Faucher-Gigu{\`e}re}, Claude-Andr{\'e} and {Quataert}, Eliot and {Boylan-Kolchin}, Michael and {Murray}, Norman and {Hayward}, Christopher C. and {Garrison-Kimmel}, Shea and {Hummels}, Cameron and {Feldmann}, Robert and {Torrey}, Paul and {Ma}, Xiangcheng and {Angl{\'e}s-Alc{\'a}zar}, Daniel and {Su}, Kung-Yi and {Orr}, Matthew and {Schmitz}, Denise and {Escala}, Ivanna and {Sanderson}, Robyn and {Grudi{\'c}}, Michael Y. and {Hafen}, Zachary and {Kim}, Ji-Hoon and {Fitts}, Alex and {Bullock}, James S. and {Wheeler}, Coral and {Chan}, T.~K. and {Elbert}, Oliver D. and {Narayanan}, Desika},
        title = "{FIRE-2 simulations: physics versus numerics in galaxy formation}",
      journal = {\mnras},
         year = 2018,
        month = oct,
       volume = {480},
       number = {1},
        pages = {800-863},
          doi = {10.1093/mnras/sty1690},
archivePrefix = {arXiv},
       eprint = {1702.06148},
 primaryClass = {astro-ph.GA},
       adsurl = {https://ui.adsabs.harvard.edu/abs/2018MNRAS.480..800H}
}

@ARTICLE{Kannan2023,
       author = {{Kannan}, Rahul and {Springel}, Volker and {Hernquist}, Lars and {Pakmor}, R{\"u}diger and {Delgado}, Ana Maria and {Hadzhiyska}, Boryana and {Hern{\'a}ndez-Aguayo}, C{\'e}sar and {Barrera}, Monica and {Ferlito}, Fulvio and {Bose}, Sownak and {White}, Simon D.~M. and {Frenk}, Carlos and {Smith}, Aaron and {Garaldi}, Enrico},
        title = "{The MillenniumTNG project: the galaxy population at z {\ensuremath{\geq}} 8}",
      journal = {\mnras},
         year = 2023,
        month = sep,
       volume = {524},
       number = {2},
        pages = {2594-2605},
          doi = {10.1093/mnras/stac3743},
archivePrefix = {arXiv},
       eprint = {2210.10066},
 primaryClass = {astro-ph.GA},
       adsurl = {https://ui.adsabs.harvard.edu/abs/2023MNRAS.524.2594K}
}

@ARTICLE{Labbe2023,
       author = {{Labb{\'e}}, Ivo and {van Dokkum}, Pieter and {Nelson}, Erica and {Bezanson}, Rachel and {Suess}, Katherine A. and {Leja}, Joel and {Brammer}, Gabriel and {Whitaker}, Katherine and {Mathews}, Elijah and {Stefanon}, Mauro and {Wang}, Bingjie},
        title = "{A population of red candidate massive galaxies  600 Myr after the Big Bang}",
      journal = {\nat},
         year = 2023,
        month = apr,
       volume = {616},
       number = {7956},
        pages = {266-269},
          doi = {10.1038/s41586-023-05786-2},
archivePrefix = {arXiv},
       eprint = {2207.12446},
 primaryClass = {astro-ph.GA},
       adsurl = {https://ui.adsabs.harvard.edu/abs/2023Natur.616..266L}
}

@ARTICLE{Wang2025,
       author = {{Wang}, Tao and {Sun}, Hanwen and {Zhou}, Luwenjia and {Xu}, Ke and {Cheng}, Cheng and {Li}, Zhaozhou and {Chen}, Yangyao and {Mo}, H.~J. and {Dekel}, Avishai and {Yang}, Tiancheng and {Wang}, Yijun and {Chen}, Longyue and {Zheng}, Xianzhong and {Cai}, Zheng and {Elbaz}, David and {Dai}, Y.-S. and {Huang}, J.-S.},
        title = "{JWST/MIRI Reveals the True Number Density of Massive Galaxies in the Early Universe}",
      journal = {\apjl},
         year = 2025,
        month = jul,
       volume = {988},
       number = {1},
          eid = {L35},
        pages = {L35},
          doi = {10.3847/2041-8213/adebe7},
archivePrefix = {arXiv},
       eprint = {2403.02399},
 primaryClass = {astro-ph.GA},
       adsurl = {https://ui.adsabs.harvard.edu/abs/2025ApJ...988L..35W}
}

@ARTICLE{Bhatawdekar2019,
       author = {{Bhatawdekar}, Rachana and {Conselice}, Christopher J. and {Margalef-Bentabol}, Berta and {Duncan}, Kenneth},
        title = "{Evolution of the galaxy stellar mass functions and UV luminosity functions at z = 6-9 in the Hubble Frontier Fields}",
      journal = {\mnras},
         year = 2019,
        month = jul,
       volume = {486},
       number = {3},
        pages = {3805-3830},
          doi = {10.1093/mnras/stz866},
archivePrefix = {arXiv},
       eprint = {1807.07580},
 primaryClass = {astro-ph.GA},
       adsurl = {https://ui.adsabs.harvard.edu/abs/2019MNRAS.486.3805B}
}

@ARTICLE{Simmonds2025,
       author = {{Simmonds}, C. and {Tacchella}, S. and {McClymont}, W. and {Curtis-Lake}, E. and {D'Eugenio}, F. and {Hainline}, K. and {Johnson}, B.~D. and {Kravtsov}, A. and {Pusk{\'a}s}, D. and {Robertson}, B. and {Stoffers}, A. and {Willott}, C. and {Baker}, W.~M. and {Belokurov}, V.~A. and {Bhatawdekar}, R. and {Bunker}, A.~J. and {Carniani}, S. and {Chevallard}, J. and {Curti}, M. and {Duan}, Q. and {Helton}, J.~M. and {Ji}, Z. and {Looser}, T.~J. and {Maiolino}, R. and {Maseda}, M.~V. and {Shivaei}, I. and {Williams}, C.~C.},
        title = "{Bursting at the seams: the star-forming main sequence and its scatter at z = 3─9 using NIRCam photometry from JADES}",
      journal = {\mnras},
         year = 2025,
        month = dec,
       volume = {544},
       number = {4},
        pages = {4551-4575},
          doi = {10.1093/mnras/staf1950},
archivePrefix = {arXiv},
       eprint = {2508.04410},
 primaryClass = {astro-ph.GA},
       adsurl = {https://ui.adsabs.harvard.edu/abs/2025MNRAS.544.4551S}
}

@ARTICLE{Clarke2024,
       author = {{Clarke}, Leonardo and {Shapley}, Alice E. and {Sanders}, Ryan L. and {Topping}, Michael W. and {Brammer}, Gabriel B. and {Bento}, Trinity and {Reddy}, Naveen A. and {Kehoe}, Emily},
        title = "{The Star-forming Main Sequence in JADES and CEERS at z > 1.4: Investigating the Burstiness of Star Formation}",
      journal = {\apj},
         year = 2024,
        month = dec,
       volume = {977},
       number = {1},
          eid = {133},
        pages = {133},
          doi = {10.3847/1538-4357/ad8ba4},
archivePrefix = {arXiv},
       eprint = {2406.05178},
 primaryClass = {astro-ph.GA},
       adsurl = {https://ui.adsabs.harvard.edu/abs/2024ApJ...977..133C}
}

@ARTICLE{Heintz2023,
       author = {{Heintz}, Kasper E. and {Brammer}, Gabriel B. and {Gim{\'e}nez-Arteaga}, Clara and {Strait}, Victoria B. and {Lagos}, Claudia del P. and {Vijayan}, Aswin P. and {Matthee}, Jorryt and {Watson}, Darach and {Mason}, Charlotte A. and {Hutter}, Anne and {Toft}, Sune and {Fynbo}, Johan P.~U. and {Oesch}, Pascal A.},
        title = "{Dilution of chemical enrichment in galaxies 600 Myr after the Big Bang}",
      journal = {Nature Astronomy},
         year = 2023,
        month = dec,
       volume = {7},
        pages = {1517-1524},
          doi = {10.1038/s41550-023-02078-7},
archivePrefix = {arXiv},
       eprint = {2212.02890},
 primaryClass = {astro-ph.GA},
       adsurl = {https://ui.adsabs.harvard.edu/abs/2023NatAs...7.1517H}
}

@ARTICLE{Hough2023,
       author = {{Hough}, Renier T. and {Rennehan}, Douglas and {Kobayashi}, Chiaki and {Loubser}, S. Ilani and {Dav{\'e}}, Romeel and {Babul}, Arif and {Cui}, Weiguang},
        title = "{SIMBA-C: an updated chemical enrichment model for galactic chemical evolution in the SIMBA simulation}",
      journal = {\mnras},
         year = 2023,
        month = oct,
       volume = {525},
       number = {1},
        pages = {1061-1076},
          doi = {10.1093/mnras/stad2394},
archivePrefix = {arXiv},
       eprint = {2308.03436},
 primaryClass = {astro-ph.GA},
       adsurl = {https://ui.adsabs.harvard.edu/abs/2023MNRAS.525.1061H}
}

@ARTICLE{Kocevski2025,
       author = {{Kocevski}, Dale D. and {Finkelstein}, Steven L. and {Barro}, Guillermo and {Taylor}, Anthony J. and {Calabr{\`o}}, Antonello and {Laloux}, Brivael and {Buchner}, Johannes and {Trump}, Jonathan R. and {Leung}, Gene C.~K. and {Yang}, Guang and {Dickinson}, Mark and {P{\'e}rez-Gonz{\'a}lez}, Pablo G. and {Pacucci}, Fabio and {Inayoshi}, Kohei and {Somerville}, Rachel S. and {McGrath}, Elizabeth J. and {Akins}, Hollis B. and {Bagley}, Micaela B. and {Bowler}, Rebecca A.~A. and {Bisigello}, Laura and {Carnall}, Adam and {Casey}, Caitlin M. and {Cheng}, Yingjie and {Cleri}, Nikko J. and {Costantin}, Luca and {Cullen}, Fergus and {Davis}, Kelcey and {Donnan}, Callum T. and {Dunlop}, James S. and {Ellis}, Richard S. and {Ferguson}, Henry C. and {Fujimoto}, Seiji and {Fontana}, Adriano and {Giavalisco}, Mauro and {Grazian}, Andrea and {Grogin}, Norman A. and {Hathi}, Nimish P. and {Hirschmann}, Michaela and {Huertas-Company}, Marc and {Holwerda}, Benne W. and {Illingworth}, Garth and {Juneau}, St{\'e}phanie and {Kartaltepe}, Jeyhan S. and {Koekemoer}, Anton M. and {Li}, Wenxiu and {Lucas}, Ray A. and {Magee}, Dan and {Mason}, Charlotte and {McLeod}, Derek J. and {McLure}, Ross J. and {Napolitano}, Lorenzo and {Papovich}, Casey and {Pirzkal}, Nor and {Rodighiero}, Giulia and {Santini}, Paola and {Wilkins}, Stephen M. and {Yung}, L.~Y. Aaron},
        title = "{The Rise of Faint, Red Active Galactic Nuclei at z > 4: A Sample of Little Red Dots in the JWST Extragalactic Legacy Fields}",
      journal = {\apj},
         year = 2025,
        month = jun,
       volume = {986},
       number = {2},
          eid = {126},
        pages = {126},
          doi = {10.3847/1538-4357/adbc7d},
archivePrefix = {arXiv},
       eprint = {2404.03576},
 primaryClass = {astro-ph.GA},
       adsurl = {https://ui.adsabs.harvard.edu/abs/2025ApJ...986..126K}
}

@ARTICLE{Yang2025,
       author = {{Yang}, Lilan and {Kartaltepe}, Jeyhan S. and {Franco}, Maximilien and {Ding}, Xuheng and {Achenbach}, Mark J. and {Arango-Toro}, Rafael C. and {Casey}, Caitlin M. and {Drakos}, Nicole E. and {Faisst}, Andreas L. and {Gillman}, Steven and {Gozaliasl}, Ghassem and {Huertas-Company}, Marc and {Jin}, Shuowen and {Liu}, Daizhong and {Magdis}, Georgios and {Massey}, Richard and {Silverman}, John D. and {Tanaka}, Takumi S. and {Yu}, Si-Yue and {Akins}, Hollis B. and {Allen}, Natalie and {Ilbert}, Olivier and {Koekemoer}, Anton M. and {McCracken}, Henry Joy and {Paquereau}, Louise and {Rhodes}, Jason and {Robertson}, Brant E. and {Shuntov}, Marko and {Toft}, Sune},
        title = "{COSMOS-Web: Unraveling the Evolution of Galaxy Size and Related Properties at 2 < z < 10}",
      journal = {\apjs},
         year = 2025,
        month = dec,
       volume = {281},
       number = {2},
          eid = {68},
        pages = {68},
          doi = {10.3847/1538-4365/ae0e1b},
archivePrefix = {arXiv},
       eprint = {2504.07185},
 primaryClass = {astro-ph.GA},
       adsurl = {https://ui.adsabs.harvard.edu/abs/2025ApJS..281...68Y}
}

@ARTICLE{Vogelsberger2014,
       author = {{Vogelsberger}, Mark and {Genel}, Shy and {Springel}, Volker and {Torrey}, Paul and {Sijacki}, Debora and {Xu}, Dandan and {Snyder}, Greg and {Nelson}, Dylan and {Hernquist}, Lars},
        title = "{Introducing the Illustris Project: simulating the coevolution of dark and visible matter in the Universe}",
      journal = {\mnras},
         year = 2014,
        month = oct,
       volume = {444},
       number = {2},
        pages = {1518-1547},
          doi = {10.1093/mnras/stu1536},
archivePrefix = {arXiv},
       eprint = {1405.2921},
 primaryClass = {astro-ph.CO},
       adsurl = {https://ui.adsabs.harvard.edu/abs/2014MNRAS.444.1518V}
}

@ARTICLE{Shuntov2025,
       author = {{Shuntov}, M. and {Ilbert}, O. and {Toft}, S. and {Arango-Toro}, R.~C. and {Akins}, H.~B. and {Casey}, C.~M. and {Franco}, M. and {Harish}, S. and {Kartaltepe}, J.~S. and {Koekemoer}, A.~M. and {McCracken}, H.~J. and {Paquereau}, L. and {Laigle}, C. and {Bethermin}, M. and {Dubois}, Y. and {Drakos}, N.~E. and {Faisst}, A. and {Gozaliasl}, G. and {Gillman}, S. and {Hayward}, C.~C. and {Hirschmann}, M. and {Huertas-Company}, M. and {Jespersen}, C.~K. and {Jin}, S. and {Kokorev}, V. and {Lambrides}, E. and {Le Borgne}, D. and {Liu}, D. and {Magdis}, G. and {Massey}, R. and {McPartland}, C.~J.~R. and {Mercier}, W. and {McCleary}, J.~E. and {McKinney}, J. and {Oesch}, P.~A. and {Renzini}, A. and {Rhodes}, J.~D. and {Rich}, R.~M. and {Robertson}, B.~E. and {Sanders}, D. and {Trebitsch}, M. and {Tresse}, L. and {Valentino}, F. and {Vijayan}, A.~P. and {Weaver}, J.~R. and {Weibel}, A. and {Wilkins}, S.~M. and {Yang}, L.},
        title = "{COSMOS-Web: Stellar mass assembly in relation to dark matter halos across 0.2 < z < 12 of cosmic history}",
      journal = {\aap},
         year = 2025,
        month = mar,
       volume = {695},
          eid = {A20},
        pages = {A20},
          doi = {10.1051/0004-6361/202452570},
archivePrefix = {arXiv},
       eprint = {2410.08290},
 primaryClass = {astro-ph.GA},
       adsurl = {https://ui.adsabs.harvard.edu/abs/2025A&A...695A..20S}
}

@ARTICLE{Brinchmann2004,
       author = {{Brinchmann}, J. and {Charlot}, S. and {White}, S.~D.~M. and {Tremonti}, C. and {Kauffmann}, G. and {Heckman}, T. and {Brinkmann}, J.},
        title = "{The physical properties of star-forming galaxies in the low-redshift Universe}",
      journal = {\mnras},
         year = 2004,
        month = jul,
       volume = {351},
       number = {4},
        pages = {1151-1179},
          doi = {10.1111/j.1365-2966.2004.07881.x},
archivePrefix = {arXiv},
       eprint = {astro-ph/0311060},
 primaryClass = {astro-ph},
       adsurl = {https://ui.adsabs.harvard.edu/abs/2004MNRAS.351.1151B}
}

@ARTICLE{Moustakas2013,
       author = {{Moustakas}, John and {Coil}, Alison L. and {Aird}, James and {Blanton}, Michael R. and {Cool}, Richard J. and {Eisenstein}, Daniel J. and {Mendez}, Alexander J. and {Wong}, Kenneth C. and {Zhu}, Guangtun and {Arnouts}, St{\'e}phane},
        title = "{PRIMUS: Constraints on Star Formation Quenching and Galaxy Merging, and the Evolution of the Stellar Mass Function from z = 0-1}",
      journal = {\apj},
         year = 2013,
        month = apr,
       volume = {767},
       number = {1},
          eid = {50},
        pages = {50},
          doi = {10.1088/0004-637X/767/1/50},
archivePrefix = {arXiv},
       eprint = {1301.1688},
 primaryClass = {astro-ph.CO},
       adsurl = {https://ui.adsabs.harvard.edu/abs/2013ApJ...767...50M}
}

@ARTICLE{Baldry2012,
       author = {{Baldry}, I.~K. and {Driver}, S.~P. and {Loveday}, J. and {Taylor}, E.~N. and {Kelvin}, L.~S. and {Liske}, J. and {Norberg}, P. and {Robotham}, A.~S.~G. and {Brough}, S. and {Hopkins}, A.~M. and {Bamford}, S.~P. and {Peacock}, J.~A. and {Bland-Hawthorn}, J. and {Conselice}, C.~J. and {Croom}, S.~M. and {Jones}, D.~H. and {Parkinson}, H.~R. and {Popescu}, C.~C. and {Prescott}, M. and {Sharp}, R.~G. and {Tuffs}, R.~J.},
        title = "{Galaxy And Mass Assembly (GAMA): the galaxy stellar mass function at z < 0.06}",
      journal = {\mnras},
         year = 2012,
        month = mar,
       volume = {421},
       number = {1},
        pages = {621-634},
          doi = {10.1111/j.1365-2966.2012.20340.x},
archivePrefix = {arXiv},
       eprint = {1111.5707},
 primaryClass = {astro-ph.CO},
       adsurl = {https://ui.adsabs.harvard.edu/abs/2012MNRAS.421..621B}
}

@ARTICLE{Somerville2015,
       author = {{Somerville}, Rachel S. and {Dav{\'e}}, Romeel},
        title = "{Physical Models of Galaxy Formation in a Cosmological Framework}",
      journal = {\araa},
         year = 2015,
        month = aug,
       volume = {53},
        pages = {51-113},
          doi = {10.1146/annurev-astro-082812-140951},
archivePrefix = {arXiv},
       eprint = {1412.2712},
 primaryClass = {astro-ph.GA},
       adsurl = {https://ui.adsabs.harvard.edu/abs/2015ARA&A..53...51S}
}

@ARTICLE{Schaye2015,
       author = {{Schaye}, Joop and {Crain}, Robert A. and {Bower}, Richard G. and {Furlong}, Michelle and {Schaller}, Matthieu and {Theuns}, Tom and {Dalla Vecchia}, Claudio and {Frenk}, Carlos S. and {McCarthy}, I.~G. and {Helly}, John C. and {Jenkins}, Adrian and {Rosas-Guevara}, Y.~M. and {White}, Simon D.~M. and {Baes}, Maarten and {Booth}, C.~M. and {Camps}, Peter and {Navarro}, Julio F. and {Qu}, Yan and {Rahmati}, Alireza and {Sawala}, Till and {Thomas}, Peter A. and {Trayford}, James},
        title = "{The EAGLE project: simulating the evolution and assembly of galaxies and their environments}",
      journal = {\mnras},
         year = 2015,
        month = jan,
       volume = {446},
       number = {1},
        pages = {521-554},
          doi = {10.1093/mnras/stu2058},
archivePrefix = {arXiv},
       eprint = {1407.7040},
 primaryClass = {astro-ph.GA},
       adsurl = {https://ui.adsabs.harvard.edu/abs/2015MNRAS.446..521S}
}

@ARTICLE{Dave2019,
       author = {{Dav{\'e}}, Romeel and {Angl{\'e}s-Alc{\'a}zar}, Daniel and {Narayanan}, Desika and {Li}, Qi and {Rafieferantsoa}, Mika H. and {Appleby}, Sarah},
        title = "{SIMBA: Cosmological simulations with black hole growth and feedback}",
      journal = {\mnras},
         year = 2019,
        month = jun,
       volume = {486},
       number = {2},
        pages = {2827-2849},
          doi = {10.1093/mnras/stz937},
archivePrefix = {arXiv},
       eprint = {1901.10203},
 primaryClass = {astro-ph.GA},
       adsurl = {https://ui.adsabs.harvard.edu/abs/2019MNRAS.486.2827D}
}

@ARTICLE{Carnall2023,
       author = {{Carnall}, A.~C. and {McLeod}, D.~J. and {McLure}, R.~J. and {Dunlop}, J.~S. and {Begley}, R. and {Cullen}, F. and {Donnan}, C.~T. and {Hamadouche}, M.~L. and {Jewell}, S.~M. and {Jones}, E.~W. and {Pollock}, C.~L. and {Wild}, V.},
        title = "{A surprising abundance of massive quiescent galaxies at 3 < z < 5 in the first data from JWST CEERS}",
      journal = {\mnras},
         year = 2023,
        month = apr,
       volume = {520},
       number = {3},
        pages = {3974-3985},
          doi = {10.1093/mnras/stad369},
archivePrefix = {arXiv},
       eprint = {2208.00986},
 primaryClass = {astro-ph.GA},
       adsurl = {https://ui.adsabs.harvard.edu/abs/2023MNRAS.520.3974C}
}

@ARTICLE{Valentino2023,
       author = {{Valentino}, Francesco and {Brammer}, Gabriel and {Gould}, Katriona M.~L. and {Kokorev}, Vasily and {Fujimoto}, Seiji and {Jespersen}, Christian Kragh and {Vijayan}, Aswin P. and {Weaver}, John R. and {Ito}, Kei and {Tanaka}, Masayuki and {Ilbert}, Olivier and {Magdis}, Georgios E. and {Whitaker}, Katherine E. and {Faisst}, Andreas L. and {Gallazzi}, Anna and {Gillman}, Steven and {Gim{\'e}nez-Arteaga}, Clara and {G{\'o}mez-Guijarro}, Carlos and {Kubo}, Mariko and {Heintz}, Kasper E. and {Hirschmann}, Michaela and {Oesch}, Pascal and {Onodera}, Masato and {Rizzo}, Francesca and {Lee}, Minju and {Strait}, Victoria and {Toft}, Sune},
        title = "{An Atlas of Color-selected Quiescent Galaxies at z > 3 in Public JWST Fields}",
      journal = {\apj},
         year = 2023,
        month = apr,
       volume = {947},
       number = {1},
          eid = {20},
        pages = {20},
          doi = {10.3847/1538-4357/acbefa},
archivePrefix = {arXiv},
       eprint = {2302.10936},
 primaryClass = {astro-ph.GA},
       adsurl = {https://ui.adsabs.harvard.edu/abs/2023ApJ...947...20V}
}

@ARTICLE{Villaescusa-Navarro2021CAMELS,
       author = {{Villaescusa-Navarro}, Francisco and {Angl{\'e}s-Alc{\'a}zar}, Daniel and {Genel}, Shy and {Spergel}, David N. and {Somerville}, Rachel S. and {Dave}, Romeel and {Pillepich}, Annalisa and {Hernquist}, Lars and {Nelson}, Dylan and {Torrey}, Paul and et al.},
        title = "{The CAMELS Project: Cosmology and Astrophysics with Machine-learning Simulations}",
      journal = {\apj},
         year = 2021,
        month = jul,
       volume = {915},
       number = {1},
          eid = {71},
        pages = {71},
          doi = {10.3847/1538-4357/abf7ba},
archivePrefix = {arXiv},
       eprint = {2010.00619},
 primaryClass = {astro-ph.CO},
       adsurl = {https://ui.adsabs.harvard.edu/abs/2021ApJ...915...71V}
}

@ARTICLE{Ni2022Astrid,
       author = {{Ni}, Yueying and {Di Matteo}, Tiziana and {Bird}, Simeon and {Croft}, Rupert and {Feng}, Yu and {Chen}, Nianyi and {Tremmel}, Michael and {DeGraf}, Colin and {Li}, Yin},
        title = "{The ASTRID simulation: the evolution of supermassive black holes}",
      journal = {\mnras},
         year = 2022,
        month = jun,
       volume = {513},
       number = {1},
        pages = {670-692},
          doi = {10.1093/mnras/stac351},
archivePrefix = {arXiv},
       eprint = {2110.14154},
 primaryClass = {astro-ph.GA},
       adsurl = {https://ui.adsabs.harvard.edu/abs/2022MNRAS.513..670N}
}

@ARTICLE{Crain2015,
       author = {{Crain}, Robert A. and {Schaye}, Joop and {Bower}, Richard G. and {Furlong}, Michelle and {Schaller}, Matthieu and {Theuns}, Tom and {Dalla Vecchia}, Claudio and {Frenk}, Carlos S. and {McCarthy}, Ian G. and {Helly}, John C. and et al.},
        title = "{The EAGLE simulations of galaxy formation: calibration of subgrid physics and model variations}",
      journal = {\mnras},
         year = 2015,
        month = jun,
       volume = {450},
       number = {2},
        pages = {1937-1961},
          doi = {10.1093/mnras/stv725},
archivePrefix = {arXiv},
       eprint = {1501.01311},
 primaryClass = {astro-ph.GA},
       adsurl = {https://ui.adsabs.harvard.edu/abs/2015MNRAS.450.1937C}
}

@ARTICLE{Torrey2014,
       author = {{Torrey}, Paul and {Vogelsberger}, Mark and {Genel}, Shy and {Sijacki}, Debora and {Springel}, Volker and {Hernquist}, Lars},
        title = "{A model for cosmological simulations of galaxy formation physics: multi-epoch validation}",
      journal = {\mnras},
         year = 2014,
        month = mar,
       volume = {438},
       number = {3},
        pages = {1985-2004},
          doi = {10.1093/mnras/stt2295},
archivePrefix = {arXiv},
       eprint = {1305.4931},
 primaryClass = {astro-ph.CO},
       adsurl = {https://ui.adsabs.harvard.edu/abs/2014MNRAS.438.1985T}
}

@ARTICLE{Schaye2008,
       author = {{Schaye}, Joop and {Dalla Vecchia}, Claudio},
        title = "{On the relation between the Schmidt and Kennicutt-Schmidt star formation laws and its implications for numerical simulations}",
      journal = {\mnras},
         year = 2008,
        month = jan,
       volume = {383},
       number = {3},
        pages = {1210-1222},
          doi = {10.1111/j.1365-2966.2007.12639.x},
archivePrefix = {arXiv},
       eprint = {0709.0292},
 primaryClass = {astro-ph},
       adsurl = {https://ui.adsabs.harvard.edu/abs/2008MNRAS.383.1210S}
}

@ARTICLE{Appleby2020,
       author = {{Appleby}, Sarah and {Dav{\'e}}, Romeel and {Kraljic}, Katarina and {Angl{\'e}s-Alc{\'a}zar}, Daniel and {Narayanan}, Desika},
        title = "{The impact of quenching on galaxy profiles in the SIMBA simulation}",
      journal = {\mnras},
         year = 2020,
        month = jun,
       volume = {494},
       number = {4},
        pages = {6053-6071},
          doi = {10.1093/mnras/staa1169},
archivePrefix = {arXiv},
       eprint = {1911.02041},
 primaryClass = {astro-ph.GA},
       adsurl = {https://ui.adsabs.harvard.edu/abs/2020MNRAS.494.6053A}
}

@ARTICLE{Donnari2019,
       author = {{Donnari}, Martina and {Pillepich}, Annalisa and {Nelson}, Dylan and {Vogelsberger}, Mark and {Genel}, Shy and {Weinberger}, Rainer and {Marinacci}, Federico and {Springel}, Volker and {Hernquist}, Lars},
        title = "{The star formation activity of IllustrisTNG galaxies: main sequence, UVJ diagram, quenched fractions, and systematics}",
      journal = {\mnras},
         year = 2019,
        month = jun,
       volume = {485},
       number = {4},
        pages = {4817-4840},
          doi = {10.1093/mnras/stz712},
archivePrefix = {arXiv},
       eprint = {1812.07584},
 primaryClass = {astro-ph.GA},
       adsurl = {https://ui.adsabs.harvard.edu/abs/2019MNRAS.485.4817D}
}

@ARTICLE{Schindler2025,
       author = {{Schindler}, Jan-Torge and {Hennawi}, Joseph F. and {Davies}, Frederick B. and {Bosman}, Sarah E.~I. and {Endsley}, Ryan and {Wang}, Feige and {Yang}, Jinyi and {Barth}, Aaron J. and {Eilers}, Anna-Christina and {Fan}, Xiaohui and {Kakiichi}, Koki and {Maseda}, Michael and {Pizzati}, Elia and {Nanni}, Riccardo},
        title = "{A little red dot at z = 7.3 within a large galaxy overdensity}",
      journal = {Nature Astronomy},
         year = 2025,
        month = nov,
       volume = {9},
       number = {11},
        pages = {1732-1744},
          doi = {10.1038/s41550-025-02660-1},
archivePrefix = {arXiv},
       eprint = {2411.11534},
 primaryClass = {astro-ph.GA},
       adsurl = {https://ui.adsabs.harvard.edu/abs/2025NatAs...9.1732S}
}

@ARTICLE{Salpete1955,
       author = {{Salpeter}, Edwin E.},
        title = "{The Luminosity Function and Stellar Evolution.}",
      journal = {\apj},
         year = 1955,
        month = jan,
       volume = {121},
        pages = {161},
          doi = {10.1086/145971},
       adsurl = {https://ui.adsabs.harvard.edu/abs/1955ApJ...121..161S}
}

@ARTICLE{LaChance2025,
       author = {{LaChance}, Patrick and {Croft}, Rupert and {Ni}, Yueying and {Chen}, Nianyi and {Matteo}, Tiziana Di and {Bird}, Simeon},
        title = "{The evolution of galaxy morphology from redshift z=6 to 3: Mock JWST observations of galaxies in the ASTRID simulation}",
      journal = {The Open Journal of Astrophysics},
         year = 2025,
        month = feb,
       volume = {8},
          eid = {20},
        pages = {20},
          doi = {10.33232/001c.129991},
archivePrefix = {arXiv},
       eprint = {2401.16608},
 primaryClass = {astro-ph.GA},
       adsurl = {https://ui.adsabs.harvard.edu/abs/2025OJAp....8E..20L}
}

@ARTICLE{Bhowmick2024,
       author = {{Bhowmick}, Aklant K. and {Blecha}, Laura and {Torrey}, Paul and {Somerville}, Rachel S. and {Kelley}, Luke Zoltan and {Vogelsberger}, Mark and {Weinberger}, Rainer and {Hernquist}, Lars and {Sivasankaran}, Aneesh},
        title = "{Growth of high-redshift supermassive black holes from heavy seeds in the BRAHMA cosmological simulations: implications of overmassive black holes}",
      journal = {\mnras},
         year = 2024,
        month = sep,
       volume = {533},
       number = {2},
        pages = {1907-1926},
          doi = {10.1093/mnras/stae1819},
archivePrefix = {arXiv},
       eprint = {2406.14658},
 primaryClass = {astro-ph.GA},
       adsurl = {https://ui.adsabs.harvard.edu/abs/2024MNRAS.533.1907B}
}

@ARTICLE{Saintonge2016,
       author = {{Saintonge}, Amelie and {Catinella}, Barbara and {Cortese}, Luca and {Genzel}, Reinhard and {Giovanelli}, Riccardo and {Haynes}, Martha P. and {Janowiecki}, Steven and {Kramer}, Carsten and {Lutz}, Katharina A. and {Schiminovich}, David and {Tacconi}, Linda J. and {Wuyts}, Stijn and {Accurso}, Gioacchino},
        title = "{Molecular and atomic gas along and across the main sequence of star-forming galaxies}",
      journal = {\mnras},
         year = 2016,
        month = oct,
       volume = {462},
       number = {2},
        pages = {1749-1756},
          doi = {10.1093/mnras/stw1715},
archivePrefix = {arXiv},
       eprint = {1607.05289},
 primaryClass = {astro-ph.GA},
       adsurl = {https://ui.adsabs.harvard.edu/abs/2016MNRAS.462.1749S}
}

@ARTICLE{Renzini2015,
       author = {{Renzini}, Alvio and {Peng}, Ying-jie},
        title = "{An Objective Definition for the Main Sequence of Star-forming Galaxies}",
      journal = {\apjl},
         year = 2015,
        month = mar,
       volume = {801},
       number = {2},
          eid = {L29},
        pages = {L29},
          doi = {10.1088/2041-8205/801/2/L29},
archivePrefix = {arXiv},
       eprint = {1502.01027},
 primaryClass = {astro-ph.GA},
       adsurl = {https://ui.adsabs.harvard.edu/abs/2015ApJ...801L..29R}
}

@ARTICLE{Li2023,
       author = {{Li}, Yang A. and {Ho}, Luis C. and {Shangguan}, Jinyi and {Zhuang}, Ming-Yang and {Li}, Ruancun},
        title = "{Panchromatic Photometry of Low-redshift, Massive Galaxies Selected from SDSS Stripe 82}",
      journal = {\apjs},
         year = 2023,
        month = jul,
       volume = {267},
       number = {1},
          eid = {17},
        pages = {17},
          doi = {10.3847/1538-4365/acd4b5},
archivePrefix = {arXiv},
       eprint = {2307.13461},
 primaryClass = {astro-ph.GA},
       adsurl = {https://ui.adsabs.harvard.edu/abs/2023ApJS..267...17L}
}

@ARTICLE{Chaikin2026,
       author = {{Chaikin}, Evgenii and {Schaye}, Joop and {Hu{\v{s}}ko}, Filip and {Lacey}, Cedric G. and {Ploeckinger}, Sylvia and {Schaller}, Matthieu},
        title = "{The importance of super-Eddington black hole accretion for the emergence of massive quiescent galaxies at high redshift}",
      journal = {\mnras},
         year = 2026,
        month = aug,
       volume = {550},
       number = {2},
          eid = {stag1148},
        pages = {stag1148},
          doi = {10.1093/mnras/stag1148},
archivePrefix = {arXiv},
       eprint = {2601.15207},
 primaryClass = {astro-ph.GA},
       adsurl = {https://ui.adsabs.harvard.edu/abs/2026MNRAS.550g1148C}
}

@ARTICLE{Graff2025,
       author = {{de Graaff}, Anna and {Setton}, David J. and {Brammer}, Gabriel and {Cutler}, Sam and {Suess}, Katherine A. and {Labb{\'e}}, Ivo and {Leja}, Joel and {Weibel}, Andrea and {Maseda}, Michael V. and {Whitaker}, Katherine E. and {Bezanson}, Rachel and {Boogaard}, Leindert A. and {Cleri}, Nikko J. and {De Lucia}, Gabriella and {Franx}, Marijn and {Greene}, Jenny E. and {Hirschmann}, Michaela and {Matthee}, Jorryt and {McConachie}, Ian and {Naidu}, Rohan P. and {Oesch}, Pascal A. and {Price}, Sedona H. and {Rix}, Hans-Walter and {Valentino}, Francesco and {Wang}, Bingjie and {Williams}, Christina C.},
        title = "{Efficient formation of a massive quiescent galaxy at redshift 4.9}",
      journal = {Nature Astronomy},
         year = 2025,
        month = feb,
       volume = {9},
        pages = {280-292},
          doi = {10.1038/s41550-024-02424-3},
archivePrefix = {arXiv},
       eprint = {2404.05683},
 primaryClass = {astro-ph.GA},
       adsurl = {https://ui.adsabs.harvard.edu/abs/2025NatAs...9..280D}
}

@ARTICLE{Turner2026,
       author = {{Turner}, Jack C. and {Roper}, Will J. and {Vijayan}, Aswin P. and {Newman}, Sophie L. and {Wilkins}, Stephen M. and {Lovell}, Christopher C. and {Liao}, Shihong and {Seeyave}, Louise T.~C.},
        title = "{The nature of high-redshift massive quiescent galaxies ─ searching for RUBIES-UDS-QG-z7 in FLARES}",
      journal = {\mnras},
         year = 2026,
        month = mar,
       volume = {546},
       number = {4},
          eid = {stag214},
        pages = {stag214},
          doi = {10.1093/mnras/stag214},
archivePrefix = {arXiv},
       eprint = {2509.16111},
 primaryClass = {astro-ph.GA},
       adsurl = {https://ui.adsabs.harvard.edu/abs/2026MNRAS.546ag214T}
}

@ARTICLE{Ludlow2019,
       author = {{Ludlow}, Aaron D. and {Schaye}, Joop and {Schaller}, Matthieu and {Richings}, Jack},
        title = "{Energy equipartition between stellar and dark matter particles in cosmological simulations results in spurious growth of galaxy sizes}",
      journal = {\mnras},
         year = 2019,
        month = sep,
       volume = {488},
       number = {1},
        pages = {L123-L128},
          doi = {10.1093/mnrasl/slz110},
archivePrefix = {arXiv},
       eprint = {1903.10110},
 primaryClass = {astro-ph.GA},
       adsurl = {https://ui.adsabs.harvard.edu/abs/2019MNRAS.488L.123L}
}

@ARTICLE{Ludlow2023,
       author = {{Ludlow}, Aaron D. and {Fall}, S. Michael and {Wilkinson}, Matthew J. and {Schaye}, Joop and {Obreschkow}, Danail},
        title = "{Spurious heating of stellar motions by dark matter particles in cosmological simulations of galaxy formation}",
      journal = {\mnras},
         year = 2023,
        month = nov,
       volume = {525},
       number = {4},
        pages = {5614-5630},
          doi = {10.1093/mnras/stad2615},
archivePrefix = {arXiv},
       eprint = {2306.05753},
 primaryClass = {astro-ph.GA},
       adsurl = {https://ui.adsabs.harvard.edu/abs/2023MNRAS.525.5614L}
}

@ARTICLE{Lijy2025,
       author = {Li, Junyao and Shen, Yue and Zhuang, Ming-Yang},
        title = "{A prevalent population of normal-mass central black holes in high-redshift massive galaxies}",
      journal = {arXiv e-prints},
         year = 2025,
        month = feb,
          eid = {arXiv:2502.05048},
        pages = {arXiv:2502.05048},
          doi = {10.48550/arXiv.2502.05048},
archivePrefix = {arXiv},
       eprint = {2502.05048},
 primaryClass = {astro-ph.GA},
       adsurl = {https://ui.adsabs.harvard.edu/abs/2025arXiv250205048L}
}

@ARTICLE{Juodzbalis2026,
       author = {{Juod{\v{z}}balis}, Ignas and {Maiolino}, Roberto and {Baker}, William M. and {Lake}, Emma Curtis and {Scholtz}, Jan and {D'Eugenio}, Francesco and {Trefoloni}, Bartolomeo and {Isobe}, Yuki and {Tacchella}, Sandro and {Bunker}, Andrew J. and {Carniani}, Stefano and {Charlot}, St{\'e}phane and {Jones}, Gareth C. and {Parlanti}, Eleonora and {Perna}, Michele and {Rinaldi}, Pierluigi and {Robertson}, Brant and {{\"U}bler}, Hannah and {Venturi}, Giacomo and {Willott}, Chris},
        title = "{JADES: comprehensive census of broad-line AGN from reionization to cosmic noon revealed by JWST}",
      journal = {\mnras},
         year = 2026,
        month = mar,
       volume = {546},
       number = {3},
          eid = {stag086},
        pages = {stag086},
          doi = {10.1093/mnras/stag086},
archivePrefix = {arXiv},
       eprint = {2504.03551},
 primaryClass = {astro-ph.GA},
       adsurl = {https://ui.adsabs.harvard.edu/abs/2026MNRAS.546ag086J}
}

@ARTICLE{Ren2025,
       author = {{Ren}, Wenke and {Silverman}, John D. and {Faisst}, Andreas L. and {Fujimoto}, Seiji and {Yan}, Lin and {Liu}, Zhaoxuan and {Tsujita}, Akiyoshi and {Aravena}, Manuel and {Davies}, Rebecca L. and {De Looze}, Ilse and {Dessauges-Zavadsky}, Miroslava and {Herrera-Camus}, Rodrigo and {Ibar}, Edo and {Jones}, Gareth C. and {Kartaltepe}, Jeyhan S. and {Koekemoer}, Anton M. and {Lin}, Yu-Heng and {Mitsuhashi}, Ikki and {Molina}, Juan and {Nanni}, Ambra and {Relano}, Monica and {Romano}, Michael and {Sanders}, David B. and {Solimano}, Manuel and {Veraldi}, Enrico and {Villanueva}, Vicente and {Wang}, Wuji and {Zamorani}, Giovanni},
        title = "{The ALPINE─CRISTAL─JWST survey: revealing less massive black holes in high-redshift galaxies}",
      journal = {\mnras},
         year = 2025,
        month = nov,
       volume = {544},
       number = {1},
        pages = {211-233},
          doi = {10.1093/mnras/staf1709},
archivePrefix = {arXiv},
       eprint = {2509.02027},
 primaryClass = {astro-ph.GA},
       adsurl = {https://ui.adsabs.harvard.edu/abs/2025MNRAS.544..211R}
}

@ARTICLE{Chen2026,
       author = {{Chen}, Longyue and {Wang}, Tao and {Sun}, Hanwen and {Xu}, Ke and {Zhou}, Luwenjia and {Yang}, Tiancheng and {Tarrasse}, Maxime and {Mo}, Houjun and {Li}, Zhaozhou and {Chen}, Yangyao and {Dekel}, Avishai and {Daddi}, Emanuele and {Ding}, Xuheng and {Giavalisco}, Mauro and {Elbaz}, David},
        title = "{A Bending in the Size─Mass Relation of Star-forming Galaxies across 0.5 < z < 6.0 at a Critical Stellar Mass of {}10$^{10}$M$_{{\ensuremath{\odot}}}$ Revealed by JWST}",
      journal = {\apjl},
         year = 2026,
        month = may,
       volume = {1002},
       number = {2},
          eid = {L40},
        pages = {L40},
          doi = {10.3847/2041-8213/ae5a93},
archivePrefix = {arXiv},
       eprint = {2603.22239},
 primaryClass = {astro-ph.GA},
       adsurl = {https://ui.adsabs.harvard.edu/abs/2026ApJ..1002L..40C}
}

@ARTICLE{Moster2011,
       author = {{Moster}, Benjamin P. and {Somerville}, Rachel S. and
                 {Newman}, Jeffrey A. and {Rix}, Hans-Walter},
        title = "{A Cosmic Variance Cookbook}",
      journal = {\apj},
         year = 2011,
        month = apr,
       volume = {731},
       number = {2},
          eid = {113},
        pages = {113},
          doi = {10.1088/0004-637X/731/2/113},
archivePrefix = {arXiv},
       eprint = {1001.1737},
 primaryClass = {astro-ph.CO}
}

@ARTICLE{Trenti2008,
       author = {{Trenti}, Michele and {Stiavelli}, Massimo},
        title = "{Cosmic Variance and Its Effect on the Luminosity
                  Function Determination in Deep High-Redshift Surveys}",
      journal = {\apj},
         year = 2008,
        month = apr,
       volume = {676},
       number = {2},
        pages = {767-780},
          doi = {10.1086/528674},
archivePrefix = {arXiv},
       eprint = {0712.0398},
 primaryClass = {astro-ph}
}

@ARTICLE{Tinker2010,
       author = {{Tinker}, Jeremy L. and {Robertson}, Brant E. and
                 {Kravtsov}, Andrey V. and {Klypin}, Anatoly and
                 {Warren}, Michael S. and {Yepes}, Gustavo and
                 {Gottl{\"o}ber}, Stefan},
        title = "{The Large-scale Bias of Dark Matter Halos:
                  Numerical Calibration and Model Tests}",
      journal = {\apj},
         year = 2010,
        month = dec,
       volume = {724},
       number = {2},
        pages = {878-886},
          doi = {10.1088/0004-637X/724/2/878},
archivePrefix = {arXiv},
       eprint = {1001.3162},
 primaryClass = {astro-ph.CO}
}

@ARTICLE{Pandya2026,
    author = {{Pandya}, Viraj and {Bryan}, Greg L. and {Makinen}, T. Lucas and {Gabrielpillai}, Austen and {Carr}, Christopher and {Fielding}, Drummond B. and {Hernquist}, Lars and {Ho}, Matthew and {Iyer}, Kartheik and {Jespersen}, Christian Kragh and {Koudmani}, Sophie and {Laska}, Marta and {Lemos}, Pablo and {Lovell}, Christopher C. and {Perez}, Lucia A. and {Robinson}, Jr., William F. and {Somerville}, Rachel S. and {Starkenburg}, Tjitske K. and {Stiskalek}, Richard and {Terrazas}, Bryan and {Voit}, G. Mark},
    title = "{Introducing sapphire: Towards Hybrid Physics-Informed, Data-Driven Modeling of Galaxy Formation}",
    journal = {ApJ},
    year = {2026},
    note = {submitted},
    eprint = {2604.06318},
    archivePrefix = {arXiv},
    primaryClass = {astro-ph.GA}
}

@article{Hewett2006,
   title={The UKIRT Infrared Deep Sky Survey ZY JHK photometric system: passbands and synthetic colours},
   volume={367},
   ISSN={0035-8711},
   url={http://dx.doi.org/10.1111/j.1365-2966.2005.09969.x},
   DOI={10.1111/j.1365-2966.2005.09969.x},
   number={2},
   journal={Monthly Notices of the Royal Astronomical Society},
   publisher={Oxford University Press (OUP)},
   author={Hewett, P. C. and Warren, S. J. and Leggett, S. K. and Hodgkin, S. T.},
   year={2006},
   month=Apr, pages={454–468} }

@ARTICLE{Bessell1990,
       author = {{Bessell}, M.~S.},
        title = "{UBVRI passbands.}",
      journal = {\pasp},
         year = 1990,
        month = oct,
       volume = {102},
        pages = {1181-1199},
          doi = {10.1086/132749},
       adsurl = {https://ui.adsabs.harvard.edu/abs/1990PASP..102.1181B}
}

@ARTICLE{Williams2009,
       author = {{Williams}, Rik J. and {Quadri}, Ryan F. and {Franx}, Marijn and {van Dokkum}, Pieter and {Labb{\'e}}, Ivo},
        title = "{Detection of Quiescent Galaxies in a Bicolor Sequence from Z = 0-2}",
      journal = {\apj},
         year = 2009,
        month = feb,
       volume = {691},
       number = {2},
        pages = {1879-1895},
          doi = {10.1088/0004-637X/691/2/1879},
archivePrefix = {arXiv},
       eprint = {0806.0625},
 primaryClass = {astro-ph},
       adsurl = {https://ui.adsabs.harvard.edu/abs/2009ApJ...691.1879W}
}

@article{Paquereau2025,
    title={Tracing the galaxy-halo connection with galaxy clustering in COSMOS-Web from $z=0.1$ to $z\sim12$},
    author={Paquereau, Louis and others},
    journal={A\&A},
    volume={692},
    pages={A128},
    year={2025},
    eprint={2501.11674},
    archivePrefix={arXiv},
    primaryClass={astro-ph.GA}
}

@article{Shuntov2025_clumpy,
    title={Clumpiness of galaxies revealed in the near-infrared with COSMOS-Web},
    author={Shuntov, Marko and others},
    journal={A\&A},
    volume={696},
    pages={A114},
    year={2025},
    eprint={2506.13881},
    archivePrefix={arXiv},
    primaryClass={astro-ph.GA}
}

@article{Ji2026PANORAMIC,
    author = {{Ji}, Zhiyuan and {Williams}, Christina C. and {Behroozi}, Peter and {Weibel}, Andrea and {Jespersen}, Christian Kragh and {Oesch}, Pascal A. and {Bezanson}, Rachel and {Whitaker}, Katherine E. and {Greene}, Jenny E. and {Brammer}, Gabriel and {Dayal}, Pratika and {Labb{\'e}}, Ivo and {Manning}, Sinclaire M. and {Rinaldi}, Pierluigi and {Xiao}, Mengyuan and {Zhang}, Yunchong},
    title = "{PANORAMIC: The Dawn of Massive Quiescent Galaxies I. Number Density and Cosmic Variance from 1000 arcmin$^2$ NIRCam Imaging}",
    journal = {ApJ},
    year = {2026},
    note = {submitted},
    eprint = {2604.05022},
    archivePrefix = {arXiv},
    primaryClass = {astro-ph.GA}
}

@article{Patel2012,
    title={The UVJ selection of quiescent and star forming galaxies: separating early and late-type galaxies and isolating edge-on spirals},
    author={Patel, Shannon G. and Holden, Bradford P. and Kelson, Daniel D. and Franx, Marijn and van der Wel, Arjen and Illingworth, Garth D.},
    journal={ApJ},
    volume={748},
    pages={128},
    year={2012},
    eprint={1107.3147},
    archivePrefix={arXiv},
    primaryClass={astro-ph.CO}
}

@article{Whitaker2012,
    title={CANDELS: Correlations of SEDs and Morphologies with Star-formation Status for Massive Galaxies at z ~ 2},
    author={Whitaker, Katherine E. and van Dokkum, Pieter G. and Brammer, Gabriel and Momcheva, Ivelina G. and Skelton, Rosalind and Franx, Marijn and Kriek, Mariska and Labb{\'e}, Ivo and Fumagalli, Mattia and Lundgren, Britt F. and Nelson, Erica J. and Patel, Shannon G. and Rix, Hans-Walter},
    journal={ApJ},
    volume={754},
    pages={108},
    year={2012},
    eprint={1204.4194},
    archivePrefix={arXiv},
    primaryClass={shturl.cc/s}
}

@article{Zuckerman2021,
    title={Reproducing the UVJ Color Distribution of Star-forming Galaxies at $0.5<z<2.5$ with a Geometric Model of Dust Attenuation},
    author={Zuckerman, Leah D. and Belli, Sirio and Leja, Joel and Tacchella, Sandro},
    journal={ApJL},
    volume={922},
    pages={L32},
    year={2021},
    eprint={2109.14721},
    archivePrefix={arXiv},
    primaryClass={astro-ph.GA},
    doi={10.3847/2041-8213/ac3831}
}

@article{Rusakov2025,
    title={Little red dots as young supermassive black holes in dense ionized cocoons},
    author={Rusakov, Roman and Maiolino, Roberto and Carniani, Stefano and others},
    journal={Nature},
    volume={649},
    pages={574--579},
    year={2025},
    eprint={2503.16595},
    archivePrefix={arXiv},
    primaryClass={astro-ph.GA},
    doi={10.1038/s41586-025-05741-1}
}

@article{Kokubo2025,
       author = {{Kokubo}, Mitsuru and {Harikane}, Yuichi},
        title = "{Challenging the Active Galactic Nucleus Scenario for JWST/NIRSpec Little Red Dot and Non─Little Red Dot Broad H{\ensuremath{\alpha}} Emitters in Light of Nondetection of NIRCam Photometric Variability and X-Ray}",
      journal = {\apj},
         year = 2025,
        month = dec,
       volume = {995},
       number = {1},
          eid = {24},
        pages = {24},
          doi = {10.3847/1538-4357/ae119e},
archivePrefix = {arXiv},
       eprint = {2407.04777},
 primaryClass = {astro-ph.GA},
       adsurl = {https://ui.adsabs.harvard.edu/abs/2025ApJ...995...24K}
}

@article{Tee2025,
    author = {Tee, Wei Leong and Fan, Xiaohui and Wang, Feige and Yang, Jinyi},
    title = {Lack of Rest-frame Ultraviolet Variability in Little Red Dots Based on HST and JWST Observations},
    journal = {ApJL},
    volume = {983},
    pages = {L26},
    year = {2025},
    doi = {10.3847/2041-8213/adc5e3}
}

@ARTICLE{BoylanKolchin2010,
       author = {{Boylan-Kolchin}, Michael and {Springel}, Volker and {White}, Simon D.~M. and {Jenkins}, Adrian},
        title = "{There's no place like home? Statistics of Milky Way-mass dark matter haloes}",
      journal = {\mnras},
         year = 2010,
        month = aug,
       volume = {406},
       number = {2},
        pages = {896-912},
          doi = {10.1111/j.1365-2966.2010.16774.x},
archivePrefix = {arXiv},
       eprint = {0911.4484},
 primaryClass = {astro-ph.CO},
       adsurl = {https://ui.adsabs.harvard.edu/abs/2010MNRAS.406..896B}
}

@ARTICLE{2019Genel_scaling,
       author = {{Genel}, Shy and {Bryan}, Greg L. and {Springel}, Volker and {Hernquist}, Lars and {Nelson}, Dylan and {Pillepich}, Annalisa and {Weinberger}, Rainer and {Pakmor}, R{\"u}diger and {Marinacci}, Federico and {Vogelsberger}, Mark},
        title = "{A Quantification of the Butterfly Effect in Cosmological Simulations and Implications for Galaxy Scaling Relations}",
      journal = {\apj},
         year = 2019,
        month = jan,
       volume = {871},
       number = {1},
          eid = {21},
        pages = {21},
          doi = {10.3847/1538-4357/aaf4bb},
archivePrefix = {arXiv},
       eprint = {1807.07084},
 primaryClass = {astro-ph.GA},
       adsurl = {https://ui.adsabs.harvard.edu/abs/2019ApJ...871...21G}
}

@ARTICLE{Chuang2024,
       author = {{Chuang}, Chen-Yu and {Jespersen}, Christian Kragh and {Lin}, Yen-Ting and {Ho}, Shirley and {Genel}, Shy},
        title = "{Leaving No Branches Behind: Predicting Baryonic Properties of Galaxies from Merger Trees}",
      journal = {\apj},
         year = 2024,
        month = apr,
       volume = {965},
       number = {2},
          eid = {101},
        pages = {101},
          doi = {10.3847/1538-4357/ad2b6c},
archivePrefix = {arXiv},
       eprint = {2311.09162},
 primaryClass = {astro-ph.GA},
       adsurl = {https://ui.adsabs.harvard.edu/abs/2024ApJ...965..101C}
}

@ARTICLE{Ito2024,
       author = {{Ito}, Kei and {Valentino}, Francesco and {Brammer}, Gabriel and {Faisst}, Andreas L. and {Gillman}, Steven and {G{\'o}mez-Guijarro}, Carlos and {Gould}, Katriona M.~L. and {Heintz}, Kasper E. and {Ilbert}, Olivier and {Jespersen}, Christian Kragh and {Kokorev}, Vasily and {Kubo}, Mariko and {Magdis}, Georgios E. and {McPartland}, Conor J.~R. and {Onodera}, Masato and {Rizzo}, Francesca and {Tanaka}, Masayuki and {Toft}, Sune and {Vijayan}, Aswin P. and {Weaver}, John R. and {Whitaker}, Katherine E. and {Wright}, Lillian},
        title = "{Size─Stellar Mass Relation and Morphology of Quiescent Galaxies at z {\ensuremath{\geq}} 3 in Public JWST Fields}",
      journal = {\apj},
         year = 2024,
        month = apr,
       volume = {964},
       number = {2},
          eid = {192},
        pages = {192},
          doi = {10.3847/1538-4357/ad2512},
archivePrefix = {arXiv},
       eprint = {2307.06994},
 primaryClass = {astro-ph.GA},
       adsurl = {https://ui.adsabs.harvard.edu/abs/2024ApJ...964..192I}
}

@ARTICLE{Matharu2024,
       author = {{Matharu}, Jasleen and {Nelson}, Erica J. and {Brammer}, Gabriel and {Oesch}, Pascal A. and {Allen}, Natalie and {Shivaei}, Irene and {Naidu}, Rohan P. and {Chisholm}, John and {Covelo-Paz}, Alba and {Fudamoto}, Yoshinobu and {Giovinazzo}, Emma and {Herard-Demanche}, Thomas and {Kerutt}, Josephine and {Kramarenko}, Ivan and {Marchesini}, Danilo and {Meyer}, Romain A. and {Prieto-Lyon}, Gonzalo and {Reddy}, Naveen and {Shuntov}, Marko and {Weibel}, Andrea and {Wuyts}, Stijn and {Xiao}, Mengyuan},
        title = "{A first look at spatially resolved star formation at 4.8 < z < 6.5 with JWST FRESCO NIRCam slitless spectroscopy}",
      journal = {\aap},
         year = 2024,
        month = oct,
       volume = {690},
          eid = {A64},
        pages = {A64},
          doi = {10.1051/0004-6361/202450522},
archivePrefix = {arXiv},
       eprint = {2404.17629},
 primaryClass = {astro-ph.GA},
       adsurl = {https://ui.adsabs.harvard.edu/abs/2024A&A...690A..64M}
}

@ARTICLE{Li2025ASPIRE,
       author = {{Li}, Zihao and {Kakiichi}, Koki and {Christensen}, Lise and {Cai}, Zheng and {Dekel}, Avishai and {Fan}, Xiaohui and {Farina}, Emanuele Paolo and {Jun}, Hyunsung D. and {Li}, Zhaozhou and {Li}, Mingyu and {Pudoka}, Maria and {Sun}, Fengwu and {Trebitsch}, Maxime and {Walter}, Fabian and {Wang}, Feige and {Yang}, Jinyi and {Zhang}, Huanian and {Zou}, Siwei},
        title = "{Insights on metal enrichment and environmental effects at z {\ensuremath{\approx}} 5─7 with JWST ASPIRE/EIGER and the chemical evolution model}",
      journal = {\aap},
         year = 2025,
        month = nov,
       volume = {703},
          eid = {A106},
        pages = {A106},
          doi = {10.1051/0004-6361/202555372},
archivePrefix = {arXiv},
       eprint = {2504.18616},
 primaryClass = {astro-ph.GA},
       adsurl = {https://ui.adsabs.harvard.edu/abs/2025A&A...703A.106L}
}

@ARTICLE{Garcia2025,
       author = {{Garcia}, Alex M. and {Torrey}, Paul and {Ellison}, Sara L. and {Grasha}, Kathryn and {Chen}, Qian-Hui and {Hemler}, Z.~S. and {Zimmerman}, Dhruv T. and {Wright}, Ruby J. and {Zovaro}, Henry R.~M. and {Nelson}, Erica J. and {Sanders}, Ryan L. and {Kewley}, Lisa J. and {Hernquist}, Lars},
        title = "{Does the fundamental metallicity relation evolve with redshift? - II. The evolution in normalization of the mass-metallicity relation}",
      journal = {\mnras},
         year = 2025,
        month = jan,
       volume = {536},
       number = {1},
        pages = {119-144},
          doi = {10.1093/mnras/stae2587},
archivePrefix = {arXiv},
       eprint = {2407.06254},
 primaryClass = {astro-ph.GA},
       adsurl = {https://ui.adsabs.harvard.edu/abs/2025MNRAS.536..119G}
}

@ARTICLE{Hirschmann2023,
       author = {{Hirschmann}, Michaela and {Charlot}, Stephane and {Somerville}, Rachel S.},
        title = "{High-redshift metallicity calibrations for JWST spectra: insights from line emission in cosmological simulations}",
      journal = {\mnras},
         year = 2023,
        month = dec,
       volume = {526},
       number = {3},
        pages = {3504-3518},
          doi = {10.1093/mnras/stad2745},
archivePrefix = {arXiv},
       eprint = {2305.03753},
 primaryClass = {astro-ph.GA},
       adsurl = {https://ui.adsabs.harvard.edu/abs/2023MNRAS.526.3504H}
}

@ARTICLE{Jespersen2025_cv,
       author = {{Jespersen}, Christian Kragh and {Steinhardt}, Charles L. and {Somerville}, Rachel S. and {Lovell}, Christopher C.},
        title = "{On the Significance of Rare Objects at High Redshift: The Impact of Cosmic Variance}",
      journal = {\apj},
         year = 2025,
        month = mar,
       volume = {982},
       number = {1},
          eid = {23},
        pages = {23},
          doi = {10.3847/1538-4357/adb422},
archivePrefix = {arXiv},
       eprint = {2403.00050},
 primaryClass = {astro-ph.GA},
       adsurl = {https://ui.adsabs.harvard.edu/abs/2025ApJ...982...23J}
}

@ARTICLE{Jespersen2025_qg,
       author = {{Jespersen}, Christian Kragh and {Carnall}, Adam C. and {Lovell}, Christopher C.},
        title = "{Explaining Ultramassive Quiescent Galaxies at 3 < z < 5 in the Context of Their Environments}",
      journal = {\apjl},
         year = 2025,
        month = jul,
       volume = {988},
       number = {1},
          eid = {L19},
        pages = {L19},
          doi = {10.3847/2041-8213/adeb7c},
archivePrefix = {arXiv},
       eprint = {2507.05340},
 primaryClass = {astro-ph.GA},
       adsurl = {https://ui.adsabs.harvard.edu/abs/2025ApJ...988L..19J}
}

@ARTICLE{Gehrels1986,
       author = {{Gehrels}, N.},
        title = "{Confidence Limits for Small Numbers of Events in Astrophysical Data}",
      journal = {\apj},
         year = 1986,
        month = apr,
       volume = {303},
        pages = {336},
          doi = {10.1086/164079},
       adsurl = {https://ui.adsabs.harvard.edu/abs/1986ApJ...303..336G}
}

@ARTICLE{Weibel2026_cv,
       author = {{Weibel}, Andrea and {Jespersen}, Christian Kragh and {Oesch}, Pascal A. and {Williams}, Christina C. and {Bezanson}, Rachel and {Brammer}, Gabriel and {Cloonan}, Aidan P. and {Dayal}, Pratika and {Hutter}, Anne and {Ji}, Zhiyuan and {Maseda}, Michael V. and {Shuntov}, Marko and {Whitaker}, Katherine E.},
        title = "{Exploring Cosmic Dawn with PANORAMIC. II. Cosmic Variance and Galaxy Clustering at z {\ensuremath{\sim}} 10}",
      journal = {\apj},
         year = 2026,
        month = jul,
       volume = {1005},
       number = {1},
          eid = {118},
        pages = {118},
          doi = {10.3847/1538-4357/ae7716},
archivePrefix = {arXiv},
       eprint = {2512.14212},
 primaryClass = {astro-ph.GA},
       adsurl = {https://ui.adsabs.harvard.edu/abs/2026ApJ..1005..118W}
}

@ARTICLE{Pacifici2023,
       author = {{Pacifici}, Camilla and {Iyer}, Kartheik G. and {Mobasher}, Bahram and {da Cunha}, Elisabete and {Acquaviva}, Viviana and {Burgarella}, Denis and {Calistro Rivera}, Gabriela and {Carnall}, Adam C. and {Chang}, Yu-Yen and {Chartab}, Nima and {Cooke}, Kevin C. and {Fairhurst}, Ciaran and {Kartaltepe}, Jeyhan and {Leja}, Joel and {Ma{\l}ek}, Katarzyna and {Salmon}, Brett and {Torelli}, Marianna and {Vidal-Garc{\'\i}a}, Alba and {Boquien}, M{\'e}d{\'e}ric and {Brammer}, Gabriel G. and {Brown}, Michael J.~I. and {Capak}, Peter L. and {Chevallard}, Jacopo and {Circosta}, Chiara and {Croton}, Darren and {Davidzon}, Iary and {Dickinson}, Mark and {Duncan}, Kenneth J. and {Faber}, Sandra M. and {Ferguson}, Harry C. and {Fontana}, Adriano and {Guo}, Yicheng and {Haeussler}, Boris and {Hemmati}, Shoubaneh and {Jafariyazani}, Marziye and {Kassin}, Susan A. and {Larson}, Rebecca L. and {Lee}, Bomee and {Mantha}, Kameswara Bharadwaj and {Marchi}, Francesca and {Nayyeri}, Hooshang and {Newman}, Jeffrey A. and {Pandya}, Viraj and {Pforr}, Janine and {Reddy}, Naveen and {Sanders}, Ryan and {Shah}, Ekta and {Shahidi}, Abtin and {Stevans}, Matthew L. and {Triani}, Dian Puspita and {Tyler}, Krystal D. and {Vanderhoof}, Brittany N. and {de la Vega}, Alexander and {Wang}, Weichen and {Weston}, Madalyn E.},
        title = "{The Art of Measuring Physical Parameters in Galaxies: A Critical Assessment of Spectral Energy Distribution Fitting Techniques}",
      journal = {\apj},
         year = 2023,
        month = feb,
       volume = {944},
       number = {2},
          eid = {141},
        pages = {141},
          doi = {10.3847/1538-4357/acacff},
archivePrefix = {arXiv},
       eprint = {2212.01915},
 primaryClass = {astro-ph.GA},
       adsurl = {https://ui.adsabs.harvard.edu/abs/2023ApJ...944..141P}
}

@ARTICLE{Papovich2023,
       author = {{Papovich}, Casey and {Cole}, Justin W. and {Yang}, Guang and {Finkelstein}, Steven L. and {Barro}, Guillermo and {Buat}, V{\'e}ronique and {Burgarella}, Denis and {P{\'e}rez-Gonz{\'a}lez}, Pablo G. and {Santini}, Paola and {Seill{\'e}}, Lise-Marie and {Shen}, Lu and {Arrabal Haro}, Pablo and {Bagley}, Micaela B. and {Bell}, Eric F. and {Bisigello}, Laura and {Calabr{\`o}}, Antonello and {Casey}, Caitlin M. and {Castellano}, Marco and {Chworowsky}, Katherine and {Cleri}, Nikko J. and {Costantin}, Luca and {Cooper}, M.~C. and {Dickinson}, Mark and {Ferguson}, Henry C. and {Fontana}, Adriano and {Giavalisco}, Mauro and {Grazian}, Andrea and {Grogin}, Norman A. and {Hathi}, Nimish P. and {Holwerda}, Benne W. and {Hutchison}, Taylor A. and {Kartaltepe}, Jeyhan S. and {Kewley}, Lisa J. and {Kirkpatrick}, Allison and {Kocevski}, Dale D. and {Koekemoer}, Anton M. and {Larson}, Rebecca L. and {Long}, Arianna S. and {Lucas}, Ray A. and {Pentericci}, Laura and {Pirzkal}, Nor and {Ravindranath}, Swara and {Somerville}, Rachel S. and {Trump}, Jonathan R. and {Urbano Stawinski}, Stephanie M. and {Weiner}, Benjamin J. and {Wilkins}, Stephen M. and {Yung}, L.~Y. Aaron and {Zavala}, Jorge A.},
        title = "{CEERS Key Paper. V. Galaxies at 4 < z < 9 Are Bluer than They Appear─Characterizing Galaxy Stellar Populations from Rest-frame {\ensuremath{\sim}}1 {\ensuremath{\mu}}m Imaging}",
      journal = {\apjl},
         year = 2023,
        month = jun,
       volume = {949},
       number = {2},
          eid = {L18},
        pages = {L18},
          doi = {10.3847/2041-8213/acc948},
archivePrefix = {arXiv},
       eprint = {2301.00027},
 primaryClass = {astro-ph.GA},
       adsurl = {https://ui.adsabs.harvard.edu/abs/2023ApJ...949L..18P}
}

@ARTICLE{Steinhardt2023,
       author = {{Steinhardt}, Charles L. and {Kokorev}, Vasily and {Rusakov}, Vadim and {Garcia}, Ethan and {Sneppen}, Albert},
        title = "{Templates for Fitting Photometry of Ultra-high-redshift Galaxies}",
      journal = {\apjl},
         year = 2023,
        month = jul,
       volume = {951},
       number = {2},
          eid = {L40},
        pages = {L40},
          doi = {10.3847/2041-8213/acdef6},
archivePrefix = {arXiv},
       eprint = {2208.07879},
 primaryClass = {astro-ph.GA},
       adsurl = {https://ui.adsabs.harvard.edu/abs/2023ApJ...951L..40S}
}

@ARTICLE{Behroozi2019,
       author = {{Behroozi}, Peter and {Wechsler}, Risa H. and {Hearin}, Andrew P. and {Conroy}, Charlie},
        title = "{UNIVERSEMACHINE: The correlation between galaxy growth and dark matter halo assembly from z = 0-10}",
      journal = {\mnras},
         year = 2019,
        month = sep,
       volume = {488},
       number = {3},
        pages = {3143-3194},
          doi = {10.1093/mnras/stz1182},
archivePrefix = {arXiv},
       eprint = {1806.07893},
 primaryClass = {astro-ph.GA},
       adsurl = {https://ui.adsabs.harvard.edu/abs/2019MNRAS.488.3143B}
}

@ARTICLE{Greene2024,
       author = {{Greene}, Jenny E. and {Labbe}, Ivo and {Goulding}, Andy D. and {Furtak}, Lukas J. and {Chemerynska}, Iryna and {Kokorev}, Vasily and {Dayal}, Pratika and {Volonteri}, Marta and {Williams}, Christina C. and {Wang}, Bingjie and {Setton}, David J. and {Burgasser}, Adam J. and {Bezanson}, Rachel and {Atek}, Hakim and {Brammer}, Gabriel and {Cutler}, Sam E. and {Feldmann}, Robert and {Fujimoto}, Seiji and {Glazebrook}, Karl and {de Graaff}, Anna and {Khullar}, Gourav and {Leja}, Joel and {Marchesini}, Danilo and {Maseda}, Michael V. and {Matthee}, Jorryt and {Miller}, Tim B. and {Naidu}, Rohan P. and {Nanayakkara}, Themiya and {Oesch}, Pascal A. and {Pan}, Richard and {Papovich}, Casey and {Price}, Sedona H. and {van Dokkum}, Pieter and {Weaver}, John R. and {Whitaker}, Katherine E. and {Zitrin}, Adi},
        title = "{UNCOVER Spectroscopy Confirms the Surprising Ubiquity of Active Galactic Nuclei in Red Sources at z > 5}",
      journal = {\apj},
         year = 2024,
        month = mar,
       volume = {964},
       number = {1},
          eid = {39},
        pages = {39},
          doi = {10.3847/1538-4357/ad1e5f},
archivePrefix = {arXiv},
       eprint = {2309.05714},
 primaryClass = {astro-ph.GA},
       adsurl = {https://ui.adsabs.harvard.edu/abs/2024ApJ...964...39G}
}

@ARTICLE{Liuyk2026,
    author = {{Liu}, Yikai and {Guo}, Hong and {Springel}, Volker and {Liu}, Kexin and {Ho}, Luis C. and {Zhu}, BoCheng},
    title = "{The VariableTNG project: mass-dependent regulation of galaxy morphology by baryonic feedback}",
    journal = {A\&A},
    year = {2026},
    note = {submitted},
    eprint = {2608.19543},
    archivePrefix = {arXiv},
    primaryClass = {astro-ph.GA}
}

\begin{appendix} 

\section{Radiative transfer procedure}
\label{sec:appendixA}
This appendix summarizes the radiative transfer and mock-observation pipeline adopted in this work. The setup closely follows established post-processing procedures applied to high-redshift cosmological simulations (e.g., \citet{Trcka22,Vijayan2022,Punyasheel2025}) and is implemented using the Monte Carlo radiative transfer code SKIRT (version 9).

All modelling choices—including stellar population synthesis models, birth-cloud treatment, dust prescriptions, and AGN templates—are adopted directly from the literature or inherited from the simulation outputs, without additional tuning to match specific observational scaling relations. Therefore, the trends discussed in the main text reflect variations in the underlying feedback parameters rather than adjustments in the radiative transfer implementation.

Below we describe the treatment of stellar emission, dust modelling, AGN contribution, and image construction in detail.
\subsection{Stellar emission}
Stellar emission constitutes the primary radiation source. For each galaxy, stellar particles older than 10 Myr are assigned simple stellar population (SSP) spectral energy distributions from \citet{bc03}, assuming a \citet{chabrier03} initial mass function. The particle properties—including initial mass, position, metallicity, and age—are directly extracted from the simulation outputs.

To balance computational efficiency and accuracy, we use $5\times10^{6}$ photon packets per galaxy. The smoothing length of each stellar particle is defined as the distance to its $N_k$-th nearest neighbour, identified via a KD-tree algorithm. We adopt $N_k=32$ \citep{Trcka22,tng50atlas24}, reducing this value dynamically when fewer neighbouring particles are available. For systems with extremely few particles, a global smoothing length based on the mean radial distance from the galaxy centre is adopted.

\subsection{Young stellar particles and birth clouds}
Stellar particles younger than 10 Myr are assumed to reside within dusty birth clouds. For these particles, we employ MAPPINGS III templates \citep{groves08}, parameterized by metallicity, compactness, ISM pressure, and photodissociation region (PDR) covering fraction. The templates are scaled according to the particle’s star formation rate, defined as its birth mass averaged over 10 Myr to match the assumptions of the SED library.

We adopt a constant ISM pressure of $P=1.38\times10^{-12}\,\mathrm{Pa}$. The compactness parameter is drawn from a lognormal distribution with mean $\log C=5$ and standard deviation 0.4 \citep{Kapoor21,Trcka22}. The PDR covering fraction evolves with particle age following
\begin{equation}
  f_{\mathrm{PDR}}(t) = e^{-t/\tau},
\end{equation}
where $\tau=3\,\mathrm{Myr}$, consistent with \citet{Trcka22}, so that younger H\,\textsc{ii} regions have higher covering fractions and the PDR clears exponentially with time. The stellar metallicity is taken directly from the simulation.

\subsection{Dust modelling}
To model dust attenuation and re-emission, we construct the three-dimensional dust distribution from the gaseous component of each galaxy. We separate interstellar medium gas from the hot circumgalactic medium using the criterion of \cite{Torrey2019}:
\begin{equation}
    \log \left( \frac{T_{\mathrm{gas}}}{\mathrm{K}} \right) < 6 + 0.25 \, \log \left( \frac{\rho_{\mathrm{gas}}}{10^{10} \, h^2 \, \mathrm{M}_{\odot}\, \mathrm{kpc}^{-3}} \right).
\end{equation}
Dust density is then computed as
\begin{equation}
    \rho_{\text{dust}} = f_{\text{dust}} \, Z_{\text{gas}} \, \rho_{\text{gas}}
\end{equation}
where $Z_{\mathrm{gas}}$ is the gas metallicity and $f_{\mathrm{dust}}$ is the dust-to-metal mass ratio.

For high-redshift galaxies, $f_{\rm dust}$ is typically lower than in local galaxies, and the evolution of the median dust-to-metal ratio with redshift has been identified in both IllustrisTNG \citep{vogelsberger20_modelc} and FLARES \citep{Vijayan2022}. We adopt the redshift-dependent dust-to-metal ratios from the Model~C calibration presented in Table~3 of \citet{vogelsberger20_modelc}. In particular, we apply the best-fitting values corresponding to $z=6$ and $z=8$, with $f_{\rm dust}=0.11$ and $f_{\rm dust}=0.06$, respectively, which are applied uniformly to all galaxies at the same redshift. The physical properties of the dust are modelled by the THEMIS dust model \citep{themis17}, which can reproduce observable properties of diffuse dust.

Because VTNG uses the moving-mesh code AREPO, gas properties are naturally defined on an unstructured Voronoi mesh. These Voronoi cells are directly imported into SKIRT and subsequently re-gridded onto a hierarchical octree grid \citep{saftly13}. We adopt refinement levels between 6 and 12, allowing higher resolution in dense regions while maintaining computational efficiency. This combination of original Voronoi inputs with octree-accelerated grids preserves the geometric characteristics of the dust distribution from the VTNG simulations while substantially enhancing the efficiency of the radiative transfer calculations. By dynamically adjusting the grid refinement, SKIRT achieves high-precision calculations in the complex dust environments of high-redshift galaxies, while optimizing computational resource usage.

\subsection{AGN contribution}
For galaxies hosting black hole particles, we include AGN emission in the radiative transfer calculations. In the standard unified model of AGNs, a dusty torus surrounds the central region, absorbs UV/optical radiation from the accretion disk, and re-emits it in the infrared \citep{Antonucci1993}. To model this process, we use the SKIRTOR clumpy torus model \citep{Stalevski2012,Stalevski2016}, which has been extensively implemented in SED modelling codes such as \textsc{CIGALE} for the decomposition of observed galaxy SEDs into stellar, dust, and AGN components \citep{Boquien2019,Yang2020}. SKIRTOR provides a large library of AGN SED templates computed with the radiative transfer code \textsc{SKIRT}, taking into account a wide variety of dusty torus structures and viewing angles. In this work, we adopt the template \texttt{t3\_p1\_q1\_oa30\_R30\_Mcl0.97\_i30\_sed.dat}, which yields a relatively strong UV contribution, due to the combination of a moderate opening angle, a clumpy structure, and the relatively low inclination angle. The definition of the SKIRTOR template parameters is given in \citet{Stalevski2012,Stalevski2016}.

The AGN is implemented as a central point source in each galaxy utilizing the \textsc{SKIRT} configuration files. The AGN template for each galaxy is scaled according to its bolometric luminosity, calculated as follows,
\begin{equation}
L_{\rm bol} = \eta \dot{M}_{\rm BH} c^2, \label{eq:Lbol}
\end{equation}
where $\eta = 0.2$ denotes the radiative efficiency of the black hole. Consequently, the AGN radiation, after passing through the torus, is treated as a template SED which propagates through the galaxy-scale dust using the \texttt{dustEmission} mode. Together with stellar light, the AGN light undergoes absorption, scattering, and re-emission by the interstellar medium.

\subsection{Instrument settings}

In the radiative transfer calculations performed with \textsc{SKIRT}, the virtual observational instruments are placed at a distance of 1\,Mpc from each galaxy. For each system, we adopt a fixed field of view of $60 \times 60$~pkpc$^2$, discretized into a $400 \times 400$ pixel image. This corresponds to a physical pixel scale of 0.15~pkpc per pixel, while the effective angular resolution depends on redshift, yielding pixel scales of approximately $0.026^{\prime\prime}$ and $0.030^{\prime\prime}$ at $z=6$ and $z=8$, respectively. The overall instrument configuration and image setup are chosen to be consistent with those adopted in the \textsc{FLARES} mock-observation framework for galaxies at $z>5$ \citep{Vijayan2022,Punyasheel2025}.

Our analysis focuses exclusively on galaxies at redshifts $z=6$ and $z=8$. At $z=6$, we select galaxies with stellar masses $M_\star > 10^{9}\,M_\odot$, while at $z=8$ we consider galaxies with $M_\star > 10^{8.5}\,M_\odot$. All galaxies are observed on the same image plane, such that differences in projected morphology arise naturally from the intrinsic three-dimensional orientations of the galaxies. 

For each galaxy, \textsc{SKIRT} outputs a full IFU-like data cube covering the wavelength range $0.1$--$16\,\mu\mathrm{m}$, sampled with 600 linearly spaced wavelength bins, providing a spatially resolved SED for every pixel across the field of view. All images are centred on the galaxy centre of potential, ensuring a consistent and physically motivated centring for all mock observations.

\subsection{Mock image construction and size measurements}
\label{app:mock_size}

We construct JWST/NIRCam mock images from the spatially resolved spectral cubes produced by \textsc{SKIRT}. For each spatial pixel, the emergent spectrum is integrated through the corresponding redshifted filter-transmission curve to produce an observer-frame image. The resulting specific intensity $I_\nu (\rm MJy/sr)$ is cosmologically dimmed by a factor of $(1+z)^{-3}$, with the emitted intensity evaluated at the corresponding rest-frame frequency. Mock images are generated in the F150W filter and are adopted for the galaxy-size measurements presented in this work. The original $60\times60,{\rm kpc}^{2}$ field of view is preserved while the images are resampled to approximately match the angular pixel scale of the CEERS mosaics ($\sim0.03,{\rm arcsec,pixel^{-1}}$). The resampled images are then convolved with the corresponding JWST/NIRCam PSF generated using \textsc{STPSF} \citep{Perrin2012}.

To reproduce the observational conditions of the CEERS imaging, we directly add each PSF-convolved, source-only image to multiple real CEERS F150W background cutouts. The backgrounds are extracted from the publicly available CEERS Data Release 0.5 NIRCam F150W mosaic for pointing 1, which has a pixel scale of $0.03''$ pixel$^{-1}$. We use the background-subtracted science image (\texttt{SCI\_BKSUB}), together with the corresponding weight (\texttt{WHT}) and background-mask (\texttt{BKGMASK}) maps. Candidate background regions are required to have at least 95\% valid weight-map coverage over the full cutout and 98\% within the central source-insertion region. Regions satisfying these coverage criteria are ranked according to their masked-pixel fractions, the presence of bright or extreme pixels, and large-scale background gradients. We select ten background regions with high weight-map coverage, low masked-pixel fractions, no prominent bright residuals, and stable background curves of growth. Within each region, a clean position satisfying these criteria is selected for source injection. For each galaxy, an identical source image is inserted into multiple CEERS background cutouts.

For each galaxy, we first measure the size in three independent CEERS background realizations. If at least two reliable measurements are obtained and they satisfy the background-stability criterion, the primary set is accepted. Otherwise, two additional background realizations are processed. The final effective radius is taken as the median of all reliable measurements available among the three or five realizations. A galaxy is retained only if at least two reliable measurements are obtained and their scatter in \(\log_{10}R_{\rm e}\) does not exceed 0.25 dex. Galaxies that still fail either requirement after all five background realizations are excluded from the final size sample.

For each realization, the local background is estimated from an annulus spanning \(90\)--\(120\) pixels from the injected position. We estimate the background in two steps. First, invalid pixels and sources flagged in the original CEERS mask are excluded, and an initial constant background is determined using \(3\sigma\) clipping. After subtracting this preliminary background, additional sources are detected directly from the background-subtracted image based on their significance relative to the local background RMS. Their masks are expanded to include the surrounding flux, and these pixels are excluded from the background annulus. The local background is then recalculated using only the remaining unmasked pixels. The final background level is subtracted from the inserted image, and the standard deviation of the remaining background pixels is adopted as the local background RMS.

Source detection is performed on the background-subtracted image using \textsc{photutils} \citep{Bradley2016}. The target is identified near the known injection position using a \(2\sigma_{\rm bkg}\) detection threshold and a minimum of three connected pixels. A smoothed 
\(1.5\sigma_{\rm bkg}\) segmentation map is then used to associate nearby fragments of the same galaxy. 
The centre is defined by the flux-weighted centroid of the detected target pixels in the noisy image. The segmentation map is used only to identify the target and determine its centre, which is not used to define the galaxy boundary. During the curve-of-growth measurement, unrelated sources and invalid pixels are masked, but all remaining pixels within the circular apertures are included. This allows genuine low-surface-brightness emission below the detection threshold to contribute to the measured flux and size.

Galaxy sizes are measured using a non-parametric circular
curve-of-growth method, with apertures increasing in steps of
\(0.5\) pixels up to a maximum radius of \(70\) pixels. Correlated background fluctuations are characterised using blank-sky curves of growth measured in the same CEERS background regions. Because the original CEERS image is available before source injection, its background-only curve of growth is used to help identify the radial interval over which the source flux reaches a stable plateau. This reduces the risk that background fluctuations shift the adopted total-flux interval to excessively large or small radii.

We first search for a strict plateau beyond both \(4\) pixels and three times the PSF half-light radius. The profile must remain consistent with the empirical background fluctuations over the plateau and show no significant renewed growth at larger radii. If no strict plateau is found, we search for a broader, approximately stable \(8\)-pixel interval beyond both \(8\) pixels and three times the PSF half-light radius. This fallback interval must have a fractional flux range no larger than \(0.20\). Profiles that remain significantly rising at the largest usable radius are rejected. The total flux is taken as the median cumulative flux of the noisy inserted image over the selected plateau interval, and \(R_{\rm e}\) is defined as the radius enclosing one half of this total flux.

An individual realization is considered reliable when the target is detected, a strict or fallback plateau is identified, the recovered centroid lies within \(8\) pixels of the injection position, and the measured effective radius is finite and positive. We additionally require the source to be unblended, to have \(S/N\geq2\), 
\({R_{\rm e}}/{R_{\rm PSF}}\geq1\), and an acceptable fraction of unmasked pixels within the total-flux aperture.

The final effective radius of each galaxy is taken as the median of all reliable measurements obtained from the three or five background realizations. At least two reliable measurements are required, and the scatter among them must satisfy \(\sigma(\log_{10}R_{\rm e})\leq0.25~{\rm dex}\). Galaxies that fail either requirement are excluded from the final size sample. This multi-background procedure accounts for the sensitivity of the recovered size to correlated noise, residual background structure, masking, and local source contamination.

\begin{figure*}
    \centering
    \includegraphics[width=\textwidth]{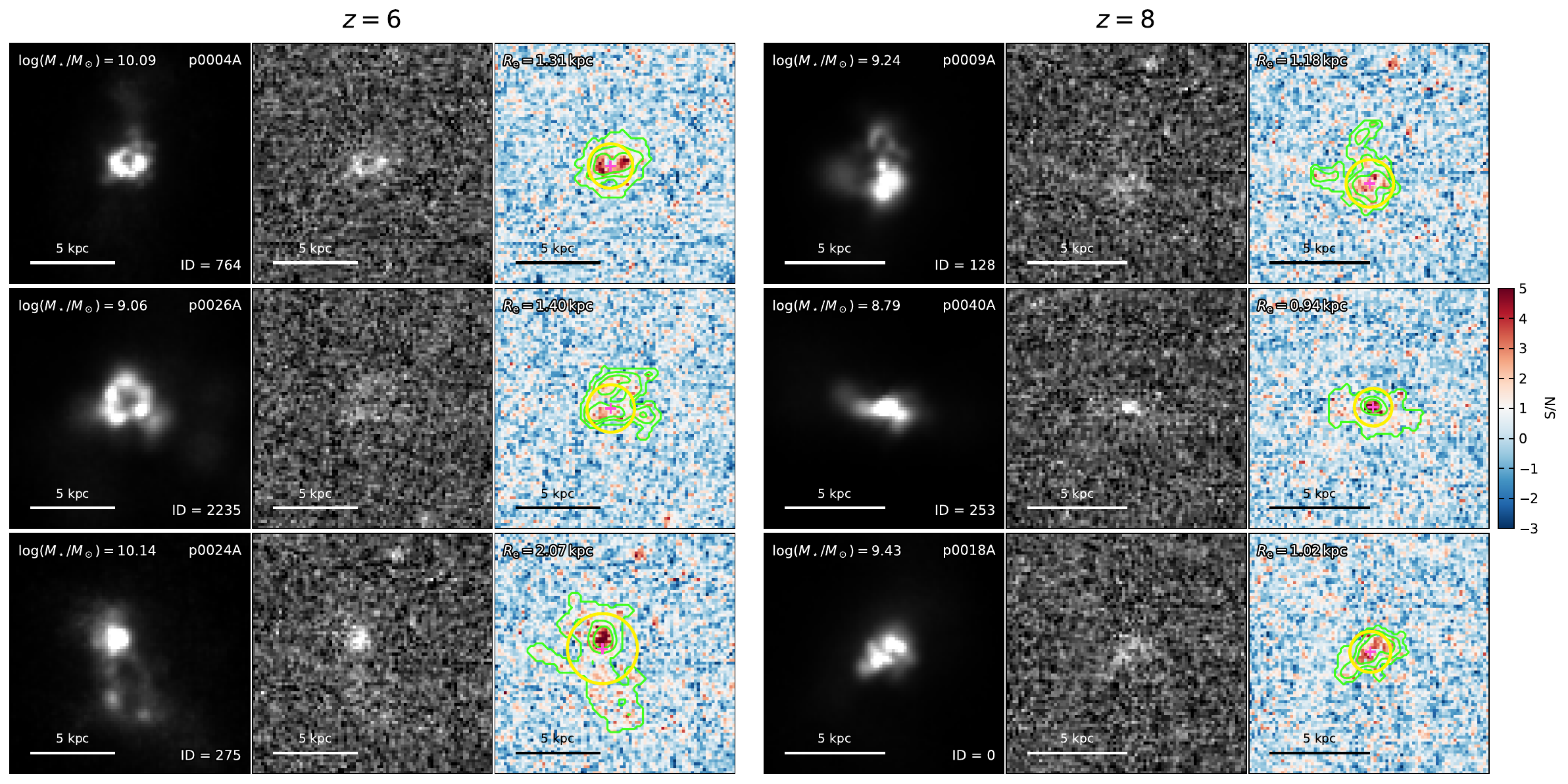}
    \caption{
    Examples of the mock-image construction and non-parametric size
    measurements for representative galaxies at \(z=6\) and \(z=8\).
    Each row contains one \(z=6\) galaxy in the left three panels and one \(z=8\) galaxy in the right three panels. For each galaxy, the panels show, from left to right, the PSF-convolved F150W image, the same source inserted into a real CEERS F150W background cutout, and the display-optimized, background-subtracted S/N map. Green curves show target-associated surface-brightness isophotes, the yellow circle indicates the recovered effective radius \(R_{\rm e}\), and the magenta cross marks the flux-weighted centre. 
    The stellar mass and simulation run are indicated in the source-only panel. Obvious masked background sources outside the protected target region are removed only in the displayed S/N maps; all size measurements are performed using the original noisy inserted images. The outermost green isophote corresponds to \(10\%\) of the peak of the smoothed target image. The isophotes are shown only for visualization and are not used to define the galaxy boundary or the curve-of-growth apertures.
    }
    \label{fig:mock_size_examples}
\end{figure*}

Figure~\ref{fig:mock_size_examples} illustrates the main stages of the mock-imaging and size-measurement procedure for representative galaxies at \(z=6\) and \(z=8\). For each galaxy, we show the PSF-convolved image, the same source after insertion into a real CEERS background, and the corresponding background-subtracted S/N map used for visualizing the recovered morphology. The examples illustrate that the curve-of-growth measurement is performed in the presence of realistic correlated noise and residual background structure. Because the circular apertures are not restricted to the source-detection pixels, low-surface-brightness emission outside the detected segmentation can also contribute to the measured flux and
size.

\section{Quantitative comparison with observations}
\label{app:chi2_results}

To complement the visual comparison with observational data, we compute a $\chi^2$-like statistical-distance metric for each scaling relation shown together with observations.
For each observational data set, we interpolate the median VTNG relation to the observed $x_{\mathrm{obs},i}$ and define
\begin{equation}
\chi^2_{\rm like} =
\sum_{i=1}^{N_{\rm obs}}
\frac{
\left[y_{\rm sim}(x_i)-y_{{\rm obs},i}\right]^2
}{
\sigma_{{\rm obs},i}^2+\sigma_{\rm sim}^2(x_i)
},
\label{eq:obs_chi2}
\end{equation}
where $y_{{\rm obs},i}$ and $\sigma_{{\rm obs},i}$ are the observed value and its quoted uncertainty, and $\sigma_{\rm sim}(x_i)$ is the simulated scatter evaluated at the same $x_i$.

For the SMF, we additionally include the correlated cosmic-variance contribution using
\begin{equation}
\chi^2_{\rm like}=\boldsymbol{\Delta}^{\rm T}\mathbf{C}^{-1}\boldsymbol{\Delta},
\qquad
C_{ij}=\delta_{ij}\left(\sigma_{{\rm obs},i}^2+\sigma_{{\rm sim},i}^2\right)
+\sigma_{{\rm cv},i}\sigma_{{\rm cv},j},
\label{eq:obs_chi2_smf}
\end{equation}
where $\Delta_i=y_{\rm sim}(x_i)-y_{{\rm obs},i}$. The last term is the cosmic-variance covariance derived in Appendix~\ref{app:cosmic_variance}, for which we adopt $\rho_{ij}=1$.

The comparison is performed in logarithmic space for quantities that span several orders of magnitude. Specifically, for the SMF we use $y=\log_{10}\Phi$, where $\Phi$ is the number density of galaxies in units of ${\rm Mpc}^{-3}$. For the SFMS we use $y=\log_{10}{\rm SFR}$, for the stellar MZR we use $y=\log_{10}(Z_\star/Z_\odot)$, for the gas-phase MZR we use $y=12+\log_{10}({\rm O/H})$, for the MSR we use \(y=\log_{10}(R_{\rm eff}/{\rm kpc})\), and for the black hole--stellar mass relation we use $y=\log_{10}(M_{\rm BH}/M_\odot)$. For the MSR, we calculate the statistical distance using the binned median VTNG relation interpolated to the observed stellar masses. Observational uncertainties are converted into the corresponding logarithmic units when necessary, and asymmetric uncertainties are symmetrized by averaging the upper and lower errors. Data points outside the stellar-mass range covered by the simulations are not included, in order to avoid extrapolation.

We report the normalized statistical distance $\chi^2_{\rm norm} = \chi^2_{\rm like}/N_{\rm obs}$.
We adopt $\chi^2_{\rm norm}$ rather than the classical reduced $\chi^2$ because no model parameters are fitted to the observational data here. The eight VTNG subgrid parameters define the simulation suite and are varied a priori, rather than being optimized separately for each observational data set. Thus, the VTNG ensemble is treated as a fixed prediction and summarized by its median relation and scatter. Our $\chi^2_{\rm norm}$ thus serves strictly as a descriptive measure of model--data tension, rather than a formal likelihood for parameter inference.

\begin{table*}
\centering
\caption{Quantitative comparison between the VTNG median relations and the observational datasets.}
\label{tab:chi2_summary}

\begin{tabular}{lccccc}
\hline
Relation & $N_{\rm obs}$ &
$\chi^2_{\rm like}$ & $\chi^2_{\rm norm}$ &
median $|\Delta|$ & mean $\Delta$ \\
\hline
\multicolumn{6}{c}{$z=6$} \\
\hline
SMF  & 41 & 97.82 & 2.39 & 0.17 & $-0.04$ \\
SMF (with CV)   & 41 & 71.25 & 1.74 & 0.17 & $-0.04$ \\
SFMS & 22 & 47.64 & 2.17 & 0.13 & 0.12 \\
Gas MZR & 62 & 205.04 & 3.31 & 0.29 & 0.31 \\
Stellar MZR & 17 & 19.48 & 1.15 & 0.02 & 0.06 \\
MSR & 43 & 977.11 & 22.72 & 0.57 & 0.56 \\
BH--$M_\star$ & 78 & 295.72 & 3.79 & 0.82 & $-0.59$ \\
\hline
\multicolumn{6}{c}{$z=8$} \\
\hline
SMF  & 21 & 77.00 & 3.67 & 0.26 & $-0.29$ \\
SMF (with CV)   & 21 & 36.66 & 1.75 & 0.26 & $-0.29$ \\
SFMS & 48 & 82.06 & 1.71 & 0.24 & 0.02 \\
Gas MZR & 54 & 212.92 & 3.94 & 0.33 & 0.38 \\
Stellar MZR & 11 & 4.63 & 0.42 & 0.20 & $-0.18$ \\
MSR & 101 & 1561.64 & 15.46 & 0.39 & 0.43 \\
BH--$M_\star$ & 12 & 38.67 & 3.22 & 0.80 & $-0.62$ \\
\hline
\end{tabular}

\tablefoot{
For each observational point, the VTNG median relation is interpolated to the same stellar mass. The offset is defined as $\Delta=y_{\rm sim}-y_{\rm obs}$, where $y=\log_{10}\Phi$ for the SMF, $\log_{10}{\rm SFR}$ for the SFMS, $12+\log_{10}({\rm O/H})$ for the gas-phase MZR, $\log_{10}(Z_\star/Z_\odot)$ for the stellar MZR, $\log_{10}{R_{\rm eff}}$ for the MSR, and $\log_{10}(M_{\rm BH}/M_\odot)$ for the BH--$M_\star$ relation. The quantity $\chi^2_{\rm norm}=\chi^2_{\rm like}/N_{\rm obs}$ is used as a descriptive tension metric rather than as a formal reduced chi-square. For the SMF, the comparison is restricted to the observational points for which the cosmic-variance covariance is available; the two SMF rows therefore use the same data points and differ only in whether this covariance is included. The median $|\Delta|$ measures the typical absolute vertical offset between the VTNG median relation and the observational data, while the mean $\Delta$ retains the sign of the offset and indicates whether the VTNG relation lies systematically above or below the observations. For the SFMS comparison, \citet{Venturi2024} is excluded from the metric because its very small quoted uncertainties would otherwise dominate the statistical distance, although it remains shown in the corresponding figure for visual comparison.
}
\end{table*}

We summarize the quantitative comparison between the VTNG median relations and the observational datasets in Table~\ref{tab:chi2_summary}. The results are  reported separately at $z=6$ and $z=8$. For each relation, we report the total $\chi^2_{\rm like}$, the normalized value $\chi^2_{\rm norm}=\chi^2_{\rm like}/N_{\rm obs}$, the median absolute offset,  and the mean offset. The offset is defined as $\Delta=y_{\rm sim}-y_{\rm obs}$, such that a negative value indicates that the VTNG median relation lies below the observational data.

The quantitative comparison largely supports the visual trends discussed above.
For the SMF, including the correlated cosmic-variance contribution reduces
$\chi^2_{\rm norm}$ from 2.39 to 1.74 at $z=6$ and from 3.67 to 1.75 at $z=8$.
The corresponding mean offsets are $-0.04$ and $-0.29$ dex, respectively.
Thus, cosmic variance reduces the overall statistical tension, although a residual discrepancy remains toward the massive end at $z=8$.

For the SFMS, the quantitative metric gives
$\chi^2_{\rm norm}=2.17$ at $z=6$ and $1.71$ at $z=8$, with mean offsets of $0.12$ and $0.02$ dex, respectively. Thus, the simulated SFMS does not show a large systematic normalization offset relative to the compiled observational data. The remaining statistical distance mainly reflects the scatter among individual observational samples and their uncertainties.

For the MZR, the gas-phase relation gives $\chi^2_{\rm norm}=3.31$ at $z=6$ and $3.94$ at $z=8$, with median absolute offsets of $0.29$ and $0.33$ dex, respectively. The corresponding mean signed offsets are $+0.31$ and $+0.38$ dex, indicating that the VTNG gas-phase MZR is systematically shifted toward higher metallicities relative to the compiled observational measurements at both redshifts. 
The stellar MZR gives $\chi^2_{\rm norm}=1.15$ at $z=6$ and $0.42$ at $z=8$, and is in better agreement with the available observational constraints.
These results are consistent with the trends discussed in
Section~\ref{sec:MZR}: the gas-phase MZR shows increased
tension in its normalization, whereas the stellar MZR remains
broadly compatible with the current measurements. The low
stellar-MZR value at $z=8$ should nevertheless be interpreted
cautiously because it is based on a smaller observational
sample, particularly at the high-mass end.

The MSR shows a strong tension among the relations considered here, with $\chi^2_{\rm norm}=22.72$ at $z=6$ and $15.46$ at $z=8$. The median absolute offsets are $0.57$ and $0.39$ dex, while the corresponding mean offsets are positive, with values of $0.56$ and $0.43$ dex. These positive offsets indicate that the VTNG median ${R_{\rm eff}}$ are systematically larger than the observational measurements at fixed stellar mass. In linear units, the mean offsets correspond to characteristic size differences of approximately factors of $3.6$ and $2.7$ at $z=6$ and $z=8$, respectively. Although the VTNG sizes are systematically larger, the simulated MSR broadly reproduces the observed increase in galaxy size with stellar mass.

The BH--$M_\star$ relation shows a large tension. The values of $\chi^2_{\rm norm}=3.79$ at $z=6$ and $3.22$ at $z=8$ are accompanied by negative mean offsets of $-0.59$ and $-0.62$ dex, respectively. This indicates that the VTNG median relation lies below the observational high-redshift AGN measurements at fixed stellar mass. As discussed in Section~\ref{sec:bh_mass}, this does not necessarily imply that the simulations fail to reproduce the full galaxy population. The observational samples are selected from active black holes and are therefore expected to probe the upper part of the intrinsic black-hole distribution. In addition, virial black-hole masses, bolometric corrections, and AGN classifications for LRD-like sources remain subject to substantial systematic uncertainties. We therefore interpret the large $\chi^2_{\rm norm}$ values as evidence that the currently observed AGN candidates occupy the upper envelope of the VTNG black-hole population, rather than as a direct one-to-one discrepancy with an unbiased galaxy sample.

\section{Mock-photometric UVJ diagnostic}
\label{app:appendix_uvj}

Additionally, we examine the rest-frame UVJ colour–colour diagram using the available \textsc{SKIRT} mock photometry. We restrict the UVJ analysis to galaxies with $\log(M_\star/M_\odot)\geq 7.5$ at $z=6$ and $z=8$, ensuring that the selected galaxies are represented by a sufficient number of stellar particles. For each galaxy, synthetic rest-frame magnitudes are derived by convolving the emergent \textsc{SKIRT} SEDs with standard filter transmission curves. We adopt the Bessell $U$ and $V$ filters \citep{Bessell1990} alongside the UKIRT/WFCAM $J$ band \citep{Hewett2006}, from which we construct the rest-frame optical colours $U-V$ and $V-J$.

The mock colours are computed using the \textsc{ExtinctionOnly} mode of \textsc{SKIRT}. This choice is appropriate for the rest-frame UV, optical, and near-infrared bands considered here, where dust primarily modifies the emergent stellar continuum through absorption and scattering. By contrast, thermal dust re-emission is expected to contribute predominantly at mid- and far-infrared wavelengths and therefore has a negligible effect on the UVJ colours analysed here.

\begin{figure*}
\centering
\includegraphics[width=0.95\textwidth]{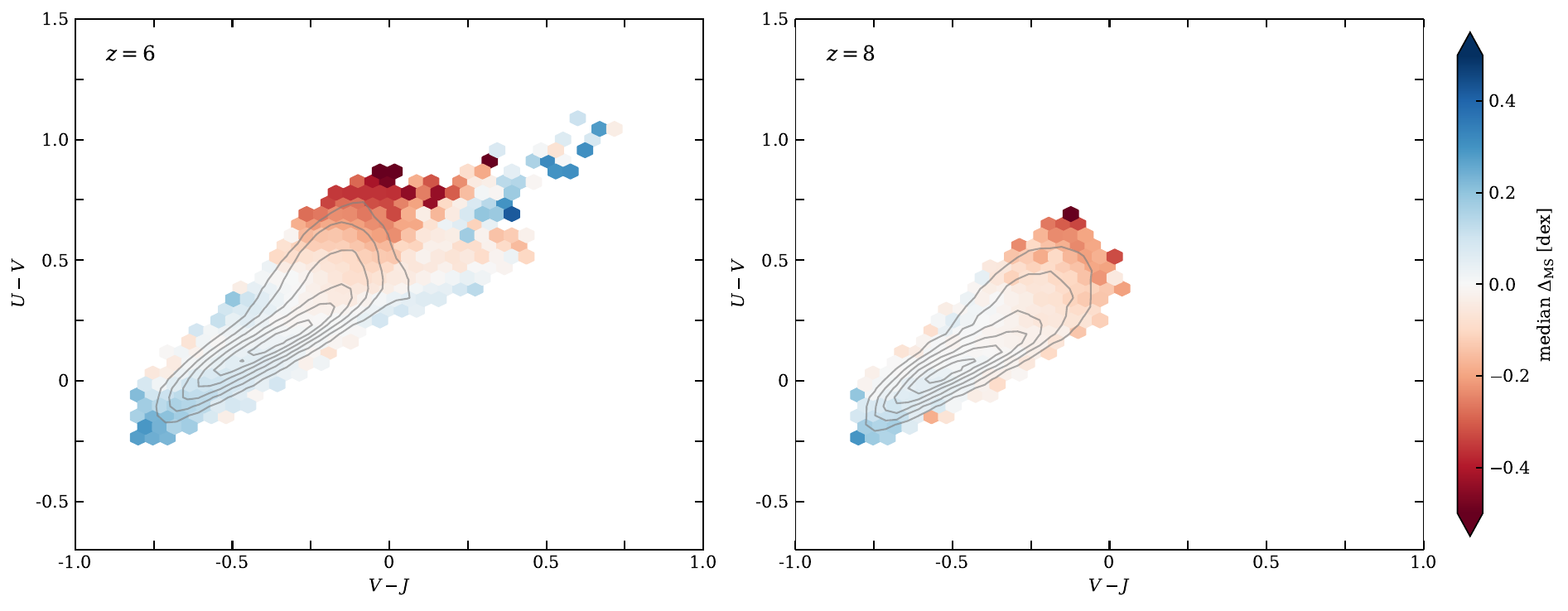}
\caption{
Rest-frame UVJ colour--colour diagram for galaxies with
$\log(M_\star/M_\odot)\geq 7.5$ at $z=6$ and $z=8$. The rest-frame colours are measured using the Bessell $U$ and $V$ bands and the UKIRT/WFCAM $J$ band. Grey contours show the smoothed projected number-density distribution in the UVJ plane, with contour levels corresponding to 5\%, 15\%, 30\%, 50\%, 70\%, and 90\% of the peak projected number density. Hexbins are colour-coded by the median $\Delta_{\rm MS}$ of galaxies in each UVJ bin. For each VTNG run, we compute 
$\Delta_{\rm MS}=\log{\rm SFR}-\log{\rm SFR}_{\rm MS}^{\rm run}$, where $\log{\rm SFR}_{\rm MS}^{\rm run}$ is the SFMS measured from that run. Only galaxies with robust $\Delta_{\rm MS}$ measurements are included in the UVJ hexbin colour mapping. Galaxies in stellar-mass bins that do not contain enough objects to define the run-specific SFMS are excluded. In total, valid $\Delta_{\rm MS}$ values are recovered for 99\% of the full UVJ sample at $z=6$ and 98\% at $z=8$.}

\label{fig:uvj_diagram}
\end{figure*}

The resulting UVJ distributions are shown in Fig.~\ref{fig:uvj_diagram}. The hexbins are colour-coded by the median $\Delta_{\rm MS}$, where $\Delta_{\rm MS}$ is measured relative to the SFMS defined for each VTNG run. The median $\Delta_{\rm MS}$ varies systematically across the main UVJ locus. Galaxies in the lower-left part of the diagram, with bluer $U-V$ and $V-J$ colours, tend to have positive or near-zero
$\Delta_{\rm MS}$, while galaxies toward the upper-right part of the locus show progressively lower $\Delta_{\rm MS}$. This behaviour is consistent with the expected interpretation of rest-frame UVJ colours: bluer colours are generally associated with actively star-forming systems, whereas redder colours reflect a combination of older stellar populations, lower specific star formation rates, and/or stronger dust attenuation \citep{Williams2009,Patel2012,Whitaker2012,Zuckerman2021}.
Over the stellar-mass and redshift range considered here, the simulated galaxies are dominated by star-forming systems, and no substantial separate quiescent population is identified in the SFMS analysis. Accordingly, we do not use the UVJ diagram to define a separate quiescent sample. Instead, the UVJ colour gradient is used as a mock-photometric consistency check, showing that the \textsc{SKIRT} mock-observation pipeline can connect intrinsic galaxy properties with observational colour diagnostics.

This demonstration also highlights the flexibility of the full \textsc{SKIRT} mock-photometry pipeline. In future work, we will extend this procedure to a broader set of directly observable diagnostics, including synthetic UV luminosity functions for direct comparison with recent observational measurements \citep{Weibel2026_cv}, as well as galaxy clustering statistics and field-to-field variance measurements \citep{Paquereau2025,Shuntov2025_clumpy,Ji2026PANORAMIC}, to further connect the simulations with high-redshift survey data and provide additional constraints on the underlying feedback parameters.

\section{Cosmic variance of the SMF of VTNG}
\label{app:cosmic_variance}

In this Appendix, we estimate the field-to-field cosmic variance of the SMF for a VTNG-sized volume. We consider both the cosmic-variance uncertainty in individual stellar-mass bins, which is used in the comparison with the observed SMFs in Section~\ref{sec:SMF},
and the integrated abundance over $7.5<\log_{10}(M_\star/M_\odot)<9.5$ used as the random-forest target. Because the VTNG simulations share identical initial conditions, this field-to-field variance is not included in the
run-to-run scatter of the VTNG ensemble. We therefore estimate cosmic variance separately in this section.

Within the standard linear-bias framework, the fractional cosmic variance of galaxies in stellar-mass bin $i$ is
\begin{equation}
    \sigma_{{\rm cv},i}^{\rm frac}(z)
    =
    b_{{\rm eff},i}(z)\,
    \sigma_{\rm dm}(V,z),
    \label{eq:app_sigma_cv}
\end{equation}
where $\sigma_{\rm dm}(V,z)$ denotes the rms fluctuation of the linear matter-density field within the simulation volume, and $b_{{\rm eff},i}$ is the effective large-scale bias of galaxies in the corresponding mass bin \citep{Trenti2008,Moster2011}. 

To estimate $\sigma_{\rm dm}$, we represent the
$L_{\rm box}=50\,{\rm Mpc/h}$ VTNG volume by an equivalent lightcone geometry with the same comoving volume. Specifically, we adopt a $30.3'\times30.3'$ footprint over $5.914<z<6.088$ at $z\simeq6$, and a $28.0'\times28.0'$ footprint
over $7.875<z<8.128$ at $z\simeq8$, chosen to match the comoving volume of the $50\,{\rm Mpc}/h$ VTNG box. We use the cosmic-variance calculator of \citet{Moster2011}, with the updated implementation of \citet{Jespersen2025_cv}, to determine the corresponding matter variance at $z=6$ and $z=8$. The resulting values are
\begin{equation}
    \sigma_{\rm dm}(z=6)=0.043,
    \qquad
    \sigma_{\rm dm}(z=8)=0.033 .
    \label{eq:app_sigma_dm_values}
\end{equation}

For each stellar-mass bin, we estimate the effective galaxy bias from
the host haloes of the galaxies in that bin. The host-halo mass is taken
from ${\tt Group\_M\_Crit200}$, and the corresponding halo bias is
calculated using the \citet{Tinker2010} fitting function. We then take
the number-weighted mean halo bias of the galaxies in each bin,
\begin{equation}
    b_{\rm eff}(z)
    =
    \frac{1}{N}
    \sum_g b_{\rm h}(M_{{\rm h},g},z),
\end{equation}
and adopt the median value across the 50 VTNG runs. For use in
Fig.~\ref{fig:SMF} and Eq.~\eqref{eq:obs_chi2_smf}, the fractional
cosmic-variance uncertainties are converted to logarithmic units,
$\sigma_{\rm cv,i}^{\rm dex}\simeq \sigma_{\rm cv,i}^{\rm frac}/\ln 10$.
Within the adopted linear-bias model, we treat the cosmic-variance fluctuations as a common-mode fluctuation across stellar-mass bins,
such that $C^{\rm cv}_{ij}=\sigma_{\rm cv,i}^{\rm dex}
\sigma_{\rm cv,j}^{\rm dex}$,
corresponding to $\rho_{ij}=1$.

For the integrated abundance used as the RF target,
$7.5<\log_{10}(M_\star/M_\odot)<9.5$, we calculate $b_{\rm eff}$
in the same way using all galaxies in the mass interval.
The resulting median effective biases are
\begin{equation}
    b_{\rm eff}=4.41 \quad (z=6),
    \qquad
    b_{\rm eff}=6.63 \quad (z=8).
\end{equation}
Combining these values with the matter fluctuations in
Eq.~(\ref{eq:app_sigma_dm_values}) gives
\begin{equation}
    \sigma_{\rm cv}^{\rm frac}
    =
    0.188 \quad (z=6),
    \qquad
    0.220 \quad (z=8).
\end{equation}
Expressed as logarithmic scatter, these correspond to
\begin{equation}
    \sigma_{\rm cv}^{\rm dex}
    =
    0.081~{\rm dex} \quad (z=6),
    \qquad
    0.095~{\rm dex} \quad (z=8).
    \label{eq:app_cv_integrated}
\end{equation}

For comparison, we measure the parameter-driven scatter in the same
integrated stellar-mass range directly from the 50 VTNG runs. For each
run $r$, we calculate the integrated number density $n_r$ and define $\sigma_{\rm param}^{\rm dex}=
\left(P_{84}[\log_{10}n_r]-P_{16}[\log_{10}n_r]\right)/2$.
We obtain
\begin{equation}
    \sigma_{\rm param}^{\rm dex}
    =
    0.035~{\rm dex} \quad (z=6),
    \qquad
    0.060~{\rm dex} \quad (z=8).
\end{equation}
Thus, the expected field-to-field cosmic variance exceeds the parameter-driven scatter of the integrated SMF abundance by factors of approximately $2.3$ at $z=6$ and $1.6$ at $z=8$.

To assess the possible impact of non-linear clustering, we use the public UniverseMachine mock lightcones \citep{Behroozi2019}. We place fixed VTNG-sized apertures within 24 mock sightlines, apply the same redshift intervals, and measure the sightline-to-sightline scatter in galaxy counts. We then compare the directly measured cosmic variance with the linear-bias prediction. \citet{Weibel2026_cv} showed that non-linear clustering becomes less important for larger fields, with the inferred bias approaching the linear prediction for field side lengths $\gtrsim20'$ at $z\sim10$. In our calculation for the VTNG-equivalent footprints, the median ratios of the measured scatter to the linear prediction are $1.10$ at $z=6$ and $1.07$ at $z=8$, while the corresponding ratios for the integrated $7.5<\log_{10}(M_\star/M_\odot)<9.5$ sample are $1.18$ and $1.11$. 
For these relatively large footprints, the contribution of enhanced small-scale clustering to the volume-averaged variance is modest. We therefore retain the linear-bias estimate as our fiducial cosmic-variance model, since adopting the full UniverseMachine scatter would not alter our qualitative conclusions.

\begingroup
\renewcommand{\thefigure}{E.\arabic{figure}}
\setcounter{figure}{0}

\begin{figure*}
    \centering
    \includegraphics[width=\textwidth]{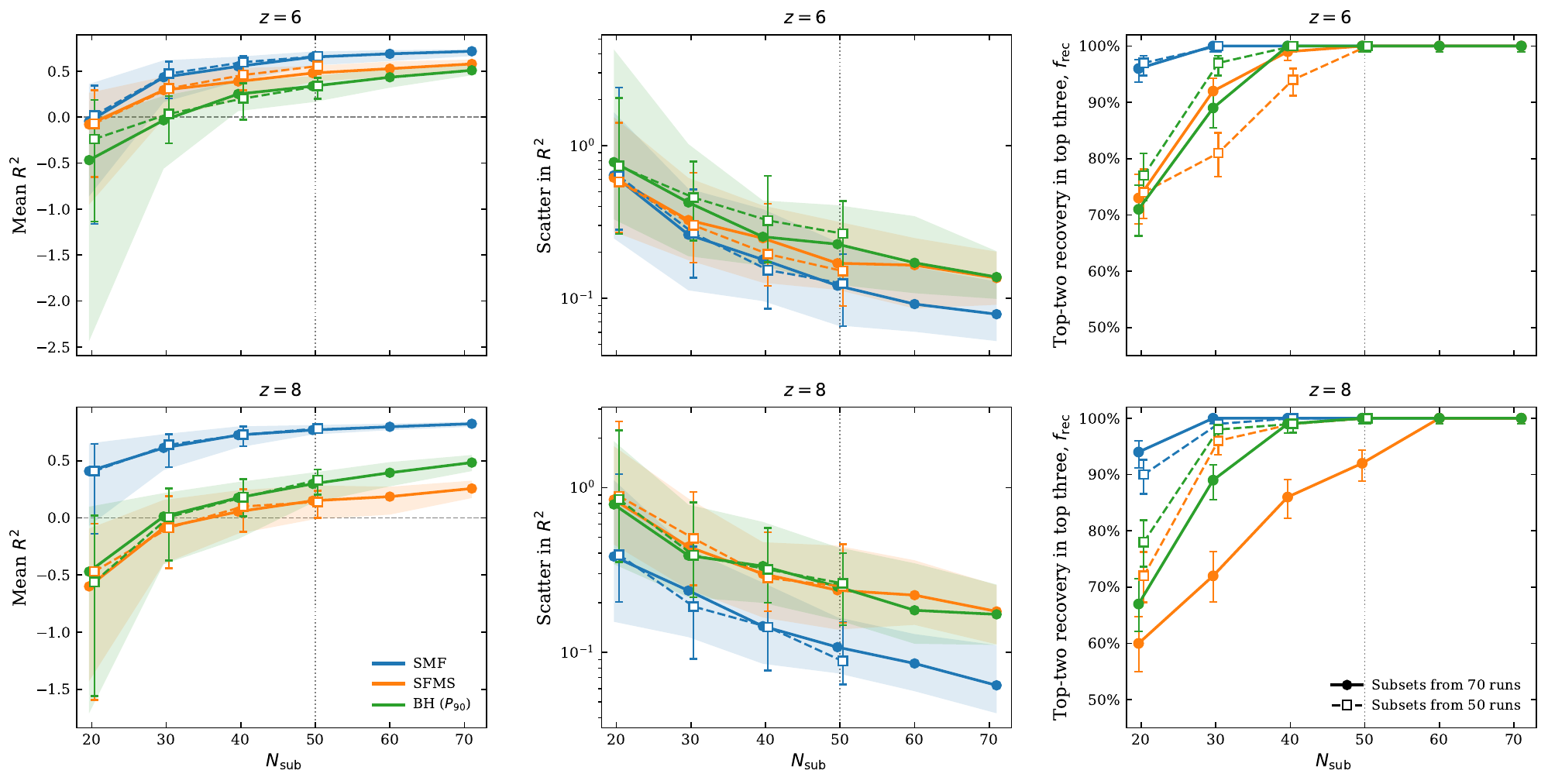}
    \caption{Dependence of the RF results on the number of simulations used for training at $z=6$ (top) and $z=8$ (bottom). Colours distinguish the SMF, SFMS scatter, and BH $P_{90}$ diagnostics. Solid curves with filled circles show random subsets drawn from the 70-simulation pool, while dashed curves with open squares show subsets restricted to the original runs 1--50. At $N_{\rm sub}=50$, the latter corresponds to the exact simulation sample used in the main analysis. The left column shows the median mean five-fold $R^2$, and the shaded regions or error bars indicate the 16th--84th percentile range across 100 repetitions. The middle column shows the median fold-to-fold standard deviation in $R^2$ on a logarithmic scale. The right column shows the recovery fraction $f_{\rm rec}$, defined as the fraction of repetitions in which the two highest-ranked parameters in the complete 70-simulation reference remain among the three highest-ranked parameters. The uncertainties on $f_{\rm rec}$ show approximately $68\%$ Wilson-score intervals. The vertical dotted line marks the $N_{\rm sub}=50$ sample size adopted in the main analysis.}
    \label{fig:rf_training_size}
\end{figure*}

\endgroup

\section{Dependence of the RF--SHAP results on the number of simulations}
\label{app:rf_robust}
\setcounter{figure}{1}

The RF analyses presented in the main text use the original set of 50 VTNG simulations, with the eight varied subgrid parameters adopted as input features. Given the limited number of simulations relative to the dimensionality of the parameter space, we test explicitly whether the predictive performance and the parameter sensitivities depend strongly on the size or composition of the training sample. Figure~\ref{fig:rf_training_size} summarizes how the predictive performance and SHAP-ranking stability vary with the size and composition of the training sample.

We perform two complementary training-size tests. First, we use
an enlarged sample containing 70 simulations and construct random
subsets with \(N_{\rm sub}=[20,\,30,\,40,\,50,\,60,\,70]\). Second, we restrict to the original runs 1--50 and draw random subsets with 
\(N_{\rm sub}=[20,\,30,\,40,\,50]\). At \(N_{\rm sub}=50\), the training sample is fixed to the exact set used in the main analysis, while only the cross-validation partition and RF random seed are varied. For each
sample size and each diagnostic, we perform 100 independent realisations using the same RF configuration and five-fold cross-validation procedure as in the main analysis.

We quantify the stability using three complementary statistics. The first is the mean \(R^2\) across the five cross-validation folds. The second is the standard deviation of \(R^2\), which quantifies the variation in predictive performance between the cross-validation folds. The third statistic measures the recovery of the leading SHAP-ranked parameters. We first identify the two highest-ranked parameters from the median global SHAP importance
obtained using the 70-simulation sample. For each repeated
analysis, we then test whether both reference parameters remain among the three highest-ranked parameters. The corresponding recovery fraction is defined as \(f_{\rm rec}=N_{\rm rec}/N_{\rm rep}\), where \(N_{\rm rep}=100\), and \(N_{\rm rec}\) is the number of repetitions in which both reference parameters are recovered among the three highest-ranked parameters. Their internal ordering is not
required to remain unchanged. The uncertainty on \(f_{\rm rec}\) is estimated using an approximately $68\%$ Wilson-score interval.

The original 50-simulation sample and random 50-simulation subsets drawn from the enlarged sample give similar predictive performance. Across the six diagnostics, their median \(R^2\) values differ by less than \(0.08\). The close agreement between the original 50-simulation sample and random 50-simulation subsets drawn from the enlarged Sobol sample indicates that the main RF results are not strongly sensitive to the particular choice of runs. This suggests that the original 50-run sample is broadly representative of the enlarged design and is sufficient to recover the leading predictive trends.

For the SMF, increasing the sample from the original 50 simulations to the 70-simulation set raises the median \(R^2\) by approximately \(8\%\) at \(z=6\) and \(5\%\) at \(z=8\). For the SFMS scatter, the corresponding increase is approximately \(4\%\) at \(z=6\), while the predictive performance remains weak at \(z=8\). The BH \(P_{90}\)
models also show improved predictive performance at both redshifts.
Overall, the enlarged sample increases the reliability of the RF predictions but does not materially change the qualitative results or the identification of the leading parameters.

The leading SHAP-ranked parameters are more stable than the
predictive scores. At $N_{\rm sub}=50$, random subsets drawn from the enlarged sample recover both reference top-two parameters within their top three in approximately 92--100 per cent of the repetitions. The original 50-simulation sample gives similarly high recovery fractions when the cross-validation partition and RF seed are varied. Thus, although the precise ordering of the lower-ranked parameters can change, the dominant parameters are already largely stable at the sample size used in the main analysis.

The main-text analysis is based on the original 50-run sample, which was the complete set available when the analysis was carried out. The additional simulations are used here to test the robustness of those results. The enlarged sample improves the statistical stability of the RF models but leaves the qualitative conclusions and the leading parameter rankings essentially unchanged. We therefore retain the 50-run results in the main text and restrict the physical interpretation to the highest-ranked parameters.

\end{appendix}

\end{document}